%% file: S2_FD_paper.tex
\documentclass[aps,prd,showpacs,amssymb,amsmath,amsfonts,superscriptaddress,
twocolumn,nofootinbib,floatfix,preprintnumbers,altaffilletter]{revtex4}
\usepackage{graphicx}
\usepackage{hyperref}
\usepackage{amssymb}
\usepackage{amsmath}
\usepackage{longtable}
\usepackage{rotating}
\usepackage{color}
\usepackage{epsfig}
\usepackage{epsf}
\usepackage{fancyhdr}

\def\min{\textrm{\mbox{\tiny{min}}}}
\def\max{\textrm{\mbox{\tiny{max}}}}

\newcommand{\ee}[1]{\!\times\!10^{#1}}

\newcommand{\F}{\mathcal{F}}
\def\ltsima{$\; \buildrel < \over \sim \;$}
\def\simlt{\lower.5ex\hbox{\ltsima}}
\def\gtsima{$\; \buildrel > \over \sim \;$}
\def\simgt{\lower.5ex\hbox{\gtsima}}
\input ourdefs.tex

\begin{document}
\pagestyle{fancy}
\rhead[]{}
\lhead[]{}
%

\title{Coherent searches for periodic gravitational waves from unknown isolated sources and Scorpius X-1: results from the second LIGO science run.}

\input{authorlist.tex}

%
\date{$$Revision: 1.420 $$ $$Date: 2006/05/31 17:36:27 $$}
\begin{abstract}
We carry out two searches for periodic gravitational waves using the most sensitive few hours of data from the second LIGO science run.
Both searches exploit fully coherent matched filtering and cover wide areas of parameter
space, an innovation over previous analyses which requires considerable
algorithm development and computational power.
The first search is targeted at isolated, previously unknown neutron stars, 
covers the entire sky in the frequency band 160--728.8~Hz, and assumes a
frequency derivative of less than $4\times10^{-10}$~Hz/s.
The second search targets the accreting neutron star in the low-mass X-ray
binary Scorpius~X-1 and covers the frequency bands 464--484~Hz and 604--624~Hz as well as the two relevant binary orbit parameters.
Due to the high computational cost of these searches we limit the analyses to
the most sensitive 10~hours and 6~hours of data respectively.
Both searches look for coincidences between the Livingston and Hanford
4-km interferometers.
Given the limited sensitivity and duration of the analyzed data set, we do not attempt deep follow-up studies. Rather we concentrate on demonstrating the data analysis method on a real data set and present our results as upper limits over large volumes of the parameter space. 
For isolated neutron stars our 95\% confidence upper limits on the
gravitational wave strain amplitude range from $6.6\times10^{-23}$ to
$1\times10^{-21}$ across the frequency band; 
For Scorpius~X-1 they range from $1.7\times10^{-22}$ to
$1.3\times10^{-21}$ across the two 20-Hz frequency bands.
The upper limits presented in this paper are the first broad-band wide parameter space upper limits on periodic
gravitational waves from coherent search techniques. The methods developed here lay the foundations for
upcoming hierarchical searches of more sensitive data which may detect astrophysical signals.
\end{abstract}
%
\maketitle

\section{Introduction}
\label{s:intro}

Rapidly rotating neutron stars are the most likely sources of
persistent gravitational radiation in the frequency band
$\approx 100\,\mathrm{Hz} - 1$ kHz. These objects may generate
continuous gravitational waves (GW) through a variety of mechanisms, including
nonaxisymmetric distortions of the star~\cite{Bildsten:1998ey,
Ushomirsky:2000ax, Cutler:2002nw, Melatos:2005ez, Owen:2005fn},
velocity perturbations in the star's fluid~\cite{Owen:1998xg,
Bildsten:1998ey,Andersson:1998qs}, and free
precession~\cite{Jones:2001yg, VanDenBroeck:2004wj}.  Regardless of
the specific mechanism, the emitted signal is a
quasi-periodic wave whose frequency changes slowly during
the observation time due to energy loss through gravitational wave emission, and possibly
other mechanisms. At an Earth-based detector the signal exhibits
amplitude and phase modulations due to the motion of the Earth 
with respect to the
source. The intrinsic gravitational wave 
amplitude is likely to be several orders of magnitude
smaller than the typical root-mean-square value of the detector noise,
hence detection can only be achieved by means of long integration
times, of the order of weeks to months.

Deep, wide parameter space searches for continuous gravitational wave 
signals are computationally bound. 
At fixed computational resources the optimal
sensitivity is achieved through 
hierarchical search schemes~\cite{BCCS98,BC00,cgk}. 
Such schemes alternate incoherent and
coherent search stages
in order to first efficiently identify statistically significant candidates and then follow them up with more sensitive, albeit computationally intensive, methods. Hierarchical search schemes 
have been investigated only theoretically, under the simplified
assumption of Gaussian and stationary instrumental noise; the 
computational costs have been estimated only on the basis of 
counts of floating point operations necessary to evaluate the relevant detection
statistic and have not taken into account additional costs 
coming {\em e.g.} from data input/output; computational savings 
obtainable through efficient dedicated numerical implementations 
have also been neglected. Furthermore, 
general theoretical investigations have not relied on the 
optimizations that can be introduced on the basis of
the specific area in parameter space at which a search is aimed.

In this paper we demonstrate and characterize the coherent stage of a hierarchical pipeline by carrying out two large parameter space coherent searches on 
data collected by LIGO during the second science run with the Livingston and Hanford 4-km interferometers. The second LIGO science run took place over the period 14 Feb. 2003 to 14 Apr. 2003. As we will show, this analysis requires careful tuning of 
a variety of search parameters and implementation choices, such as 
the tilings of the parameter space, the selection of the data,
and the choice of the coincidence windows, that are difficult to 
determine on purely theoretical grounds. 
This paper complements the study presented in~\cite{S2Hough} where  
we reported results obtained by applying an incoherent 
analysis method~\cite{hough04} to data taken during the same science run.
Furthermore, here we place upper limits on
regions of the parameter space that have never been explored before.

The search described in this paper has been the test-bench for the
core science analysis that the Einstein@home \cite{EatH} project is carrying out now. The development of 
analysis techniques such as the one described here, together with the 
computing power of Einstein@home in the context of a hierarchical search scheme, will allow the deepest searches for continuous gravitational waves.

In this paper the same basic pipeline
is applied to and tuned for two different searches: 
(i) for signals from isolated sources over the whole sky and the frequency band 160~Hz -- 728.8~Hz, and (ii) for a signal from the low-mass X-ray
binary Scorpius X-1 (Sco X-1) over orbital parameters and in the frequency bands 464~Hz -- 484~Hz and 
604~Hz -- 624~Hz. It is the first time that a coherent analysis is carried
out over such a wide frequency band, using data in coincidence and
(in one case) for a rotating neutron star in a binary system; the only 
other example of a somewhat similar analysis is 
an all-sky search over two days of data from the Explorer 
resonant detector over a 0.76~Hz band around 922~Hz~\cite{astone,astone:03,astone:05}.

The main scope of the paper is to illustrate an analysis method by
applying it to two different wide parameter spaces. In fact, based on
the typical noise performance of the detectors during the run, which
is shown in Fig.~\ref{f:s2strains}, and the amount of data that we
were able to process in $\approx 1$ month with our computational
resources (totalling about 800 CPUs over several Beowulf clusters) we
do not expect to detect gravitational waves. For isolated neutron
stars we estimate (see Section~\ref{s:targets} for details) that
statistically the strongest signal that we expect from an isolated
source is $\simlt 4\times 10^{-24}$ which is a factor $\simgt 20$
smaller than the dimmest signal that we would have been able to
observe with the present search.  For Scorpius X-1, the signal is
expected to have a strength of at most $\sim 3\times 10^{-26}$ and our
search is a factor $\sim 5000$ less sensitive.  The results of the
analyses confirm these expectations and we report upper limits for
both searches.

The paper is organized as follows: in Section~\ref{s:S2} we describe
the instrument configuration during the second science run and the
details of the data taking. In Section~\ref{s:targets} we review the current
astrophysical understanding of neutron stars as gravitational wave
sources, including a somewhat novel statistical argument that
the strength of the strongest such signal that we can expect to receive does
not exceed $h_0^{\max} \approx 4 \times 10^{-24}$. We also
detail and motivate the choice of parameter spaces explored in this
paper. In Section~\ref{s:sources} we review the signal model
and discuss the search area
considered here. In Section~\ref{s:analysis} we describe the 
analysis pipeline. In Section~\ref{s:results} we present and discuss the results of the analyses. In
Section~\ref{s:concl} we recapitulate the most relevant results in the wider context and
provide pointers for future work.

\section{Instruments and the second science run}
\label{s:S2}

Three detectors at two independent sites comprise the Laser
Interferometer Gravitational Wave Observatory, or LIGO.  Detector
commissioning has progressed since the fall of 1999, interleaved with
periods in which the observatory ran nearly
continuously for weeks or months, the so-called ``science runs''.  
The first science run (S1) was made in concert with the gravitational wave detector GEO600; 
results from the analysis of those data were presented
in~\cite{S1PulPaper,S1InspiralPaper,S1BurstPaper,S1StochPaper},
while the instrument status was detailed in~\cite{S1NIM}.  Significant
improvements in the strain sensitivity of the LIGO interferometers (an
order of magnitude over a broad band) culminated in the second science
run (S2), which took place from February 14 to April 14, 2003.
Details of the S2 run,
including detector improvements between S1 and S2 can be found in
\cite{LSC:05-S2TD}, Section IV of~\cite{S2-GRB}, and Section II of~\cite{S2Hough} and~\cite{S2-burst}. 

\begin{figure}
\centering
\includegraphics[width=7cm]{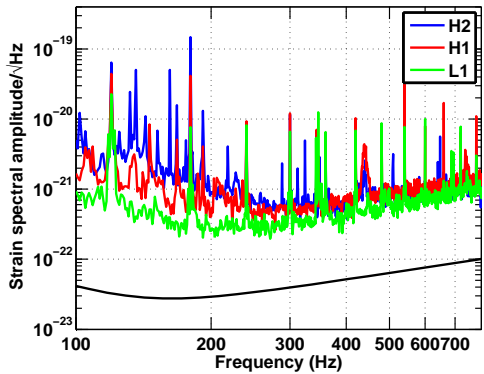}
\caption{
Typical one-sided amplitude spectral densities of detector noise during the
second science run, for the three LIGO instruments. 
The solid black line is the design sensitivity for the two 4-km
instruments L1 and H1.}
\label{f:s2strains}
\end{figure}

Each LIGO detector is a recycled Michelson interferometer with Fabry-Perot
arms, whose lengths are defined by suspended mirrors that double as
test masses.  Two detectors reside in the same vacuum in Hanford, WA, 
one (denoted H1) with 4-km armlength and one with 2-km (H2), while a
single 4-km counterpart (L1) exists in Livingston Parish, LA.  Differential
motions are sensed interferometrically, and the resultant sensitivity
is broadband (40~Hz -- 7~kHz), with spectral disturbances such
as 60~Hz power line harmonics evident in the noise spectrum (see
Fig.~\ref{f:s2strains}).  Optical resonance, or ``lock", in a given detector
is maintained by servo loops; lock may be interrupted by, for example,
seismic transients or poorly conditioned servos.  S2 duty cycles,
accounting for periods in which lock was broken and/or detectors were
known to be functioning not at the required level, 
were 74\% for H1, 58\% for H2, and 37\% for L1.
The two analyses described in this paper used a small
subset of the data from the two most sensitive instruments during S2, 
L1 and H1; the choice of the
segments considered for the analysis is detailed in 
Sec.~\ref{ss:dataset}. 

The strain signal at the interferometer output is reconstructed from the
error signal of the feedback loop which is used to control the
differential length of the arms of the instrument. Such a process---known
as calibration---involves the injection of continuous, constant amplitude
sinusoidal excitations into the end test mass control systems, which are
then monitored at the measurement error point. The calibration process
introduces uncertainties in the amplitude of the recorded signal that
were estimated to be $\simlt 11\%$ during S2~\cite{S2caldoc}. In addition,
during the run artificial pulsar-like signals were injected into the
data stream by  physically moving the mirrors of the Fabry-Perot cavity. 
Such ``hardware injections'' were used to validate the
data analysis pipeline and details are presented in 
Sec.~\ref{ss:hardinj}.

\section{Astrophysical sources}

\label{s:targets}

We review the physical mechanisms of periodic gravitational wave
emission and the target populations of the two searches described in
this paper.  We also compare the sensitivity of these searches to
likely source strengths.

\subsection{Emission mechanisms}

In the LIGO frequency band there are three predicted mechanisms for
producing periodic gravitational waves, all of which involve neutron
stars or similar compact objects: (1) nonaxisymmetric distortions of
the solid part of the star~\cite{Bildsten:1998ey, Ushomirsky:2000ax,
Cutler:2002nw, Melatos:2005ez, Owen:2005fn}, (2) unstable $r$-modes in the
fluid part of the star~\cite{Owen:1998xg, Bildsten:1998ey,
Andersson:1998qs}, and (3) free precession of the whole
star~\cite{Jones:2001yg, VanDenBroeck:2004wj}.

We begin with nonaxisymmetric distortions. These could not exist in a
perfect fluid star, but in realistic neutron stars such distortions
could be supported either by elastic stresses or by magnetic fields.  The
deformation is often expressed in terms of the ellipticity
\begin{equation}
\label{ellipticity definition}
\epsilon = \frac{I_{\rm xx} - I_{\rm yy} }{I_{\rm zz}},
\end{equation}
which is (up to a numerical factor of order unity) the $m=2$
quadrupole moment divided by the principal moment of inertia.  A
nonaxisymmetric neutron star rotating with frequency $\nu$ emits
periodic gravitational waves with amplitude
\begin{equation}
\label{h0 definition}
h_0 = \frac {4\pi^2G} {c^4} \frac {I_{\rm zz}f^2} {d} \epsilon,
\end{equation}
where $G$ is Newton's gravitational constant, $c$ is the speed of
light, $I_{\rm zz}$ is the principal moment of inertia of the object, $f$
(equal to $2\nu$) is the gravitational wave frequency, and $d$ is
the distance to the object.
Equation~(\ref{h0 definition}) gives the strain amplitude of a
gravitational wave from an optimally oriented source [see Eq.~(\ref{eq:ht})
below].

The ellipticity of neutron stars is highly uncertain.  The maximum ellipticity that can be supported by a neutron
star's crust is estimated to be~\cite{Ushomirsky:2000ax}
\begin{equation}
\label{NS ellipticity}
\epsilon_{\rm \max} \approx 5\times10^{-7} \,\left(\frac{\sigma}{10^{-2}}\right) \, ,
\end{equation}
where $\sigma$ is the breaking strain of the solid crust.  The
numerical coefficient in Eq.~(\ref{NS ellipticity}) is small mainly because
the shear modulus of the inner crust (which constitutes most of
the crust's mass) is small, in the sense that it is about $10^{-3}$
times the pressure.
Eq.~(\ref{NS ellipticity}) uses a fiducial breaking strain of
$10^{-2}$ since that is roughly the upper limit for the best
terrestrial alloys.  However, $\sigma$ could be as high as $10^{-1}$
for a perfect crystal with no defects~\cite{Kittel}, or several orders
of magnitude smaller for an amorphous solid or a crystal with many
defects.

Some exotic alternatives to standard neutron stars feature solid
cores, which could support considerably larger
ellipticities~\cite{Owen:2005fn}.
The most speculative and highest-ellipticity model is that of a solid
strange-quark star, for which
\begin{equation}
\label{SS ellipticity}
\epsilon_{\rm \max} \approx 4\times10^{-4} \,\left(\frac{\sigma}{10^{-2}}\right) \, .
\end{equation}
This much higher value of $\epsilon_{\rm \max}$ is mostly due to the
higher shear modulus, which for some strange star models can be almost
as large as the pressure.
Another (still speculative but more robust) model is the hybrid star, which
consists of a normal neutron star outside a solid core of mixed quark
and baryon matter, which may extend from the center to nearly the
bottom of the crust.  For hybrid stars,
\begin{equation}
\label{HS ellipticity}
\epsilon_{\rm \max} \approx 9\times10^{-6}\,\left(\frac{\sigma}{10^{-2}}\right),
\end{equation}
although this is highly dependent on the poorly known range of
densities occupied by the quark-baryon mixture.
Stars with charged meson condensates could also have solid cores with
overall ellipticities similar to those of hybrid stars.

Regardless of the maximum ellipticity supportable by shear stresses,
there is the separate problem of how to reach the maximum.
The crust of a young neutron star probably cracks as the neutron star
spins down, but it is unclear how long it takes for gravity to smooth
out the neutron star's shape.  Accreting neutron stars in binaries
have a natural way of reaching and maintaining the maximum
deformation, since the accretion
flow, guided by the neutron star's magnetic field, naturally produces
``hot spots'' on the surface, which can imprint themselves as lateral
temperature variations throughout the crust.
Through the temperature dependence of electron capture, these variations
can lead to ``hills'' in hotter areas which extend down to the dense inner
crust, and with a reasonable temperature variation the ellipticity might
reach the maximum elastic value~\cite{Bildsten:1998ey}.
The accreted material can also be held up in mountains on the surface by
the magnetic field itself:
The matter is a good conductor, and thus it crosses field lines relatively
slowly and can pile up in mountains larger than those supportable by
elasticity alone~\cite{Melatos:2005ez}.
Depending on the field configuration, accretion rate, and temperature, the
ellipticity from this mechanism could be up to $10^{-5}$ even for ordinary
neutron stars.

Strong internal magnetic fields are another possible cause of
ellipticity~\cite{Cutler:2002nw}.  Differential rotation immediately
after the core collapse in which a neutron star is formed can lead to
an internal magnetic field with a large toroidal part.
Dissipation tends to drive the symmetry axis of a toroidal
field toward the star's equator, which is the orientation that
maximizes the ellipticity.  The resulting ellipticity is
\begin{equation}
\label{epsilon_B}
\epsilon \approx \left\{ \begin{array}{ll}
1.6 \times 10^{-6}~\left(\frac{B}{10^{15}\,\mathrm{G}}\right)  & \quad B < 10^{15}\,\mathrm{G},  \\
1.6 \times 10^{-6}~\left(\frac{B}{10^{15}\,\mathrm{G}}\right)^2 & \quad B > 10^{15}\,\mathrm{G},
                \end{array}
\right.
\end{equation}
where $B$ is the root-mean-square value of the toroidal part of the field averaged
over the interior of the star. Note that this mechanism requires
that the external field be much smaller than the internal field, since
such strong external fields will spin a star out of the LIGO frequency
band on a very short timescale.

An alternative way of generating asymmetry is the $r$-modes, fluid
oscillations dominated by the Coriolis restoring force.
These modes
may be unstable to growth through gravitational radiation reaction
(the CFS instability) under astrophysically realistic conditions.
Rather than go into the many
details of the physics and  astrophysics, we refer the reader to a recent
review~\cite{Stergioulas:2003yp} of the literature and summarize here
only what is directly relevant to our search:
The $r$-modes
have been proposed as a source of gravitational waves from newborn
neutron stars~\cite{Owen:1998xg} and from rapidly accreting neutron
stars~\cite{Bildsten:1998ey, Andersson:1998qs}.
The CFS instability of the $r$-modes
in newborn neutron stars is probably not a good candidate for
detection because the emission is very short-lived, low amplitude, or
both.
Accreting neutron stars (or quark stars) are a better prospect for a
detection of $r$-mode gravitational radiation because the emission may be
long-lived with a duty cycle near unity~\cite{Wagoner:2002vr,
Andersson:2001ev}.

Finally we consider free precession, i.e.\ the wobble of a neutron star
whose symmetry axis does not coincide with its rotation axis.
A large-amplitude wobble would produce~\cite{Jones:2001yg}
\be
h_0 \sim 10^{-27} \left(\frac{\theta_{\rm w}}{0.1}\right) 
\left(\frac{1~{\rm kpc}}{d}\right)\left(\frac{\nu}{500~{\rm Hz}}\right)^2
\ee
where $\theta_{\rm w}$ is the wobble amplitude in radians.
Such wobble may be longer lived than previously
thought~\cite{VanDenBroeck:2004wj}, but the amplitude is still small enough
that such radiation is a target for second generation interferometers such as Advanced LIGO.

In light of our current understanding of emission mechanisms, the most
likely sources of detectable gravitational waves are isolated neutron
stars (through deformations) and accreting neutron stars in binaries
(through deformations or $r$-modes).

\subsection{Isolated neutron stars}

The target population of this search is isolated rotating compact stars that have not been observed electromagnetically.  Current models of stellar
evolution suggest that our Galaxy contains of order $10^9$ neutron
stars, while only of order $10^5$ are active pulsars.  Up to now only
about 1500 have been observed~\cite{ATNF}; there are numerous reasons
for this, including selection effects and the fact that many have
faint emission.  Therefore the target population is a large fraction
of the neutron stars in the Galaxy.

\subsubsection{Maximum expected signal amplitude at the Earth}
\label{ss:maximum_expected_amplitude}

Despite this large target population and the variety of GW emission
mechanisms that have been considered, one can make a robust argument,
based on energetics and statistics, that the amplitude of the
strongest gravitational-wave pulsar that one could reasonably hope to
detect on Earth is bounded by $h_0 \alt 4 \times 10^{-24}$. The
argument is a modification of an observation due to Blandford (which
was unpublished, but credited to him in Thorne's review in~\cite{300Yrs}).

The argument begins by assuming, very optimistically, that all neutron
stars in the Galaxy are born at very high spin rate and then spin down
principally due to gravitational wave emission.
For simplicity we shall also assume that all neutron stars follow the same
spin-down law $\dot{\nu}(\nu)$ or equivalently $\dot{f}(f)$, although this
turns out to be unnecessary to the conclusion.
It is helpful to express the spin-down law in terms of the spin-down
timescale
\be\label{eq:taugw}
\tau_{\rm gw}(f) \equiv \frac{f}{|4\dot f(f)|}.
\ee
For a neutron star with constant ellipticity, $\tau_{\rm gw}(f)$ is the time
for the gravitational wave frequency to drift down to $f$ from some
initial, much higher spin frequency---but the argument does not place any
requirements on the ellipticity or the emission mechanism.
A source's gravitational wave amplitude $h_0$ is then related to
$\tau_{\rm gw}(f)$ by
\be\label{eq:blandford1}
h_0(f) = d^{-1} \sqrt{ 5GI_{\rm zz} \over 8c^3 \tau_{\rm gw}(f) }.
\ee
Here we are assuming that
the star is not accreting, so that the angular momentum loss to GWs
causes the star to slow down. The case of accreting neutron stars is dealt
with separately, below.

We now consider the distribution of neutron stars in space and frequency.
Let $N(f)\Delta f$ be the number of Galactic neutron stars in the frequency
range $[f-\Delta f/2, f+\Delta f/2]$.
We assume that the birthrate has been roughly constant over about the last
$10^9$ years, so that this distribution has settled into a statistical
steady state: $dN(f)/dt = 0$.
Then $N(f)\dot f$ is just the neutron star birthrate $1/\tau_{\rm b}$, where
$\tau_{\rm b}$ may be as short as 30 years.
For simplicity, we model the spatial distribution of neutron stars in 
our Galaxy as that of a 
uniform cylindrical disk, with radius $R_G \approx 10\,$kpc and 
height $H \approx 600\,$pc.
Then the density $n(f)$ of neutron stars near the Earth, 
in the frequency
range $[f-\Delta f/2, f+\Delta f/2]$, is just
$ n(f)\Delta f = (\pi R_G^2 H)^{-1} N(f)\Delta f$.

Let $\hat N(f,r)$ be that portion of $N(f)$ due to neutron stars whose
distance from Earth is less than $r$.
For $H/2 \alt r \alt R_G$, we have
\bea\label{eq:r-dist}
{d\hat N(f,r) \over dr} &=& 2\pi r H n(f) \\ 
&=& 2 N(f)\,\frac{r}{R^2_G}
\eea
(and it drops off rapidly for $r \agt R_G$).
Changing variables from $r$ to $h_0$ using Eqs.~(\ref{eq:taugw}) and
(\ref{eq:blandford1}), we have
\be\label{eq:dnfdh0}
{d\hat N(f,h_0) \over dh_0} ={3\over 2}
\frac{5 G I_{zz}}{c^3 \tau_b R^2_G}f^{-1}h_0^{-3} \, .
\ee
Note that the dependence on the poorly known $\tau_{gw}(f)$ has dropped out
of this equation. This was the essence of Blandford's observation.

Now consider a search for GW pulsars in the frequency range
$[f_{\min},f_{\max}]$. Integrating the distribution in
Eq.~(\ref{eq:dnfdh0}) over this band, we obtain the distribution of
sources as a function of $h_0$:
\be\label{eq:dndh0}
{dN_{\rm band} \over d h_0} = 
\frac{5 G I_{\rm zz}}{c^3 \tau_b R^2_G}\,h_0^{-3}\,\ln \left(\frac{f_{\max}}{f_{\min}}\right) \, .
\ee 
The amplitude $h_0^{\max}$ of the strongest source is implicitly given by
\be\label{eq:half}
\int_{h_0^{\max}}^{\infty} {dN_{\rm band} \over d h_0} dh_0 = \frac{1}{2} \, .
\ee 
That is, even given our optimistic assumptions about the neutron star
population, there is only a fifty percent chance of seeing a source as
strong as $h_0^{\max}$.
The integral in Eq.~(\ref{eq:half}) is trivial; it yields
\be
h_0^{\max} = \left[\frac{5 G I_{\rm zz}}{c^3 \tau_b R^2_G}\ln \left(\frac{f_{\max}}{f_{\min}}\right)\right]^{1/2} \, .
\ee
Inserting
$\big[{\rm ln}(f_{\max}/f_{\min})\big]^{1/2} \approx 1$ (appropriate for 
a typical broadband search, as conducted here), and adopting as
fiducial values $I_{\rm zz} = 10^{45}{\rm g\,cm}^2$, 
$R_G = 10\,$kpc, and $\tau_b = 30\,$yr,   
we arrive at
\be
\label{eq:cutler1}
h_0^{\max} \approx 4 \times 10^{-24}\, .
\ee
This is what we aimed to show.

We now address the robustness of some assumptions in the argument.
First, the assumption of a universal spin-down function $\tau_{\rm gw}(f)$ was
unnecessary, since $\tau_{\rm gw}(f)$ disappeared from Eq.~(\ref{eq:dnfdh0})
and the subsequent equations that led to $h_0^{\max}$.
Had we divided neutron stars into different classes labelled by $i$ and
assigned each a spin-down law $\tau^i_{\rm gw}(f)$ and birthrate $1/\tau_b^i$,
each would have contributed its own term to $d\hat{N}/dh_0$ which would
have been independent of $\tau_{\rm gw}^i$ and the result for $h_0^{\max}$ would
have been the same.

Second, in using Eq.~(\ref{eq:r-dist}), we have in effect assumed that
the strongest source is in the distance range $H/2 \alt r \alt R_G$.
We cannot evade the upper limit by assuming that the neutron stars have
extremely long spin-down times (so that $r < H/2$) or extremely short ones
(so that the brightest is outside our Galaxy, $r > R_G$).
If the brightest sources are at $r < H/2$
(as happens if these sources have long spin-down times, $\tau_{\rm gw} \agt  \tau_b (2 R_G/H)^2$), then
 our estimate of $h_0^{\max}$
only decreases, because at short distances the spatial distribution of
neutron stars becomes approximately spherically symmetric instead of planar
and the right hand sides of Eqs.~(\ref{eq:r-dist}) and~(\ref{eq:dnfdh0})
are multiplied by a factor $2r/H < 1$.
On the other hand, if $\tau_{\rm gw}(f)$ (in the LIGO range) is much shorter 
than $\tau_{\rm b}$, then the probability that such an object exists inside
our Galaxy is $ \ll 1$.
For example, a neutron star with $\tau_{\rm gw}(f) = 3\,$ yr located at 
$r = 10\,$kpc would have
$h_0 = 4.14 \times 10^{-24}$, but the probability of currently 
having a neutron
star with this (or shorter) $\tau_{\rm gw}$ 
is only $\tau_{\rm gw}/\tau_{\rm b} \lesssim 1/10$.

Third, we have implicitly assumed that each neutron star spins down only
once. In fact, it is clear that some stars in binaries are ``recycled''
to higher spins by accretion, and then spin down again.
This effectively increases the neutron star birth rate (since for
our purposes the recycled stars are born twice), but since the fraction of
stars recycled is very small the increase in the effective birth rate is also small.

\subsubsection{Expected sensitivity of the S2 search}

Typical noise levels of LIGO during the
	S2 run were approximately $[S_h(f)]^{1/2} \approx 3 \times 10^{-22}~{\rm
	Hz}^{-1/2}$, where $S_h$ is the strain noise power spectral density, as shown in Fig.~\ref{f:s2strains}.
Even for a {\it known} GW pulsar with an average sky position, inclination angle, polarization,
and frequency, the amplitude of the signal that we could detect in Gaussian
stationary noise with a false alarm rate of 1\% and a false dismissal rate
of 10\% is~\cite{S1PulPaper} 
\be
\label{uls}
\langle h_0(f) \rangle = 11.4 \sqrt{\frac{S_h(f)}{T_\mathrm{obs}}},
\ee
where $T_\mathrm{obs}$ is the integration time and the angled brackets indicate an
average source. In all-sky searches for pulsars with {\it unknown} parameters, the
amplitude $h_0$ must be several times greater than this to rise
convincingly above the background.  Therefore, in $T_{obs} = 10$ hours of S2
data, signals with amplitude $h_0$ below about $10^{-22}$
would not be detectable. This is a factor $\approx 25$ greater than the
$h_0^{\max}$ of Eq.~(\ref{eq:cutler1}), so our S2 analysis is unlikely to
be sensitive enough to reveal previously unknown pulsars.

The sensitivity of our search is further restricted by the template bank, 
which does not include the effects of signal spin-down for reasons of computational cost.
Phase mismatch between the signal and matched filter causes the detection
statistic (see Sec.~\ref{ss:Fstat}) to decrease rapidly for GW frequency
derivatives $\dot f$ that exceed
\be
\mathrm{max}[\dot{f}] = \frac{1}{2} T_{obs}^{-2} = 4 \times 10^{-10}\,\left(\frac{T_\mathrm{obs}}{10\,\mathrm{h}}\right)^{-2} 
{\rm Hz}\,{\rm s}^{-1} \, .
\label{eq:maxSpindown}
\ee
Assuming that all of the spin-down of a neutron star is due to gravitational
waves (from a mass quadrupole deformation), our search is restricted to pulsars
with ellipticity $\epsilon$ less than
\be\label{eq:esd}
\epsilon_{\rm sd} = \bigg(\frac{5  c^5 \, \mathrm{max}[\dot{f}] }{32 \pi^4  G\,
I_{\rm zz}\, f^5}\bigg)^{1/2} \, .
\ee
This limit, derived from combining the quadrupole formula for GW luminosity
\begin{equation}
{dE \over dt} = {1\over10} {G\over c^5} (2\pi f)^6 I_{\rm zz}^2 \epsilon^2
\end{equation}
(the first factor is 1/10 instead of 1/5 due to time averaging of the signal) 
with the kinetic energy of rotation
\begin{equation}
\label{kinetic energy}
E = \frac{1}{2}\,\pi^2 f^2 I_{\rm zz} \, ,
\end{equation}
(assuming $f=2\nu$) takes the numerical value
\be
\epsilon_{\rm sd} = 9.6 \times 10^{-6} \bigg(\frac{10^{45} {\rm
g\,cm}^2}{I_{\rm zz}}\bigg)^{1/2}\bigg(\frac{ 300~{\rm Hz}}{f}\bigg)^{5/2}
\ee
for our maximum $\dot f$.

\begin{figure}[t]
\centering
\includegraphics[width=8cm]{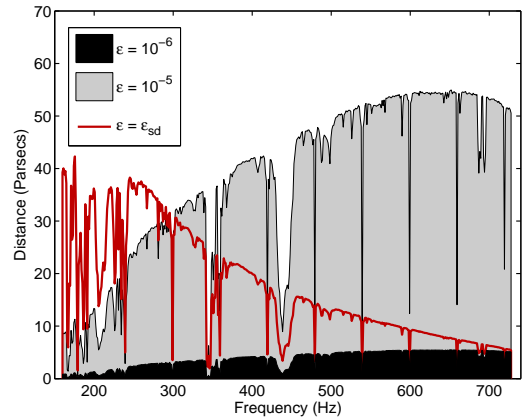}
\caption{
Effective average range (defined in the text) of our search as a function
of frequency for three ellipticities:
$10^{-6}$ (maximum for a normal neutron star), $10^{-5}$ (maximum for a
more optimistic object), and $\epsilon_{sd}$, the spin-down limit defined in
the text. Note that for sources above 300 Hz the reach of the search is limited by the maximum spin-down value of a signal that may be detected without loss of sensitivity.}
\label{f:sky_expected}
\end{figure}

The curves in Fig.~\ref{f:sky_expected} are obtained by combining Eqs.~(\ref{h0 definition}) and (\ref{uls})\footnote{
Note that the value of $h_0$ derived from Eq.~\ref{uls} yields a value of the detection statistic $2\F$ for an average source as seen with a detector at S2 sensitivity and over an observation time of 10 hours, of about $21$, which is extremely close to the value of $20$ which is used in this analysis as threshold for registering candidate events. Thus combining Eqs.~(\ref{h0 definition}) and (\ref{uls}) determines the smallest amplitude that our search pipeline could detect (corresponding to a signal just at the threshold), provided appropriate follow-up studies of the registered events ensued.
} and solving for the distance $d$ for different values of the ellipticity, using an average value for noise in the detectors during the S2 run. The curves show the average distance, in the sense of the definition~(\ref{uls}), at which a source may be detected.
 
The dark gray region shows that a GW pulsar with $\epsilon = 10^{-6}$ could
be detected by this search only if it were very close, less than $\sim 5$ parsecs away.
The light gray region shows the distance at which a GW pulsar with $\epsilon = 10^{-5}$ could
be detected if templates with sufficiently large spin-down values were searched. However,
{\it this} search can detect such pulsars 
only below
$300~{\rm Hz}$, because above $300~{\rm Hz}$ a GW pulsar with $\epsilon = 10^{-5}$
spins down too fast to be detected with the no-spin-down templates used.
The thick line indicates the distance limit for the
(frequency-dependent) maximum value of epsilon that could be detected
with the templates used in this search.
At certain frequencies below $300~{\rm Hz}$, a GW pulsar could be seen
somewhat farther away 
than $30~{\rm pc}$, but only if it has $\epsilon > 10^{-5}$. 
Although $\epsilon_{sd}$ and the
corresponding curve were derived assuming a quadrupolar deformation as the emission mechanism,
the results would be similar for other mechanisms. Equation~(\ref{kinetic energy}) includes an
implicit factor $f^2/(2\nu)^2$, which results in $\epsilon_{sd}$ and the corresponding range
(for a fixed GW frequency $f$) being multiplied by $f/(2\nu)$, which is $1/2$ for free precession
and about $2/3$ for $r$-modes. Even for a source with optimum inclination angle and polarization,
the range increases only by a factor $\approx 2$. The distance to the closest known pulsar
in the LIGO frequency band, PSR~J0437$-$4715, is about 140~pc~\cite{ATNF}. The distance to the closest
known neutron star, RX~J185635$-$3754, is about 120~pc~\cite{Walter:2002uq}. Therefore this search
would be sensitive only to nearby previously unknown objects.

While we have argued that a detection would be very unlikely, it
should be recalled that Eq.~(\ref{eq:cutler1}) was based on a
statistical argument.
It is always possible that there is a GW-bright
neutron star that is much closer to us than would be expected from a
random distribution of supernovae (for example due to recent star
formation in the Gould belt as considered in~\cite{Palomba:2005na}).  It is also possible that a ``blind''
search of the sort performed here could discover some previously
unknown class of compact objects not born in supernovae.

More importantly, future searches for previously undiscovered rotating
neutron stars using the methods presented here will be much more
sensitive.  The goal of initial LIGO is to take a year of data at
design sensitivity. With respect to S2, this is a factor 10 improvement in the amplitude strain noise at most frequencies.
The greater length of the
data set will also increase the sensitivity to pulsars by a factor of
a few (the precise value depends on the combination of coherent and
incoherent analysis methods used).  The net result is that initial
LIGO will have $h_0$ reduced from the S2 value by a factor of 30 or more
to a value comparable to $ h_0^{\max} \approx 4 \times 10^{-24}$ of Eq.~(\ref{eq:cutler1}).

\subsection{Accreting neutron stars}

\subsubsection{Maximum expected signal amplitude at Earth}

The robust upper limit in Eq.~(\ref{eq:cutler1}) refers only to
non-accreting neutron stars, since energy conservation plays a crucial
role.  If accretion replenishes the star's angular momentum, a
different but equally robust argument (i.e., practically independent
of the details of the emission mechanism) can be made regarding the
maximum strain $h_0^{\max}$ at the Earth. In this case
$h_0^{\max}$ is set by the X-ray luminosity of the brightest X-ray
source.

The basic idea is that if the energy (or angular momentum) lost to GWs
is replenished by accretion, then the strongest GW emitters are those
accreting at the highest rate, near the Eddington limit.  Such systems
exist: the low-mass X-ray binaries (LMXBs), so-called since the
accreted material is tidally stripped from a low-mass companion star.
The accreted gas hitting
the surface of the neutron star is heated to $10^8$~K and emits
X-rays.  As noted several times over the
years~\cite{Papaloizou:1978, Wagoner:1984pv, Bildsten:1998ey},
if one assumes that spin-down from GW emission is in equilibrium with
accretion torque, then the GW amplitude $h_0$ is directly
related to the X-ray luminosity:
\be
h_0 \approx 5 \times 10^{-27}\bigg(\frac{300 ~{\rm Hz}}{\nu}\bigg)^{1/2}
\bigg(\frac{F_{\rm{x}}}{10^{-8} {\,\rm erg\,cm}^{-2}{\,\rm s}^{-1} }\bigg)^{1/2} \, ,
\ee
where $F_{\rm{x}}$ is the X-ray flux.
In the 1970s when this connection was first proposed, there was no
observational support for the idea that the LMXBs are strong GW
emitters.  But the spin frequencies of many LMXBs are now known, and
most are observed to cluster in a fairly narrow range of spin frequencies
270~Hz $\alt \nu \alt$ 620~Hz~\cite{Chakrabarty:2003kt}.  Since most
neutron stars will have accreted enough matter to spin them up to near
their theoretical maximum spin frequencies, estimated at $\sim
1400$~Hz, the observed spin distribution is hard to explain without
some competing mechanism, such as gravitational radiation, to halt the
spin-up.  Since the gravitational torque scales as $\nu^5$,
gravitational radiation is also a natural explanation for why the spin
frequencies occupy a rather narrow window: a factor $32$ difference in
accretion rate leads to only a factor $2$ difference in equilibrium
spin rate~\cite{Bildsten:1998ey}.

If the above argument holds, then the accreting neutron star brightest in
X-rays is also the brightest in gravitational waves. Sco X-1, which
was the first extrasolar X-ray source discovered, is the
strongest persistent X-ray source in the sky.  Assuming equilibrium
between GWs and accretion, the gravitational wave strain of Sco X-1 at
the Earth is
\be\label{scolimit}
h_0 \approx  3 \times 10^{-26} \left(\frac{540~{\rm Hz}}{f}\right)^{1/2} \, ,
\ee
which should be detectable by second generation interferometers. 
The gravitational wave strains from other accreting neutron stars are
expected to be lower.

\subsubsection{Expected sensitivity of S2 search for Sco X-1}

The orbital parameters of Sco X-1 are poorly constrained by present 
(mainly optical) observations and large uncertainties affect the determination of
the rotation frequency of the source (details are provided in Section~\ref{s:s:sco}). 
The immediate implication for a coherent search for gravitational waves from such a 
neutron star is that a very large number of discrete templates are required to cover
the relevant parameter space, which in turn dramatically increases the computational 
costs~\cite{Dhurandhar:2000sd}. The optimal
sensitivity that can be achieved with a coherent search is therefore set primarily by
the length of the data set that one can afford to process (with fixed computational
resources) and the spectral density of the detector noise.
As we discuss in Section~\ref{s:s:sco}, the maximum span of the observation time set by 
the computational burden of the Sco X-1 pipeline (approximately one week on $\approx 100$ 
CPUs) limits the observation span to 6 hours. 

\begin{figure}
\centering \includegraphics[width=8.3cm]{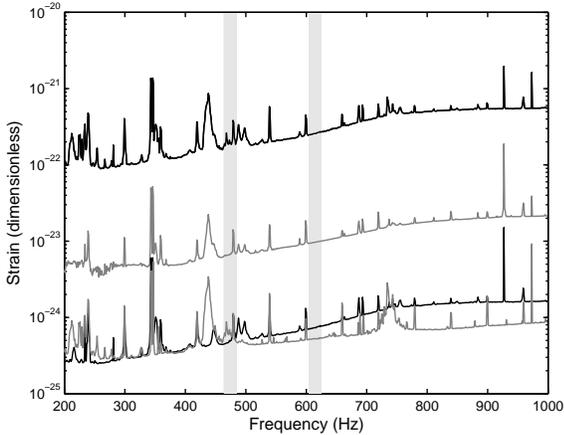} 
\caption{Here we show the expected upper limit sensitivity of the S2 Sco X-1 search.
The upper black curve represents the expected sensitivity of the
S2 analysis based on an optimally selected 6 hr dataset (chosen specifically
for our search band).
The gray curve (second from the top) shows the sensitivity in 
the hypothetical case in which \emph{all}
of the Sco X-1 system parameters are known exactly making Sco X-1 a 
single filter target and the entire S2 data set is analyzed. 
Both curves are based on a $95\%$ confidence upper limit.  
The remaining curves represent $\sqrt{{S_h(f)}/{T_{\rm obs}}}$ for
L1 (black) and H1 (gray); $S_\mathrm{h}(f)$ is the typical noise spectral
density that characterizes the L1 and H1 data, and $T_{\rm obs}$ 
is the actual observation time (taking into
account the duty cycle, which is different for L1 and H1) for each instrument.
}
\label{f:ExpSens_sco}
\end{figure}

The overall sensitivity of the search that we are describing is determined by each stage of the
pipeline, which we describe in detail in Section~\ref{ss:pipeline}. Assuming that the noise in the
instrument can be described as a Gaussian and stationary process (an assumption which however breaks
down in some frequency regions and/or for portions of the observation time) we can statistically 
model the effects of each step of the analysis and estimate the sensitivity of the search.
The results of such modelling through the use of Monte Carlo simulations are shown in 
Fig~\ref{f:ExpSens_sco} where we give the expected upper limit sensitivity of the search implemented for the
analysis.  We contrast this with the hypothetical case in which the Sco X-1 parameters are known 
perfectly making it a single filter target for the whole duration of the S2 run.  The dramatic 
difference (of at least an order of magnitude) between the estimated sensitivity curves of these two 
scenarios is primarily due to the large parameter space we have to search. This has two consequences, 
which contribute to degrading the sensitivity of the analysis: (i) we are computationally limited by 
the vast number of templates that we must search and therefore must reduce the observation to a 
subsection of the S2 data, and (ii) sampling a large number of independent locations increases the 
probability that noise alone will produce a high value of the detection statistic.

We note that the S2 
Sco X-1 analysis (see Section~\ref{s:s:sco}) is a factor 
of $\approx 5000$ less sensitive than the characteristic amplitude given in Eq.~(\ref{scolimit}).
In the hypothetical case in which Sco X-1 is a single filter target and 
we are able to analyze the entirety of S2 data, then we are still a factor 
$\sim 100$ away.  However, as mentioned in the introduction, the search reported in this paper will be one of the stages
of a more sensitive ``hierarchical pipeline'' that will allow us to achieve quasi-optimal
sensitivity with fixed computational resources.

\section{Signal Model}
\label{s:sources}

\subsection{The signal at the detector}
\label{ss:signal_at_the_detector}

We consider a rotating neutron star with equatorial coordinates
$\alpha$ (right ascension) and $\delta$ (declination). Gravitational
waves propagate in the direction $\hat{{k}}$ and the star spins
around an axis whose direction, assumed to be constant, is identified
by the unit vector $\hat{{s}}$.

The strain $h(t)$ recorded at the interferometer output at detector time $t$ is:
\bea\label{eq:ht}
h(t) & = & h_0\,\Bigl[\frac{1}{2}\,\left(1 + \cos^{2}\iota\right)
F_{+}(t;\alpha,\delta,\psi)~\cos\Phi(t)
\nonumber
\\
&&  \quad\quad+ \cos\iota\,F_{\times} (t;\alpha,\delta,\psi)\,\sin  \Phi(t)
\Bigr]\,,
\label{e:h}
\eea
where $\psi$ is the polarization angle, defined as $\tan \psi = [(\hat{{s}}
\cdot \hat{{k}})\,(\hat{{z}} \cdot \hat{{k}}) -
(\hat{{s}} \cdot \hat{{k}})]/\hat{{k}} \cdot
(\hat{{s}} \times \hat{{z}})$, $\hat{z}$ is the direction to the north 
celestial pole, and 
$\cos\iota = \hat{{k}} \cdot \hat{{s}}$. Gravitational wave 
laser interferometers are all-sky
monitors with a response that depends on the source location in the sky and
the wave polarization: this is encoded in the (time dependent) antenna
beam patterns $F_{+,\times}(t;\alpha,\delta,\psi)$. The term  $\Phi(t)$ 
in Eq.~(\ref{e:h}) represents the phase of the received gravitational signal.

The analysis challenge to detect weak quasi-periodic continuous gravitational waves stems from the Doppler shift of the gravitational phase $\Phi(t)$ due to the relative motion between the detector and the source. It is convenient
to introduce the following times: $t$, the time measured at the
detector; $T$, the solar-system-barycenter (SSB) coordinate time; and
$t_p$, the proper time in the rest frame of the pulsar\footnote{Notice that 
our notation for the three different times is different from the established
conventions adopted in the radio pulsar community, {\em e.g.}~\cite{TW:89}.}.

The timing model that links the detector time $t$ to
the coordinate time $T$ at the SSB is:
\beq
T = t + \frac{\vec{r} \cdot \hat{{n}}}{c} +
\Delta_{E\odot} - \Delta_{S\odot}\,,
\label{e:time}
\eeq
where $\vec{r}$ is a (time-dependent) vector from the SSB to the
detector at the time of the observations, $\hat{{n}}$ is a unit
vector towards the pulsar (it identifies the source position in the
sky) and $\Delta_{E\odot}$ and $\Delta_{S\odot}$ are the solar system
Einstein and Shapiro time delays, respectively~\cite{TW:89}. For an
isolated neutron star $t_p$ and $T$ are equivalent up to an additive
constant. If the source is in a binary system, as it is the case for
Sco X-1, significant additional accelerations are involved, and a
further transformation is required to relate the proper time $t_p$
to the detector time $t$. 
Following~\cite{TW:89}, we have:
\beq
T - T_0 = t_p + \Delta_R + \Delta_E + \Delta_S
\eeq
where $\Delta_R$, the Roemer time delay, is analogous to the solar
system term $(\vec{r} \cdot \hat{{n}})/c$; $\Delta_E$ and
$\Delta_S$ are the orbital Einstein and Shapiro time delay, analogous
to $\Delta_{E\odot}$ and $\Delta_{S\odot}$; and $T_0$ is an arbitrary (constant)
reference epoch. For the case of Sco X-1, we consider a circular orbit
for the analysis (cf Section~\ref{s:s:sco} for more details)
 and therefore set
$\Delta_E = 0$. Furthermore, the binary is non-relativistic and from 
the source parameters we estimate $\Delta_S < 3\,\mu\mathrm{s}$ which is
negligible. For a circular orbit, the Roemer time delay is
simply given by
\be
\Delta_R = \frac{a_{\rm{p}}}{c}\,\sin(u + \omega)
\label{e:roemer}
\ee
where $a_\mathrm{p}$ is the radius of the neutron star orbit projected on the line of sight, $\omega$ the argument of the periapsis and $u$ the
so-called eccentric anomaly; for the case of a circular orbit $u =
2\pi (t_p - t_{p,0})/P$, where $P$ is the period of the binary and $t_{p,0}$ is a constant reference time, conventionally referred to as the ``time of periapse passage''.

In this paper we consider gravitational waves whose {\em intrinsic} frequency
drift is negligible over the integration time of the searches (details are
provided in the next section), both for the
blind analysis of unknown isolated neutron stars and Sco X-1. The phase model is simplest in this case and given by:
\be
\Phi(t_p) = 2\pi f_0\, t_p + \Phi_0 ,
\label{e:phase}
\ee
where $\Phi_0$ is an overall constant phase term and $f_0$ is the frequency of the gravitational wave at the reference time. 

\subsection{Parameter space of the search}
\label{ss:parameter_space}

The analysis approach presented in this paper is used for two
different searches with different search parameters. Both searches require exploring a three dimensional
parameter space, made up of two ``position parameters'' (whose nature is different
for the two searches) and the unknown frequency of the signal. 
For the all-sky blind analysis aimed
at unknown isolated neutron stars one needs to Doppler correct the phase of
the signal for any given point in the sky, based on the angular 
resolution of the instrument over the observation time, and so a search
is performed on the sky coordinates $\alpha$ and $\delta$. For the
Sco X-1 analysis, the sky location of the system is known, however the system is in a binary orbit with poorly measured orbital elements; thus, one needs to search over a range
of orbital parameter values. The frequency search parameter 
is for both searches the $f_0$ defined by Eq.~(\ref{e:phase}), where the reference time 
has been chosen to be the time-stamp of the first sample of the data set. 
The frequency band over which the two analyses are carried out is also 
different, and the choice is determined by astrophysical and practical 
reasons.  As explained in Sec.~\ref{ss:dataset}, the data set in H1 does not coincide in time with the L1 data set for either of the analyses. Consequently a signal with a non-zero frequency derivative would appear at a different frequency template in each data set. However, for the maximum spin-down rates considered in this search, and given the time lag between the two data sets, the maximum difference between the search frequencies happens for the isolated objects search and amounts to $0.5$ mHz. We will see that the frequency coincidence window is much larger than this and that when we discuss spectral features in the noise of the data and locate them based on template-triggers at a frequency $f_0$, the spectral resolution is never finer than $0.5$ mHz. So for the practical purposes of the present discussion we can neglect this difference and will often refer to $f_0$ generically as the signal's frequency. 

\subsubsection{Isolated neutron stars}

The analysis for isolated neutron stars covers the entire sky and we have
restricted the search to the frequency
range $160$--$728.8$~Hz. The low frequency end of the band was 
chosen because the depth of our search degrades significantly below 160~Hz,
see Fig.~\ref{f:sky_expected}. The choice of the high frequency limit at 728.8 
is primarily determined by the computational burden of the analysis, which
scales as the square of the maximum frequency that is searched for.

In order to keep the computational costs at a reasonable 
level ($< 1$ month on $\simlt$ 800 CPUs), no
explicit search over spin-down parameters was carried out. The length of
the data set that is analyzed is approximately 10 hours, thus no loss of
sensitivity is incurred for sources with spin-down rates smaller than 
$4\times 10^{-10}~{\rm Hz}\,\mathrm{s}^{-1}$; see Eq.~(\ref{eq:maxSpindown}). This is 
a fairly high spin-down rate compared to those measured in isolated
radio pulsars; however it does constrain the sensitivity for sources 
above 300 Hz, as can be seen from Fig.~\ref{f:sky_expected}.

\subsubsection{Sco X-1}
\label{s:s:sco}

Sco X-1 is  a neutron star in a $18.9$ h orbit around a low mass $(\sim 0.42\,M_{\odot})$ companion 
at a distance $r = 2.8 \pm 0.3$ kpc from Earth. In this section we review our present knowledge 
of the source parameters that are relevant for gravitational wave observations. Table~\ref{t:Sco-param} 
contains a summary of the parameters and the associated uncertainties that define the search 
area. In what follows we will assume the 
observation time to be 6 hours. This is approximately what was adopted for the analysis presented 
in this paper. We will justify this choice at the end of the section. 

\begin{table*}
\begin{center}
\begin{tabular}{lcc}
\hline
\hline
right ascension                       &  $\alpha$     &  $16\mathrm{h}\,19\mathrm{m}\,55.0850\mathrm{s}$  \\
declination                           &  $\delta$     &  $-15^{o}\,38'\,24.9''$        \\
proper motion (east-west direction)    & $\mu_{x}$          &  $-0.00688 \pm 0.00007\,\mathrm{arcsec}\,\mathrm{yr}^{-1}$        \\
proper motion  (north-south direction)   & $\mu_{y}$          &  $0.01202 \pm 0.00016\,\mathrm{arcsec}\,\mathrm{yr}^{-1}$                   \\
distance                              & $d$           & $2.8 \pm 0.3$ kpc \\
orbital period                        &  $P$                   & $68023.84 \pm 0.08\, \mathrm{sec}$                    \\
time of periapse passage              & $\bar{T}$                 & $731163327 \pm 299\,\mathrm{sec}$   \\ 
projected semi-major axis             &  $a_{\rm p}$        & $1.44$ $\pm 0.18\, \mathrm{sec}$ \\
eccentricity                          &  $e$                   &   $ < 3\times 10^{-3}$           \\
QPOs frequency separation            &  \multicolumn{2}{c}{ $237 \pm 5\,\mathrm{Hz} \le \Delta\nu_{\rm QPO} \le 307 \pm 5\,\mathrm{Hz}$} \\ 
\hline
\end{tabular}
\end{center}
\caption{The parameters of the low-mass X-ray binary Scorpius X-1.  The quoted measurement errors are  all 1-$\sigma$. 
  We refer the reader to the text for details and references.}
\label{t:Sco-param}
\end{table*}

The most accurate determination of the Sco X-1 sky position comes from Very Long Baseline Array (VLBA) observations~\cite{vlba,Bradshawetal:99} and is reported in Table~\ref{t:Sco-param}. The overall error on the source location is $\sim 0.5$ arcsec, which is significantly smaller than the $\sim 100$ arcsec sky resolution associated with a 6 hour GW search. Hence we assume the position of Sco X-1 ({\em i.e.}\ the barycenter of the binary system) to be exactly known and we ``point'' (in software) at that region of the sky.

Three parameters describe the circular orbit of a star in a binary system: 
the orbital period ($P$), the projection of the 
semi-major axis of the orbit $a_{\rm p}$ (which for $e = 0$ corresponds 
to the projected radius of the orbit) and the location of the star on the 
orbit at some given reference time.  For eccentric orbits this is usually
parameterized by the time of periapse passage (or time of periastron).  In 
the case of a circular orbit we define the \emph{orbital phase reference time}
 $\bar{T}$ as the time at which the star crosses the ascending 
node as measured by an observer at the SSB.  This is equivalent to setting 
the argument of periapse (the angle between the ascending node and the 
direction of periapsis) to zero.

In the case of Sco X-1, $P$ is by far the most 
accurately determined parameter~\cite{GWL:75}, and 
over a 6 hour search it can be considered known because the 
loss of signal-to-noise ratio (SNR) introduced by matching two templates with any 
value of $P$ in the range of Table~\ref{t:Sco-param} is negligible. 
$P$ becomes a search parameter, requiring multiple filters, 
only for coherent integration times $\simgt 10^6$~s. The major 
orbital parameters with the largest uncertainties are the projected semi-major 
axis of the orbit along the line of sight, $a_{\rm p}$, and the orbital phase reference time. 
The large uncertainty on $a_{\rm p}$ is primarily due to the poor determination 
of the orbital velocity ($ 40\pm 5\,\mathrm{km}
\,s^{-1}$ \cite{SC:02}). The uncertainty on the 
orbital phase reference time is due to the difficulty in locating the Sco X-1 
low-mass companion on the orbit. The search therefore requires a discrete grid of filters in the $(a_{\rm p},\bar{T})$ space.

We assume that Sco X-1 is in a 
circular orbit, which is what one expects for a semi-detached binary system 
and which is consistent with the best fits of the orbital parameters~\cite{Steeghs_priv}. 
However, orbital fits for 
models with $e\ne 0$ were clearly dominated 
by the noise introduced by the geometry of the Roche lobe~\cite{Steeghs_priv}. 
Over an integration time of $\sim 6$ h, the eccentricity needs to be 
smaller than $\sim 10^{-4}$ in order for the detection statistic $\F$ 
to be affected less than $1\%$; for $e \approx 10^{-3}$, losses of the 
order of $10\%$ are expected and  
are consistent with the results presented later in the paper. Unfortunately 
current observations are not able to constrain $e$ to such levels of accuracy: 
in this paper we adopt the strategy of analyzing the data under the assumption 
$e = 0$, and we quantify (for a smaller set of the parameter space) the 
efficiency of the pipeline in searching for gravitational waves emitted by 
a binary with non zero eccentricity; in other words, we quote upper limits for different 
values of the eccentricity that are obtained with non-optimal search templates.

The last parameter we need to search for is the frequency of the
gravitational radiation $f$. The rotation frequency $\nu$ of Sco X-1
is inferred from the difference of the frequency of the kHz
quasi periodic oscillations (QPOs). Unfortunately this frequency difference is not constant, and
over a $4$ day observation~\cite{vdk97} has shown a very pronounced
drift between $237 \pm 5\,\mathrm{Hz}$ to $307\pm 5\,\mathrm{Hz}$,
where the errors should be interpreted as the $1 \sigma$
values~\cite{vdk00}.  This drifting of QPO frequency separation was
found to be positively correlated to the inferred mass accretion
rate. It is also important to stress that there is a still unresolved
controversy as to whether the adopted model that links $\nu$ to the
difference of the frequency of the kHz QPOs is indeed the correct one,
and if it is valid for all the observed LMXBs. Moreover, the
gravitational wave frequency $f$ is related to $\nu$ in a different
way, depending on the model that is considered: $f = 2\nu$ if one
considers nonaxisymmetric distortions and $f = (4/3) \nu$ if one
considers unstable $r$-modes.  It is therefore clear that a search for
gravitational waves from Sco X-1 should assume that the frequency is
essentially unknown and the whole LIGO sensitivity band (say from
$\approx 100$ Hz to $\approx 1$~kHz) should be considered.  Because of
the heavy computational burden, such a search requires a different
approach (this search is currently in progress). For the analysis
presented in this paper, we have decided to confine the search to GWs
emitted by nonaxisymmetric distortions ($f = 2\nu$) {\em and} to
constrain the frequency band to the two 20~Hz wide bands (464--484~Hz and 604--624~Hz) that bound
the range of the drift of $\nu$, according to currently accepted
models for the kHz QPOs.
The total computational time for the analysis can be split into two parts: (i) the {\em search time} $T_{\rm search}$ needed to search 
the data and, if no signal is detected, (ii) the {\em upper limit time} $T_{\rm inj}$ required to repeatedly inject 
and search for artificially 
generated signals for the purposes of setting the upper limits.  Let $T_{\rm span}$ be the span
of the data set which is analyzed, that is the difference between the time stamps of the 
first and last data point in the time series. Let  $T_{\rm obs}$ be the effective duration of the data set containing non-zero data points. The definitions imply $T_{\rm obs} \le T_{\rm span}$, and for data with no gaps $T_{\rm obs} = T_{\rm span}$. For 
a search confined to a period (sufficiently) shorter than the orbital period of the source, the two 
computational times are:

\begin{eqnarray}
  T_{\rm search}&\approx&90~\mathrm{hrs}\times\Big(\frac{\Delta f}{40~
  \mathrm{Hz}}\Big)\,\Big(\frac{\Delta \bar{T}}{598~
  \mathrm{s}}\Big)\,
\Big(\frac{0.1}{\mu}\Big)^{{3}/{2}}\,\Big(\frac{100}{N_{\rm cpu}}\Big) \nonumber
\\
  &\times&\frac{1}{2}\,\sum_{L1,H1}\Big(\frac{T_{\rm span}}{6~\mathrm{hrs}}\Big)^{7}\,
\Big(\frac{T_{\rm obs}}{T_{\rm span}}\Big)
\label{eq:tsearch}
\end{eqnarray} 
\begin{eqnarray}
 T_{\rm inj}&\approx&55 ~\mathrm{hrs}\times\Big(\frac{N_{\rm
 trials}}{5000}\Big)\,\Big(\frac{N_{h_{0}}}{20}\Big)\,\Big(\frac{100}{N_{\rm cpu}}\Big)
\nonumber\\
  &\times& \frac{1}{2}\,\sum_{L1,H1}\Big(\frac{T_{\rm span}}{6~\mathrm{hrs}}\Big)
\Big(\frac{T_{\rm obs}}{T_{\rm span}}\Big)
 \label{eq:tinject}
\end{eqnarray}
where $\Delta f$ is the search frequency band, $\Delta \bar{T}$ the search range for the time of periapse passage, $\mu$ is the template bank 
mismatch, and $N_{\rm cpu}$ is the number of $\sim 2$ GHz CPUs available~\cite{lsc-grid}.  
The quantities $N_{\rm trials}$ and ${N_{h_{0}}}$ are the number of artificial signals injected per value of $h_{0}$ and the 
number of different values of $h_{0}$ injected, respectively.
Note the steep dependency of 
the search time $T_{\rm search}$ on the maximum observation time span $T_{\rm span}$.  The contributing factors to this scaling are the increasing 
number of orbital and frequency filters, $N_{\rm orb}$ and $N_{\rm freq}$ respectively, with observation time span, where 
$N_{\rm orb}\propto T_{\rm span}^{5}$ and $N_{\rm freq}\propto T_{\rm obs}$.  There is also a linear scaling of computational time
with $T_{\rm span}$ (corrected by the factor ${T_{\rm obs}}/{T_{\rm span}}$ that takes into account only the non-zero data points) due to increased data volume being analyzed. From Eqs.~(\ref{eq:tsearch}) and ~(\ref{eq:tinject}) it is therefore clear that if one wants to 
complete the full analysis over a period $\simlt 1$ week the choice $T_{\rm span} = 6$ h is appropriate.

\section{Analysis of the data}
\label{s:analysis}

In this section we describe the analysis strategy 
and its implementation 
on the data collected during the second science run
by the two 4-km LIGO interferometers. 

The inner core of the analysis is built on the frequency-domain matched-filter approach that we 
applied to the data collected during the first science run
to place an upper limit on gravitational radiation from 
PSR J1939+2134~\cite{S1PulPaper}.
However, this analysis is considerably more complex with respect to ~\cite{S1PulPaper} because
(i) the search is carried out over a large number of templates 
(either over sky position or source orbital parameters), (ii) the data are
analyzed in coincidence between two interferometers in order
to reduce the false alarm probability and thereby improve the 
overall sensitivity of the search, and 
(iii) the upper limit is derived from the maximum {\it joint} significance of coincident templates.

\subsection{The detection statistic}
\label{ss:Fstat}

The optimal detection statistic (in the maximum likelihood sense) to 
search coherently for quasi-monochromatic signals is the so-called
$\F$-statistic\footnote{We
would like to stress that this statistic is completely unrelated
to the $F$-statistic described in statistical textbooks to test the null
hypothesis for two variances drawn from distributions with the
same mean.} introduced in~\cite{JKS:98}. This statistic can be extended in a straightforward manner to the case of a signal from a pulsar in a binary system.

In the absence of signal, $2\,\F$ is distributed according to a (central) $\chi^2$ distribution 
with 4 degrees of freedom and the relevant probability density function is given by
\be
p_0(2\F) = \frac{2\F}{4}e^{-\frac{2 \F}{2}}\,.
\label{e:p0}
\ee
We define the false alarm probability of $2\F$ as
\be
P_0(2\mathcal{F}) = \int_{2\mathcal{F}}^\infty \, p_0(2\mathcal{F}')\,d(2\mathcal{F}')\,.
\label{eq:fa}
\ee 
In the presence of a signal, $2\F$ follows a non-central $\chi^2$ distribution with 4 degrees of freedom and non-centrality parameter $\rho^2$; the associated probability density function is
\be
p_{1}(2 \F) = \frac{1}{2}e^{-(2 \F+\rho^2)/2}\sqrt{\frac{2
\F}{\rho^2}}
I_{1}(\sqrt{ 2\F\,\rho^2})\,,
\label{e:p1}
\ee
where $I_1$ is the modified Bessel function of the first kind 
of order one and 
\be
\rho^2 = \frac{2}{S_h(f)}\,\int_{0}^{T_{\rm obs}} h^2(t)\,dt\,.
\label{e:rho2}
\ee
The expected value of $2\F$ is $4 + \rho^2$.
From Eq.~(\ref{e:rho2}) it is clear that the detection statistic is proportional to the square of the amplitude of the gravitational wave signal, $h_0^2$, given by Eq.~(\ref{h0 definition}).

\subsection{Pipeline}
\label{ss:pipeline}

\begin{figure}
\centering \includegraphics[height=11.3cm,width=8cm]{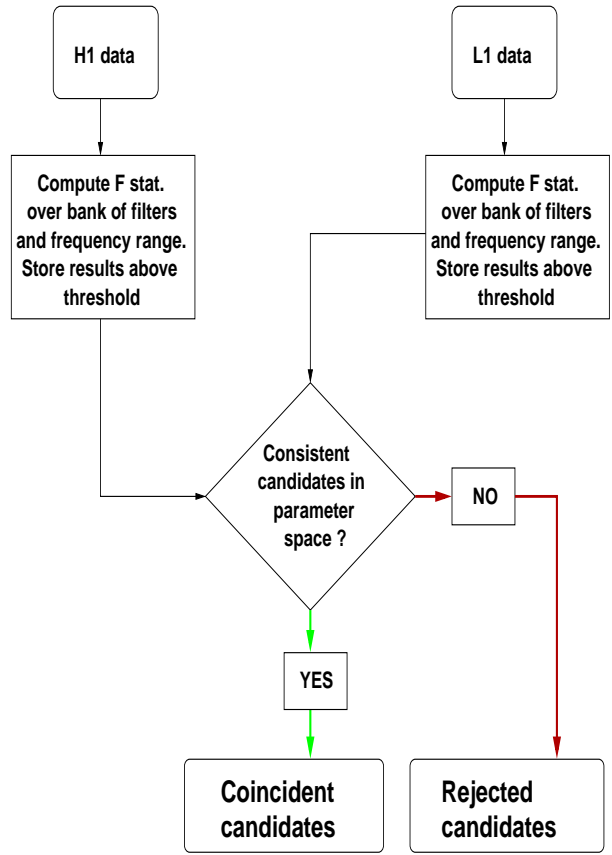} 
\caption{
Workflow of the pipeline. 
}
\label{f:pipeline}
\end{figure}

The search pipeline is schematically illustrated
in Fig.~\ref{f:pipeline}. A template bank is set up for each search
covering the parameter space under inspection. For both analyses
the template bank is three-dimensional: it covers right
ascension and declination for the unknown isolated pulsar search,
and the orbital phase reference time and the projection of the orbital semi-major 
axis for the Sco X-1 analysis. In addition in both cases we search for the unknown
gravitational wave frequency. 

The data stream is treated in exactly the same way
for each search: the full search frequency band is divided into smaller
($\sim 1$ Hz) {\em sub-bands}\footnote{Each of the sub-bands corresponds to the frequency region which the loudest candidate over the sky or the orbital parameters is maximized over and should be of comparable width with respect to typical noise floor variations. The precise bandwidth size is dictated by convenience in the computational set-up of the analyses, reflecting a different distribution of the computational load among the various nodes of the computer clusters used.}, 
the $\F$-statistic is computed at every point in the
template bank, and lists of candidate templates are produced. Search
template values are recorded when the detection statistic exceeds
the value $2\F=20$, and we will refer to them as {\em registered templates}. Note that we will also refer to these templates as ``events'', by analogy with the time-domain matched filtering analysis.

In the frequency domain search conducted in
\cite{S1PulPaper} we searched a single template in four detectors. Here we search a total of $5\times 10^{12}$ templates in each detector for the isolated pulsars and $3\times 10^{10}$ templates overall in two detectors for Sco X-1. 
In order to reduce the number of {\it recorded} templates, only the maximum of the detection statistic over a frequency interval at fixed values of the remaining
template parameters is stored. The frequency interval over which this maximization is performed is based on the
maximum expected width of the detection statistic for an actual
signal. 

For each recorded template from one detector, the list of templates from 
the other detector is scanned for
template(s) close enough so that an
astrophysical signal could have been detected in both templates. The
criteria used to define ``closeness'' are different for the two
searches and will be described in sections
\ref{sss:coincidence_isolated} and \ref{sss:coincidence_scox1}. This
procedure yields a third list of templates that are what we refer to
as the {\it coincident templates}. These are template values for which
the $\F$-statistic is above threshold {\it and} such that they could be
ascribed to the same physical signal in both data streams. The coincident templates are then ranked according to their {\it joint significance}.

The {\it most significant coincident template} is identified in each
$\sim 1$ Hz sub-band; we also refer to this as the {\it loudest event}
for that frequency sub-band. An upper limit on the value of $h_0$ from
a population of isolated sources, or from a family of possible
source parameters in the Sco X-1 system, is placed in each
frequency sub-band based
on its loudest coincident event.
Following~\cite{S1PulPaper}, this is done by injecting in the real data a set of fake
signals at the same level of significance as the loudest measured
event and by searching the data with the same pipeline as was used in the
analysis. The upper limit procedure is described in Sec.~\ref{ss:ULproc}.

\subsection{Selection of the data set}
\label{ss:dataset}

The data input to
the search is in the form of short time baseline Fourier transforms of
the time-domain data. 
At fixed observation time
the computational cost of a search increases linearly with the number
of short Fourier transforms (SFTs) employed. Hence, the longer the time baseline of the
SFT, the less computationally intensive the search. There are two
constraints to making the SFT time baseline long: i) the noise is
estimated on the SFT time scale and thus it should be reasonably
stationary on such timescale; and ii) the signal-to-noise ratio of a
putative source will be significantly degraded if the Doppler
modulation during the SFT time baseline is of the order of
$1/T_{\rm{SFT}}$. For the S2 data set and a search extending to about
750~Hz we chose 30 minutes as the time baseline for the SFTs of the search for signals from isolated sources. The Sco X-1 search was carried out using 60 s long SFTs due to the more significant acceleration
produced by the orbital motion.

As described in \cite{S1PulPaper} the SFT data is normalized by the noise spectral amplitude.
This quantity is estimated
for each SFT from the actual data near to the frequency bin of interest. 
In \cite{S1PulPaper} we used a simple average over
frequencies around the target search frequency. That approach worked
well because in the vicinity of the target frequency there were no
spectral disturbances. But clearly we cannot count on this being the
case while searching over several hundred Hz. We have thus adopted a
spectral running median estimate method \cite{mohanty02b,
  mohanty02a,badri}. In the isolated pulsar search we have chosen a very
conservative window size of $50$ 1800~s-baseline-SFT bins ($27.8$ mHz) corresponding to
a little under twice the number of terms used in the demodulation
routine that computes the detection statistic through the integrals (108) and (109) 
of \cite{JKS:98}.
We estimate the noise at every bin as the median computed on 25+25+1
values, corresponding to the 25 preceding bins, the bin itself, and
the 25 following bins.  If an outlier in the data were due to a
signal, our spectral estimate would be insensitive to it, and thus we
would be preserving it in the normalized data.
A window size of $50$ 60~s-baseline-SFT frequency bins ($0.833$ Hz) was also used for the upper frequency band 
of the Sco X-1 search, $604$--$624$ Hz.  Due to the presence of some large 
spectral features in the lower band, $464$--$484$ Hz, a window size of 
$25$ 60~s-baseline-SFT bins ($0.417$ Hz) was used in an attempt to better track the noise
floor. Noise disturbances are evident in Fig.~\ref{f:sconoise}, where we 
show (with frequency resolution $1/60$ Hz) the average noise spectral density
of the data set used in the analysis. Notice that the lower frequency band
presents numerous spectral features, especially in H1; moreover a 
strong and broad (approximately 2~Hz) excess noise in both
detectors is evident around 480~Hz, which corresponds to a harmonic 
of the 60~Hz power line.

The reconstruction of the strain from the output of the interferometer
is referred to as the calibration.  Details regarding
the calibration for the S2 run can be found in \cite{S2caldoc}. Both
analyses presented here use a calibration performed in the frequency
domain on SFTs of the detector output. The SFT-strain $h(f)$ is computed by constructing a response function
$R(f,t)$ that acts on the interferometer output $q (f)$: 
$h(f)=R(f,t)\,q(f)$. Due mainly to changes in the amount of light
stored in the Fabry-Perot cavities of the interferometers, the
response function, $R(f,t)$, varies in time. These variations are
measured using sinusoidal excitations injected into the instrument.
Throughout S2, changes in the response were computed every 60
seconds.  The SFTs used were 30 minutes long for the
isolated pulsar analysis and an averaging
procedure was used to estimate the response function for each SFT.
For the binary search, which uses 60 s SFTs, a linear interpolation was
used, since the start times of the SFTs do not necessarily correspond to
those at which the changes in the response were measured.

The observation time chosen for the two searches is significantly less
than the total observation time of S2, due to computational cost
constraints: about 10 hours and 6 hours for the isolated pulsar and Sco X-1
searches, respectively. We picked the most sensitive
data stretches covering the chosen observation times; the
criteria used to select the data sets are described below and the
differences reflect the different nature of the searches.

\subsubsection{Data selection for the isolated neutron star search} 

\begin{figure}
\centering \includegraphics[width=9.0cm]{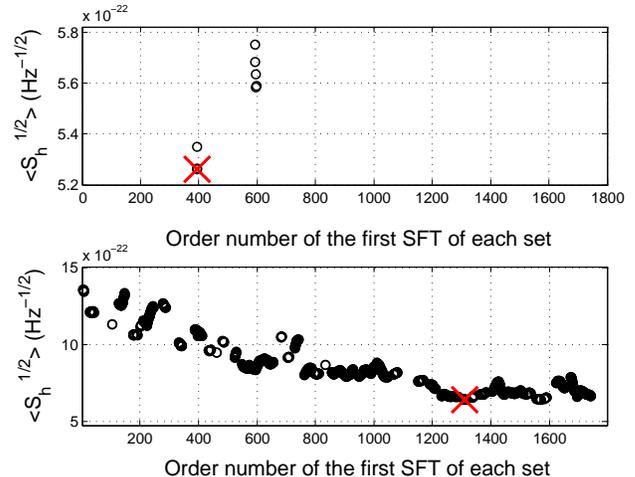} 
\caption{
The average of the noise over various 1~Hz sub-bands as described in the 
text for different sets of 20 SFTs from data of the L1 detector (top plot) 
and H1 detector (bottom plot). The x-axis
labels the order number of the first SFT in each set. SFT \#1 is the 
first SFT of the run. Neighboring sets only differ by one SFT. The cross indicates the data set chosen for the search for signals from isolated objects.
}
\label{f:IsolatedDataSetSelection}
\end{figure}

Since the blind search
for isolated neutron stars is an all-sky search, the most sensitive
data are chosen based only on the noise performance of the
detectors. The sensitivity is evaluated as an average of the
sensitivity at different frequencies in the highest sensitivity band
of the instrument. In particular,
the noise is computed in six sub-bands that span the lowest 300~Hz range to
be analyzed. The sub-bands are 1~Hz wide, with lowest frequencies
respectively at 162~Hz, 219~Hz, 282~Hz, 338~Hz, 398~Hz and 470~Hz. These
sub-bands were chosen in regions free of spectral disturbances and the
average power in these frequency regions can be taken as a measure of
the noise floor there. Even though the search band extends up to 730~Hz,
 we have chosen these reference sub-bands to lie below 500~Hz, because
this is the most sensitive frequency range of our instruments. We
construct sets of 20 SFTs (10 hrs of data), with the constraint that
the data employed in each set does not span more than 13 hours. This
constraint stems from computational requirements: the spacing used for
the template grid in the sky shrinks very fast with increasing {\it spanned}
observation time. If the data contains no gaps then each 10 hr set differs from the previous only by a single SFT. For H1 we are able to construct 892 such sets, for L1
only 8, reflecting the rather different duty cycle in the two instruments. This is obvious from the plots of Fig.~\ref{f:IsolatedDataSetSelection}: For H1 we were able to cover with sets of 20 SFTs the entire run in a fairly uniform way. For L1 it was possible to find sets of nearly-contiguous SFTs only in the first and second quarter of the run. We finally compute the average over the different frequency sub-bands and we pick
the set for which this number is the smallest. 
Figure~\ref{f:IsolatedDataSetSelection} shows this average over the frequency bands and the cross points to the lowest-noise SFT-set.

\begin{figure}
\centering \includegraphics[width=9.0cm]{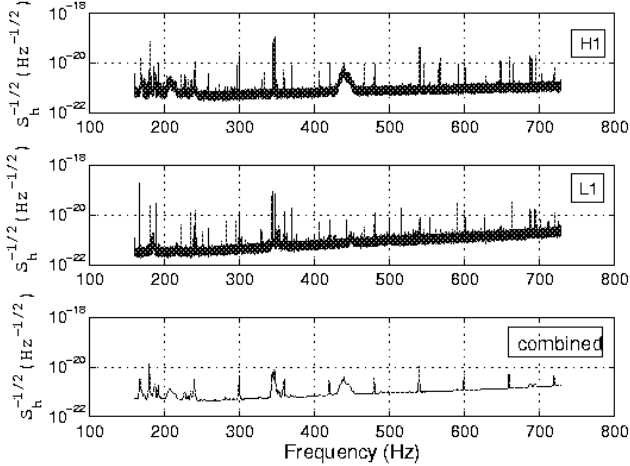} 
\caption{The average amplitude spectral density $\sqrt{S_{\rm h}(f)}$ of the data of the two detectors used for the isolated pulsars analysis. The bottom plot is the average over 1.2 Hz wide sub-bands of the average of the top two plots. The frequency resolution of the top two plots is $1\over 1800$ Hz. The frequency resolution of the bottom plots is 2160 times coarser.}
\label{f:PSDs}
\end{figure}

The data sets chosen were: for H1 20 30-minute SFTs starting at GPS time 733803157
that span 10 hours, and for L1 20 30-minute SFTs starting at GPS time 732489168
spanning 12.75 hours. 
The plots of Fig.~\ref{f:PSDs} show the average power spectral density of this data set for the two detectors separately (top two plots) and the average of the same data 
over 1.2 Hz wide sub-bands and over the two detectors (bottom plot). 

\subsubsection{Data selection for the Sco X-1 search} 
\label{sss:scodataselection}

\begin{figure}
\centering \includegraphics[width=8cm]{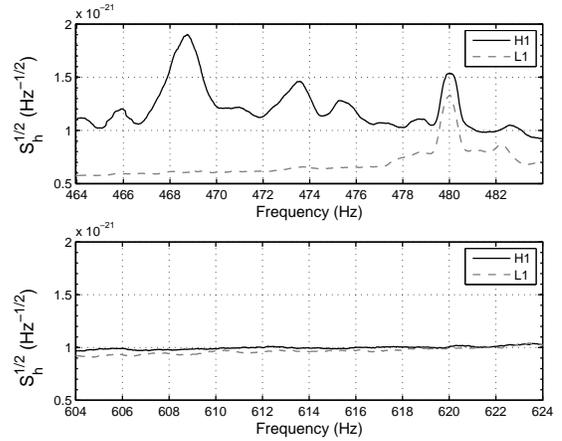}
\caption{The amplitude spectral density $\sqrt{S_{\rm h}(f)}$ of the optimally 
chosen data sets for L1 and H1 and for both frequency bands averaged over each 
60 sec SFT. The solid black and dashed gray lines correspond to H1 and L1, respectively.  
Note that the lower band contains a feature common to both detectors, 
a $60$~Hz power line harmonic at $480$~Hz with a width of $\sim 2$ Hz.  
The H1 data set also contains a variety of other features in the lower 
band, some equally as large as the power line harmonic.  The upper band
is comparatively clean with no visible features.}
\label{f:sconoise}
\end{figure}

We choose to analyze in each detector the most sensitive S2 data set 
which does not span more than 6 hours, which we have fixed based on computational cost constraints. To 
rank the sensitivity of a data 
set we use the figure of merit
\begin{equation}\label{e:Q}
  Q(\vec{T})=\sqrt{\frac{5\langle S_{h}(\vec{T})\rangle }{\left[A(\vec{T})+B(\vec{T})\right]T_{\rm obs}}} \, ,
\end{equation}
where $\vec{T}=\{T_{\rm start},T_{\rm obs}, T_{\rm span}\}$ identifies the data set --- the
time of the first data point $T_{\rm start}$, the effective time containing 
non-zero data points $T_{\rm obs}$, and 
the total span of the observation (including data gaps) $T_{\rm span}$ --- and
$\langle S_{h}(\vec{T})\rangle$ is the noise spectral
density averaged over the frequency search bands and the data set.  Also note that 
$T_{\rm obs}$ is a function of both $T_{\rm start}$ and $T_{\rm span}$.  The two 
functions $A(\vec{T})$ and $B(\vec{T})$ are the integrals of the amplitude 
modulation factors and take into account the change of sensitivity of the 
instruments for the Sco X-1 location in the sky as a function of the time at
which the observation takes place (explicit expressions for $A$ and $B$ are 
given in~\cite{JKS:98}). In our calculation of $Q$ we take into account the
presence of data gaps over $T_{\rm span}$ and 
we average over the unknown angles $\iota$
and $\psi$. From Eq.~(\ref{e:rho2}) it is straightforward to recognize 
that $Q^2$ is simply related to the non-centrality parameter $\rho^2$ for a signal amplitude $h_0$ by:
\begin{equation}
  \langle \rho(\vec{T})^{2} \rangle_{\iota,\psi}=\frac{\left[A(\vec{T})+B(\vec{T})\right]T_{\rm obs}}{5\langle S_{h}(\vec{T})\rangle} \, h_{0}^{2} \, .
\end{equation}
The parameter $Q$ is therefore a faithful measure of the sensitivity of a
given data set for the Sco X-1 search: it combines the effects of the variation
with time of the detectors' noise level, duty cycle and the (angle averaged)
sensitivity to the specified sky position. Note that for $T_{\rm span} \simlt 1$~day, 
the location of the source is a strong factor in the choice of the
optimal data set; for longer observation times the quantity $A + B$ becomes constant. 
By tuning the choice of the data set to exactly the Sco X-1 sky position we have achieved a gain in 
sensitivity $\simgt 2$ compared to selecting the data set only based on the noise level.

\begin{figure}
\centering \includegraphics[width=8cm]{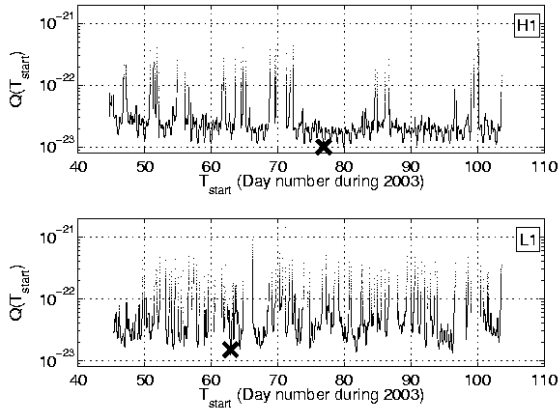} 
\caption{
The sensitivity of the LIGO interferometers during S2 for a search targeted for 
Sco X-1.
The plots shows the evolution of the sensitivity quality factor $Q$, Eq.~(\ref{e:Q}), as 
a function of observation start time;
each point corresponds to a maximum observation time span of $6$
hours.  Due to intermittent loss of lock during the S2 run each $6$
hour span can contain significantly less than $6$ hours of data.  Note
that H1's sensitivity appears more consistent than that of L1.  This
is due to the lower variation in the H1 noise during the run and the
$74\%$ duty cycle compared to L1's $37\%$.  The periodic structures,
more visible in the H1 curve, are caused by the daily variation of the detectors' antenna pattern due
to the Earth's spin.
The crosses indicate the start times corresponding to the
best sensitivity and therefore chosen for the analysis.}
\label{f:LLO_Q}
\end{figure}

We compute the figure of merit $Q$ for all possible choices of data
segments with $T_{\rm span} \le 6$ hours; the values of $Q$ for the
whole S2 are shown in Fig.~\ref{f:LLO_Q}. The data sets that we
select for the analysis begin at GPS time 732059760, end at GPS time
732081371, and span 21611 s with 359 60-second SFTs for H1, and begin at GPS time 730849195, end at GPS time 730867950, with 190 SFTs and 18755 s for L1. Notice that $T_{\rm span}$ is different for the two detectors (which has an impact on the choice of the orbital template
banks for the two instruments), and that the L1 and H1 data sets are
not coincident in time; based on the relatively short observation time and therefore coarse frequency resolution of the search the maximum spin up/down of the source due to accretion would change the signal frequency by only $\sim 0.1$ frequency bins, which is negligible.

\subsection{Template banks}
\label{ss:filters}

We employ different schemes for placing search templates for the isolated pulsar search and for the Sco X-1 search.
The optimal strategy for laying a filter bank is through
a metric approach, and this is used for the orbital parameter grid employed in the Sco X-1 analysis.
By contrast, the search 
for signals from isolated objects uses a sub-optimal grid (in terms
of computational efficiency, but not for the purpose of recovering
signal-to-noise ratio) to cover the sky; a full metric approach was not developed for this search at the time that this analysis was performed.

To refer to the templates we will equivalently use the term ``template'' and ``filter''.

\subsubsection{Isolated neutron stars}
\label{sss:bank_isolated}

\begin{figure}
\centering \includegraphics[width=8.0cm]{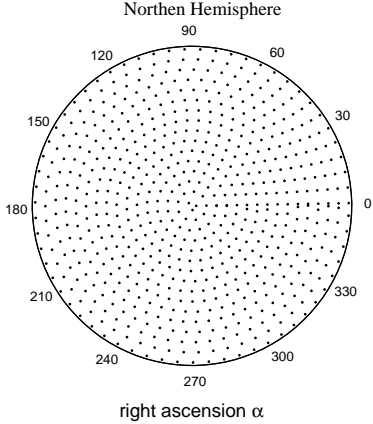} 
\caption{
The sky grid that we have used is of the kind shown here, 25 times
more dense. The angular distance between points at the equator in this
plot is 0.1 radians. The figure is a projection of the grid on the Northern hemisphere. Distance along the radial direction is proportional to the cosine of the declination.}
\label{f:allskygrid}
\end{figure}

Two independent grids are employed: one in sky position and one in frequency.
The grid in frequency is uniform with a spacing $\Delta f_0 = 3.472
\times 10^{-6}$~Hz which is about a factor of $8$ smaller than the
inverse of the observation time. To cover the sky 
we choose an isotropic grid with equatorial spacing of $0.02$ rad. Such 
a grid covers the celestial sphere with just under $31500$ patches of approximately equal
surface area. The number of templates in right ascension $\alpha$ at
any given declination $\delta$ is proportional to $\cos\delta$. At fixed
$\alpha$ the spacing in $\delta$ is constant, and equal to $0.02$ rad.
For illustration purposes Fig.~\ref{f:allskygrid} shows an
under-sampled grid of this type.

The grid is chosen based on the maximum expected degradation in the
detection statistic due to the mismatch between the actual position of
a putative source and the template grid. This effect is measured by
Monte Carlo simulations.

The simulations consist of series of searches of signals
at random locations in the sky with position templates uniformly
randomly displaced  from the signal's source position by between $0$ and half a grid step in both the $\alpha$ and the
$\delta$ direction.

\begin{figure}
\centering \includegraphics[height=7.0cm]{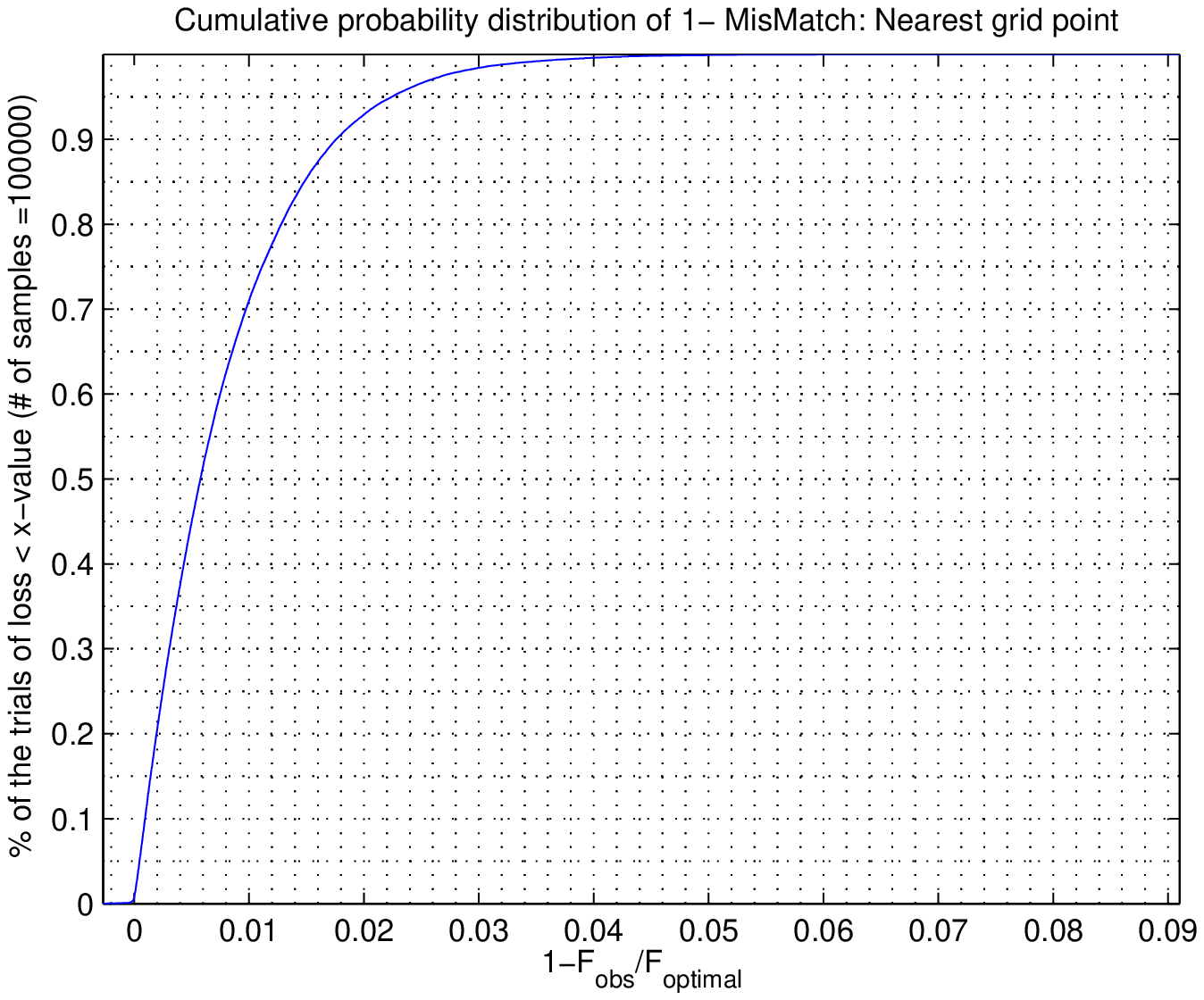} 
\caption{
Fraction of trials injected in L1 
where the ratio
$1.0-{\F_{\rm obs}/\F_{\rm optimal}}$ is smaller than the value on the $x-$axis.
Here in this plot $\F_{\rm obs}$ is the $\F$-statistic at the grid
point nearest to the signals' source location. In 99\% of the trials the mismatch is smaller than 4\%. }
\label{f:gridResL1ngp}
\end{figure}

\begin{figure}
\centering \includegraphics[height=7.0cm]{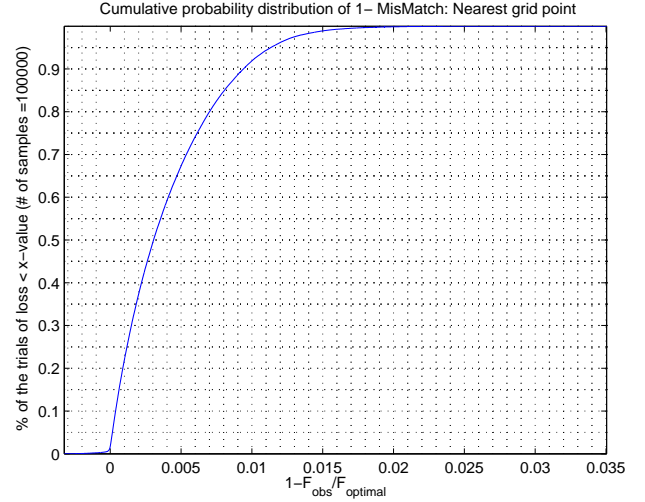} 
\caption{
Fraction of trials injected in H1 
where the ratio
$1.0-{\F_{\rm obs}/\F_{\rm optimal}}$ is smaller than the value on the
$x-$axis. Here in this plot $\F_{\rm obs}$ is the $\F$-statistic at the
point nearest to the signals' source location. In 99\% of the trials the mismatch is smaller than 2\%.}
\label{f:gridResH1ngp}
\end{figure}

We base the selection of the grid size on the properties of the
signals
and the simulations are therefore performed in the absence of noise
(e.g.\ see \cite{OS:99}). The grid spacing is chosen in such a way that the
expected loss in signal-to-noise ratio due to the signal-template mismatch
is a few percent. The results are summarized in Fig.s~\ref{f:gridResL1ngp} and \ref{f:gridResH1ngp}. 

The smaller the maximum mismatch between a signal and a template the more correlated are the filters in the bank. 
We have estimated the effective number of independent templates
 from the $2\F$ average loudest event found in single interferometer searches such as the one described 
here, in pure Gaussian and stationary noise in 1.2 Hz sub-bands: 45.7 for L1 and 41.7 for H1. These translate into an effective number of independent templates which is a factor of $\sim 4$ and $\sim 26$ smaller for L1 and H1 respectively, than the actual number of templates in the sky grid that we are using for the search. The number of independent templates was estimated as 
$1 / {P_0({\F}^*)}$, 
where ${\cal F}^*$ is the measured loudest value of the detection statistic and $P_0$ is the false alarm probability defined in Eq.~(\ref{eq:fa}). This is consistent with our Monte Carlo simulations (Fig.s~\ref{f:gridResL1ngp} and \ref{f:gridResH1ngp}) where we can see that, with the same grid, 50\% of the sky is covered in H1 with a mismatch that is always smaller than 0.5\% whereas in L1 50\% of the sky is covered with a mismatch which is about twice as large. This means that the grid ``covers'' H1 data parameter space with more redundancy than it covers L1.
 
The main reason for the difference in coverage is the fact that the spanned observation time of the data set used for the L1 detector is longer than that for H1, and the resolution in sky position is highly dependent on the spanned observation time of the data set.

\subsubsection{Sco X-1}
\label{sss:bank_scox1}

The analysis for Sco X-1 requires us to search over two orbital
parameters and the gravitational wave frequency, as defined in a given 
reference frame that we choose to be the rest frame of the source 
(plus a correction due to the constant motion of the center of mass of the 
binary system).  
In order to optimally\footnote{Although we use the
metric approach to lay templates in the parameter space, we use a
simple square grid which is non-optimal.  Using a hexagonal grid
would reduce the number of templates by $\sim 30\%$.} cover the parameter
space we consider the metric approach introduced in \cite{OWEN96} in the
context of binary inspirals and 
applied to pulsar searches in \cite{BCCS98,Dhurandhar:2000sd}.  We define the
mismatch $\mu$ between a signal, described by the parameter vector 
${\vec \lambda}=\{f_{0},a_{\rm p},\bar{T}\}$, and a template described by
${\vec \lambda} + \Delta \vec{\lambda}$ as\footnote{Note that we
use the power spectrum ${\cal P}$ to define the mismatch $\mu$ and therefore
the metric $g_{\alpha \beta}$, but we use the $\F$-statistic in the
actual search.  As the $\F$-statistic is an optimally weighted sum of
power spectra we would expect that the template bank is also as
effective used with the $\F$-statistic as for the power spectrum. This was
tested with extensive software signal injections.}
\ba
\mu(\vec{\lambda},\Delta\vec{\lambda}) & = &
1 - \frac{{\cal P}(\vec{\lambda},\Delta \vec{\lambda})}
{{\cal P}(\vec{\lambda},\vec{0})} 
\\ \nonumber
  & = &
g_{\alpha\beta}(\vec{\lambda}) {\Delta\lambda}^{\alpha}
{\Delta\lambda}^{\beta}+O(\Delta\lambda^{3}),
\label{e:mu}
\ea
where the power spectrum ${\cal P}$ is given by
\be
{\cal P}(\vec{\lambda},\Delta \vec{\lambda})=\Bigl| ~\int_{0}^{T_{\rm span}}e^{-i\Delta \Phi(t;\vec{\lambda},\Delta \vec{\lambda})}\,dt ~ \Bigr|^{2}\,;
\ee
$\Delta\Phi(t;\vec{\lambda}, \Delta {\vec\lambda}) = \Phi(t; \vec{\lambda}) - \Phi(t; \vec{\lambda} + \Delta {\vec\lambda})$ is the difference between the signal and template phase and $\alpha,\beta = 0,1,2$ label the search parameters (we follow the convention that the index $0$ labels frequency).
The metric on the parameter space is given by~\cite{BCCS98}
\be
g_{\alpha\beta} = \langle\partial_{\alpha}\Delta\Phi \partial_{\beta}\,
\Delta\Phi\rangle-\langle\partial_{\alpha}\Delta\Phi\rangle\,\langle
\partial_{\beta}\Delta\Phi\rangle\,,
\label{e:gab}
\ee
where $\partial_\alpha \Delta\Phi \equiv \partial\Delta\Phi/\partial \Delta \lambda^\alpha$ and is evaluated at $\Delta {\vec\lambda} = 0$, $ \langle ... \rangle$ stands for the time average over $T_{\rm span}$, and the gravitational wave phase is defined 
in Eq.~(\ref{e:phase}). Treating the frequency as a continuous
variable (we discuss later in this section the consequences 
of the fact that the frequency is in practice discrete) and 
``projecting out'' the search frequency dimension of the metric yields 
a 2-dimensional reduced metric only on the orbital parameters,
\begin{equation}\label{gammametric}
\gamma_{jk}=g_{jk}-\frac{g_{0j}g_{k0}}{g_{00}}\,,\quad j,k = 1,2\,.
\end{equation} 
By using the metric $\gamma_{jk}$ we take advantage of the
correlations between the frequency and the orbital parameters --- so
that a mismatch in orbital parameters can be compensated by a mismatch
in frequency --- and we therefore reduce the number of orbital
templates required to cover the parameter space. In the actual
analysis, we carry out a coordinate
transformation from $a_{\rm p}$ and $\bar{T}$ to two ``search coordinates''
in order to obtain constant
spacing and orientation of the filters over the whole parameter space,
which simplifies the numerical implementation of the grid.

The frequency, however, is sampled discretely and therefore cannot compensate exactly a mismatch in orbital parameters: it produces a mismatch $\mu_f$ only in $f$ between a template and a signal. In order to choose the appropriate frequency spacing $\Delta f$ we consider the $g_{00}$ component of the metric $g_{\alpha \beta}$ --- we treat de facto $f$ as a 1-dimensional uncorrelated dimension in the parameter space --- and obtain:
\begin{equation}
\label{e:freqres_sco}
  \Delta f\approx\frac{2\sqrt{3\mu_{f}}}{\pi T_{\rm span}}\,.
\end{equation}
The optimal method for dividing up the {\em total}
mismatch in detection statistic is to split them up equally amongst
the dimensions of a parameter space.  Therefore to achieve the
required maximal overall mismatch of $10\%$ we use a $6.6\%$ mismatch
in orbital templates and a $3.3\%$ mismatch in frequency.  This sets
the frequency resolution to $\Delta
f=1/(5T_{\rm span})$ corresponding to $9.285051\times 10^{-6}$~Hz for H1
and $1.066382\times 10^{-5}$~Hz for L1.

\begin{figure}
\centering \includegraphics[width=8.3cm]{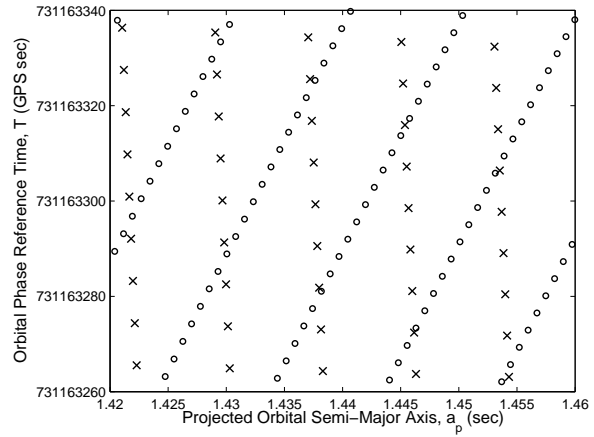}
\caption{
Here we show a small section ($\approx 1/50^{th}$) of the total
orbital parameter space.  The crosses and circles represent template locations
used to search the L1 and H1 data sets, respectively. 
This particular template bank was
generated for a maximum search frequency of $465$~Hz.
Note that the templates are {\em not} uniformly spaced in the
$(a_{\rm p},\bar{T})$ parameter space, although they appear nearly so in the
limited region shown here.}
\label{f:orbitaltemplates}
\end{figure}

The grid in the 2-dimensional space of the orbital templates is
computed using $\gamma_{jk}$, Eq.~(\ref{gammametric}). In practice,
the filter spacing is determined primarily by $T_{\rm span}$ and the
grid orientation is determined by the location of the source in its orbit during the
observation.  These effects are shown in
Fig.~\ref{f:orbitaltemplates}; although the observation spans differ
by only $\sim15\%$ between L1 and H1, the density of filters for H1 is
twice that for L1, as the number of orbital templates in the regime
$T_{\rm span}<P$ scales as $T_{\rm span}^{5}$.  The source location
within the orbit differs by $\approx2.22$ radians between the L1 and
H1 observation periods and correlations between the two orbital
parameters are therefore different between the two detectors resulting
in template banks that are clearly non-aligned.  One further step to
optimize the search is to generate separate orbital template banks for
each $1$~Hz frequency sub-band, because the grid density increases as
$(f_{0}^{\max})^{2}$, where $f_{0}^{\max}$ is the maximum search frequency. This approach allows an overall gain $\approx
30\%$ in computational speed in comparison to using a single template
bank with a maximum frequency parameter $f_{0}^{\max}=624$~Hz for the whole analysis.

The number of orbital templates used for each 1~Hz sub-band ranges
from 3391 to 3688 in the $464$--$484$~Hz band and from 5738 to 6107 in
the $604$--$624$~Hz band for the L1 analysis.  The number of frequency filters per 1~Hz band is 93,775;
therefore the number of trials used to cover the parameter space is
in the range $3.2\times 10^8 - 5.7\times 10^{8}$. 
The corresponding numbers for the H1 analysis are $6681$--$7236$, $11309$--$12032$, 
and 108,055, respectively, corresponding to a total number of trials for each 1~Hz sub-band
in the range $7.2\times 10^{8}$ -- $1.3\times 10^{9}$.

%
%
%
%
%
%

\subsection{The single detector search}
\label{ss:singleIFO}

As described in Sec.~\ref{ss:pipeline} and Fig.~\ref{f:pipeline} the data 
from each detector is searched over the entire parameter space, by computing
the $\F$-statistic for each template in frequency and either position
in the sky (for the all sky search) or orbital parameters (for the Sco X-1 search).
In both cases we store results only for those templates that yield a 
value of the detection statistic that exceeds the threshold $2\F_{{\rm{thr}}}=20$. 
This choice is based on limitations on the size of the output files of
the search. 

Our template banks are highly correlated; thus in order to decrease the
number of frequency templates that we store, we treat as correlated the
template frequencies which are sufficiently ``close to each other''. In particular, we do not register 
as separate templates, templates which differ only by such frequencies.
The frequency interval that defines how close frequencies have to be in order to be ascribed 
to the same template
is estimated based on the full width at half maximum of the $\F (f_0)$ 
curve for a representative sample of the parameter space and for random mismatches between signal and template, as
would occur in an actual search in the presence of a signal. The resulting frequency intervals are a few times $10^{-4}$ Hz.

The following information is stored for each template above 
threshold: the frequency $f_0$ at which the value of $\F$ is maximum, the
values of $\alpha$ and $\delta$ for the template, the total width in
search frequency bins of the points associated with the maximum, the
mean value and the standard deviation of $\F$ over all those points and
the value of $2\F$ at the maximum. The same information is stored in
the case of the Sco X-1 search, with the orbital parameters $a_{p}$ and $\bar{T}$ instead of the sky position parameters $\alpha$ and $\delta$.

The computational load for the searches is divided among independent
machines, each searching a small frequency region over the entire
parameter space. For the isolated pulsar search each CPU analyzes a
$60$~mHz search band. The processing time for both data streams
and the entire sky is typically about 6 hours on a 2~GHz class 
computer. The typical size of the output, after compression, from a 
single detector search is around 3 MBytes. For the Sco X-1 search,
typically an individual
machine searches $0.1$~Hz.  The
equivalent run time on a $60$~mHz search band on the entire Sco X-1
orbital parameter space is approximately $9$ hours.
Although the
Sco X-1 orbital templates are fewer than the sky position templates,
the two searches are comparable in computational time because the Sco
X-1 search uses a greater number of (shorter) SFTs.  For the
particular data sets selected for this analysis it should be noted that
with L1's shorter spanned observation time and fewer SFTs, the
computational load is primarily due to the H1 search.  The output
from the search in total, including both detectors and all search
bands, comprises $\approx1$ GByte of results, corresponding to around $700$ kBytes
per $60$~mHz band for the entire orbital parameter space.

\begin{figure}
\centering \includegraphics[width=8.3cm]{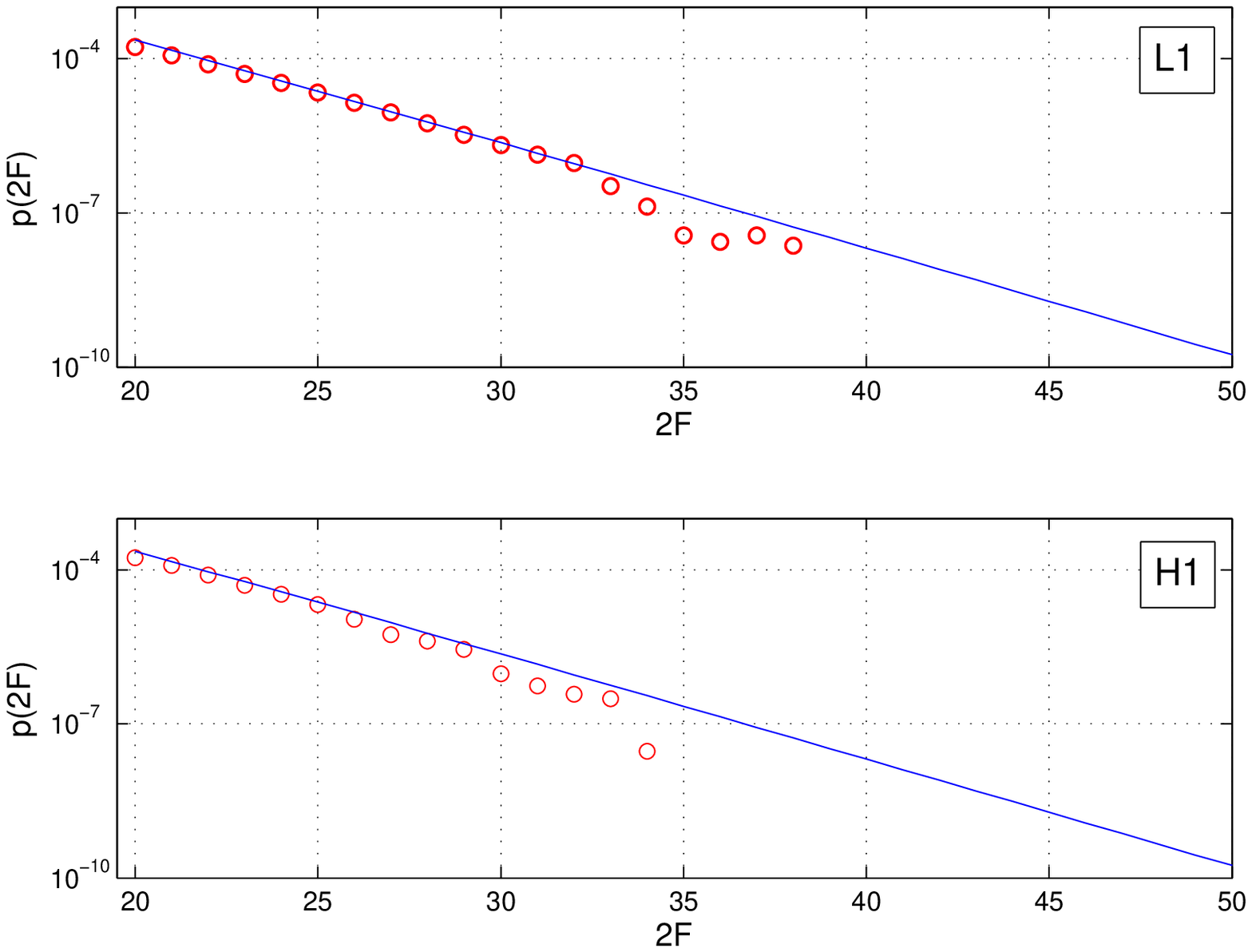} 
\centering \includegraphics[width=8.3cm]{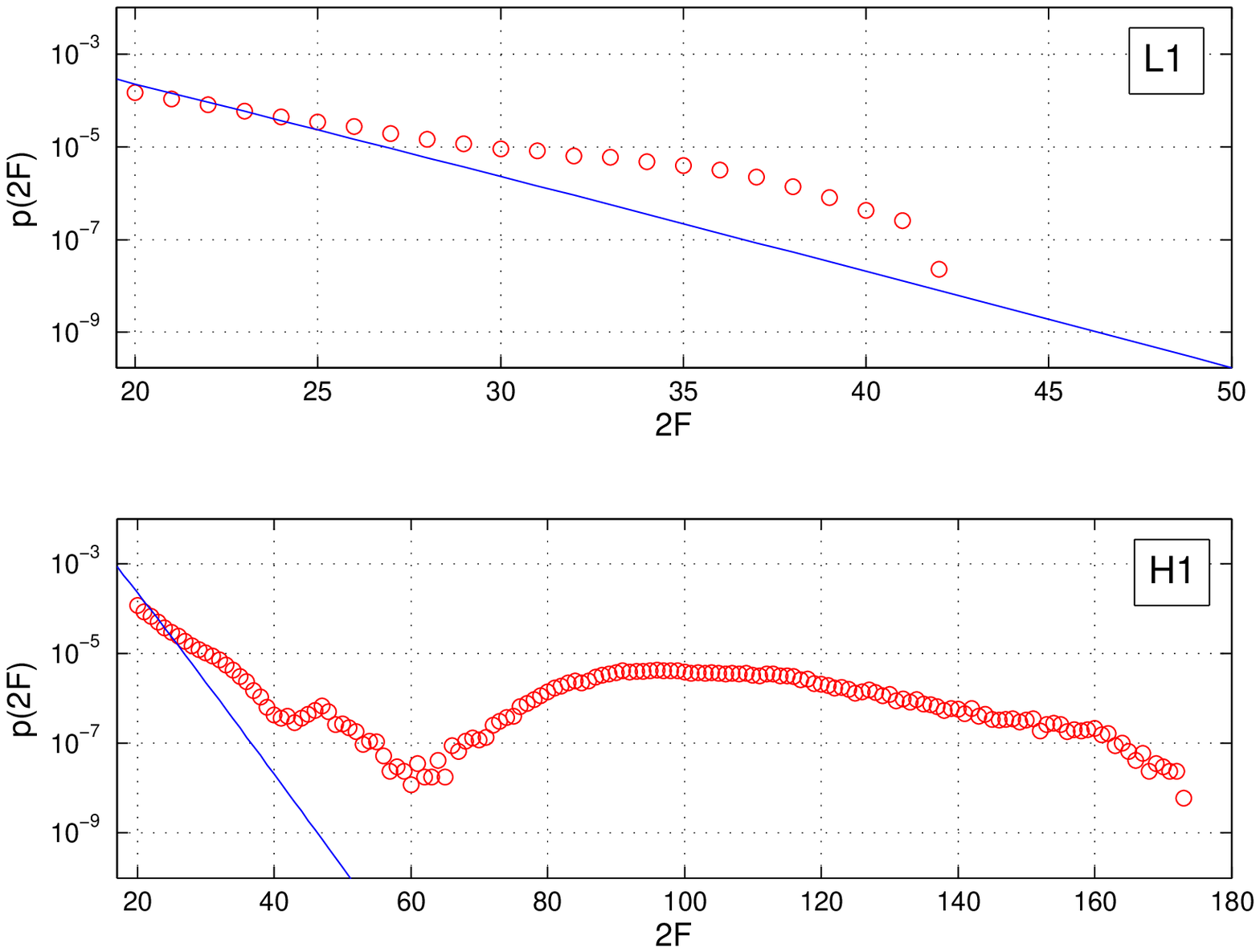} 
\caption{
The circles show the distribution of $2\F$ values for the
templates registered after the single detector all-sky search. The solid line
 shows the expected distribution for Gaussian stationary white noise. The top two plots refer to the band $247.06$--$247.12$~Hz. The bottom two plots show the same
distributions for the $329.56$--$329.62$~Hz band. The expected
distribution is dominated for high values of $2\F$
by an exponential term, as is evident from the linear behavior on
a semilog scale. In the clean $247$~Hz band, the
theoretical and the experimental distributions agree very well.}
\label{f:F247and329distr}
\end{figure}

\begin{figure}
\centering \includegraphics[width=8.3cm]{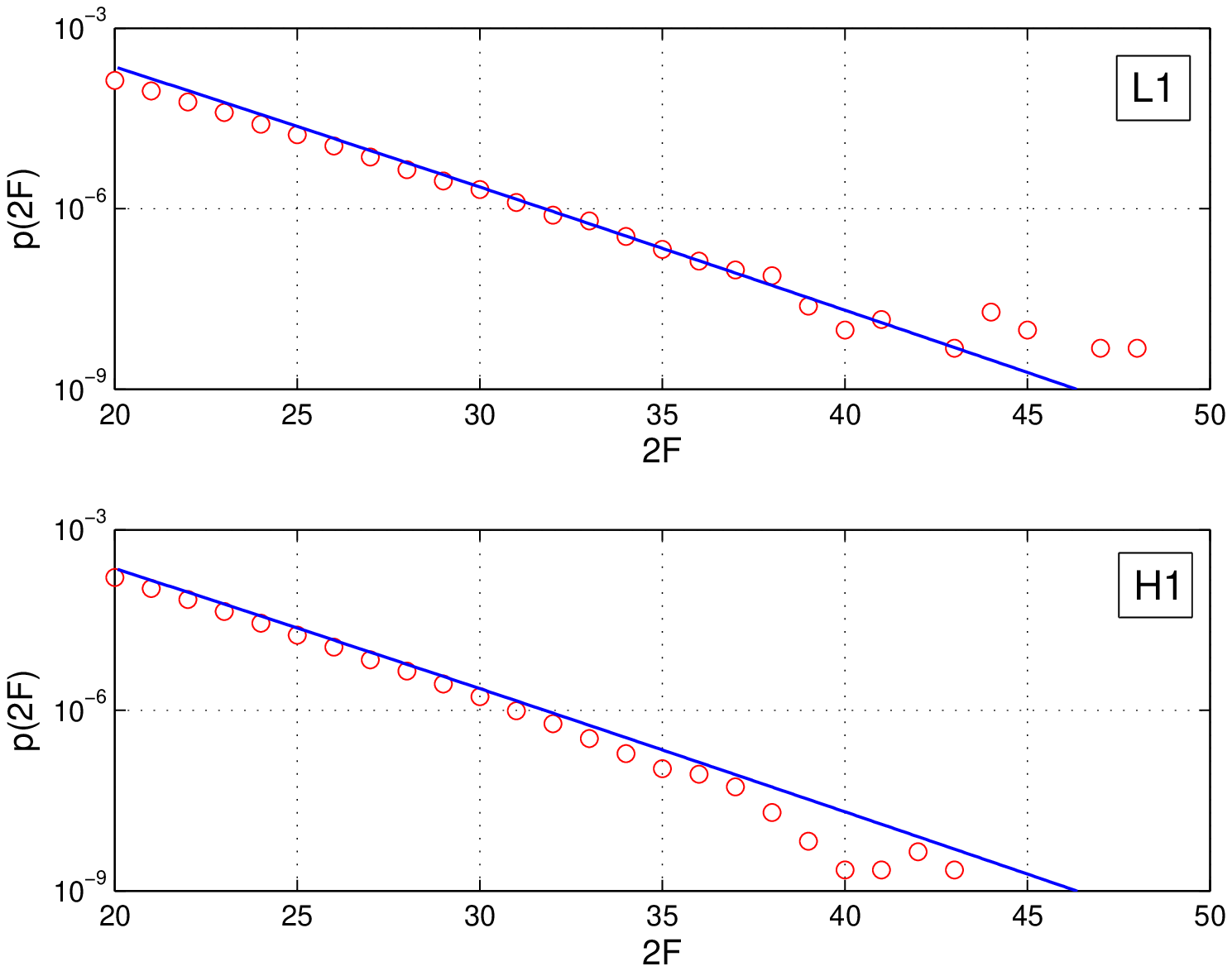} 
\centering \includegraphics[width=8.3cm]{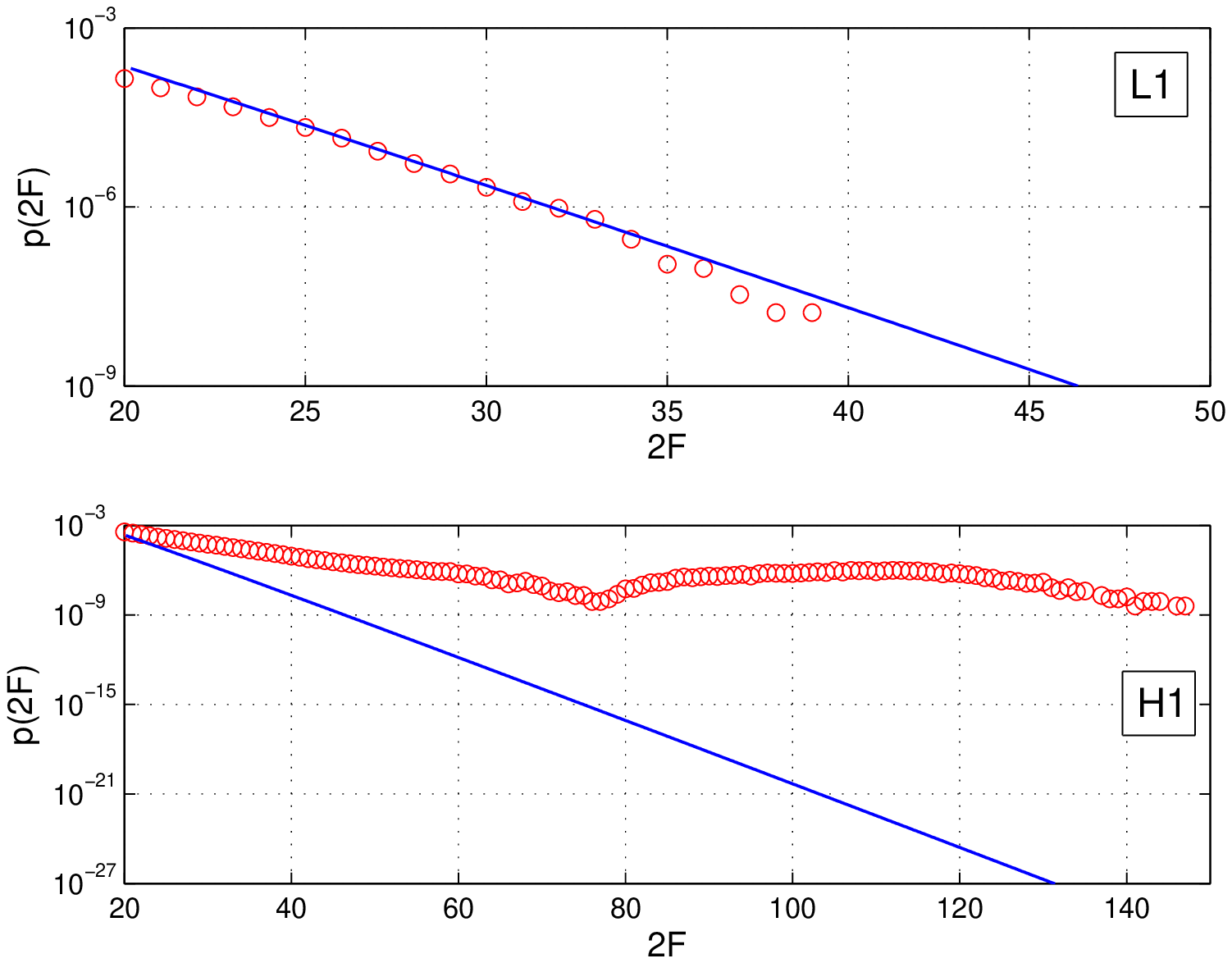}
\caption{
The circles show the distributions of $2\F$ values derived from the single detector Sco X-1 searches. The top two
plots show the distributions for a clean sub-band in both detectors, $619.0$--$620.0$~Hz.  The solid curve represents the theoretical expected
distribution.  The two bottom plots show the same
distributions for the sub-band $465$--$466$~Hz.  In this band the H1 results
are dominated by large values of $2\F$.}
\label{f:2Fdist_scox1_good}
\end{figure}

Figures~\ref{f:F247and329distr} and~\ref{f:2Fdist_scox1_good} show the distribution of $2\F$ values
of the registered templates for sub-bands in reasonably
clean spectral regions in both instruments (around $247.1$~Hz and
$619.5$~Hz respectively) in the top two plots, and in
less clean regions in the H1 data (around $329.6$~Hz and $465.5$~Hz respectively) in the bottom two plots. In the top two plots,
the distributions of $2\F$ values closely follow the expected $p_0(2\F)$ distribution, Eq.~(\ref{eq:fa}).
This is not surprising in regions free of evident
disturbances, as already shown in \cite{S1PulPaper}.  
Note that the highest $2\F$ values in the clean bands (top two plots) are higher in 
Fig.~\ref{f:2Fdist_scox1_good} than in Fig.~\ref{f:F247and329distr}. This is due to fact that the Sco X-1 search has more templates than the all-sky isolated search.

\subsection{Coincidence analysis}
\label{ss:coincidence}

The next stage of the analysis 
compares the two lists of values of $2 {\cal F}$ that 
lie above the threshold $2
{\cal F}_{\rm thr} = 20$ compiled for each detector. We require that
given a template in L1, say, there exists a template in H1 such that
their locations in parameter space are consistent with a physical
signal having triggered them both. If this is the case, the relevant
values of the detection statistic are kept (the two filters are
regarded as ``in coincidence''), otherwise they are ``rejected'' and
removed from the lists. This procedure is identical for both searches,
but the consistency criteria are different due to the different 
signals that are searched for.  This strategy is effective at reducing
the false alarm rate if the noise in the two data streams is uncorrelated.
In practice, the data are also populated by
a forest of lines present both in L1 and H1, such as 16~Hz harmonics from the data acquisition system and the
60~Hz power line harmonics, and this procedure does not eliminate them. However it does eliminate the non-Gaussian uncorrelated outliers which also are in the data. We find that the typical sensitivity improvement in
$h_0$ resulting from the coincidence stage is comparable for both
searches and in the range $10\%$ -- $20\%$, depending on the frequency
sub-band.  

An additional criterion to
identify coincident templates could be based on comparing the values of $2{\cal
F}$ produced by the two filters; however, as $2{\cal F}$ is already
maximized over the nuisance parameters $\psi$ and $\iota$, and the
integration time of the analyses is shorter than 1 day, it is in
practice difficult to introduce an ``amplitude consistency cut'' that is
simultaneously stringent and safe. For this reason we have not
included this requirement in the coincidence stage of this search (see
however the discussion in Section~\ref{ss:isolatedResults}).

The coincident templates are then sorted in order of descending {\em joint significance}. If we indicate
with $2\mathcal{F}_{L1}$ and $2\mathcal{F}_{H1}$ the values of the detection statistic
for a pair of templates in coincidence, we define their joint significance as:
\begin{equation}
  s(2\mathcal{F}_{L1},2\mathcal{F}_{H1}) = 1 - P_0(2\mathcal{F}_{L1})\,P_0(2\mathcal{F}_{H1})\,,
\label{eq:joint_signi}
\end{equation}
where $P_0(2\mathcal{F})$, defined in Eq.~(\ref{eq:fa}), is the single detector false alarm probability
for $2\F$, under the assumption that the noise is Gaussian and stationary.
We consider the {\it loudest} coincident template pair as that yielding the largest
value of joint significance. In practice, in the 
numerical implementation we rank events according to $-\left\{ \log \left [ P_0(\F_{\rm{L1}})\right] + \log \left [ P_0 (\F_{\rm{H1}} )\right ] \right\}$ with $\log \left[ P_0(\F)\right ] = \log \left (1 + \F \right) - \F.$

In the remainder of the section we provide details about the specific
implementation of the coincidence stage for the two analyses.

\subsubsection{Isolated neutron stars}
\label{sss:coincidence_isolated}

The candidate events that survive the coincidence stage are those 
present in both detectors' 
data sets and lie in locations of the parameter space that are consistent with a common signal. For the isolated search the coincidence windows are
$1$~mHz in frequency $f_0$ and $0.028$ rad angular distance in
position on the celestial sphere. These coincidence window values were
derived from the results of the Monte Carlo simulations described in
Sec. \ref{ss:singleIFO}. More specifically $0.028$ rad represents a
mismatch between sky positions of at most 1 grid point.
The value of $1$~mHz is derived
from the results of Monte Carlo simulations by requiring a null false
dismissal rate.

\begin{figure}
\centering \includegraphics[width=9.0cm]{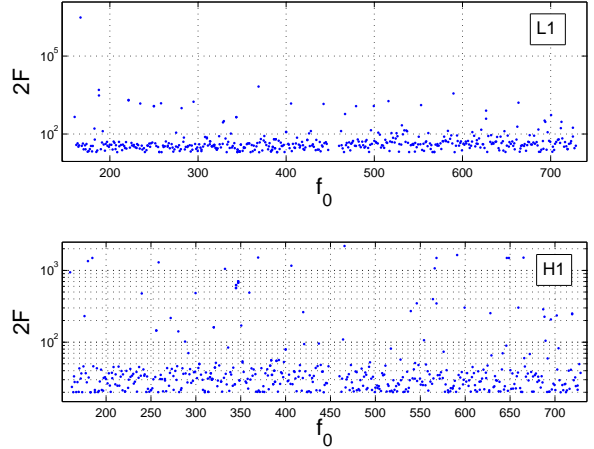} 
\caption{
These plots show two different views of the values of the detection
statistic $2\F$ of the loudest coincident template-couples (one for
every $1.2$~Hz sub-band) from the isolated pulsar search. The mean value
between the threshold and $2\F=100$ is $39.5$ for the L1 data and $32.2$ for
the H1 data.}
\label{f:Lcoinc}
\end{figure}

\begin{figure}
\centering \includegraphics[width=8.3cm]{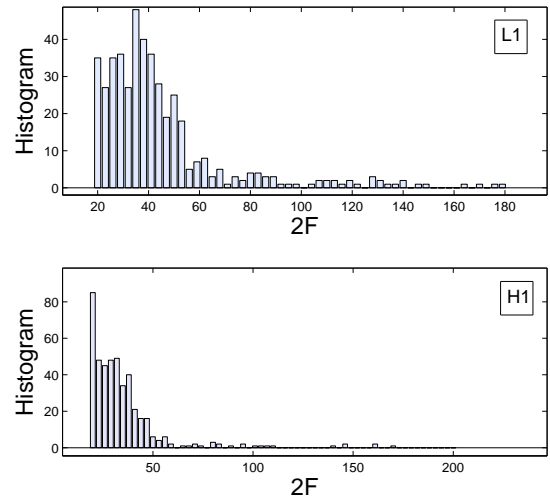} 
\caption{
These plots show the 
distributions of the values plotted in Fig.~\ref{f:Lcoinc}.
The mean value of these distributions depends on the
volume of the parameter space that the search extends over. In this
case it is the whole sky in $1.2$~Hz frequency sub-bands.}
\label{f:LcoincDistr}
\end{figure}

As described in Sec.~\ref{ss:filters}, the all-sky isolated search, if
performed on Gaussian white stationary noise, would yield
single-interferometer loudest templates in $1.2$~Hz sub-bands with
mean $2\F$ values of $45.7$ for L1 and $41.7$ for H1. The
difference in these mean values is due to the different data sets used for the two searches (the time spanned by the L1 data set is longer than that of the H1 data set) and by the different location of the
detectors on Earth and to the non-uniform antenna pattern of the
detectors. In this search, after having excluded outliers with $2\F > 100$,  we measure mean values of the loudest templates 
of $52.2$ for L1 and $46.6$ for H1\footnote{The $2\F > 100 $ cut has been made only when computing the mean values reported above in order
to eliminate large outliers that would have dominated the mean; see Fig.s~\ref{f:Lcoinc} and \ref{f:LcoincDistr}.}. This corresponds to an increased level 
of spectral contamination in the real data with respect to Gaussian stationary 
noise. This is not surprising at all---even a simple visual inspection of the spectra reveals that they are contaminated by several ``lines'' (also see the discussion in Sec.~VI~D in \cite{S2Hough}). 

After the coincidence step the mean value of $2\F$ for the loudest event is
$39.5$ for the L1 data and $32.2$ for H1. If one compares these values
with the mean values {\it before} coincidence, $52.2$ for L1 and $46.6$
for H1, one recognizes that the coincidence step yields an improvement in
$h_0$ sensitivity of $15\%$ and $20\%$ for L1 and H1 
respectively (remember from Eq.~(\ref{e:rho2}) that $2\F\propto h_0^2$). In Gaussian stationary noise the expected improvement is 
$11\%$ and $17\%$ for L1 and H1 
respectively. Thus, and again not surprisingly, the coincidence step plays a greater role on real data, which is affected by uncorrelated non-Gaussian disturbances.

\begin{figure}
\centering \includegraphics[width=9.0cm]{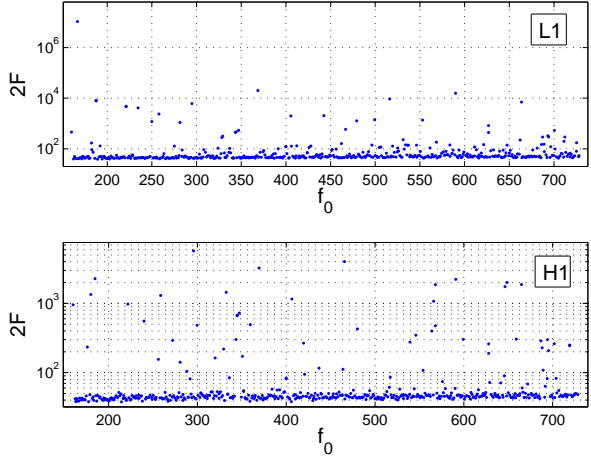} 
\caption{These plots show loudest $2\F$ values in the single detector searches
in each 1.2~Hz sub-band.}
\label{f:Floudest}
\end{figure}

\begin{figure}
\centering \includegraphics[width=8.3cm]{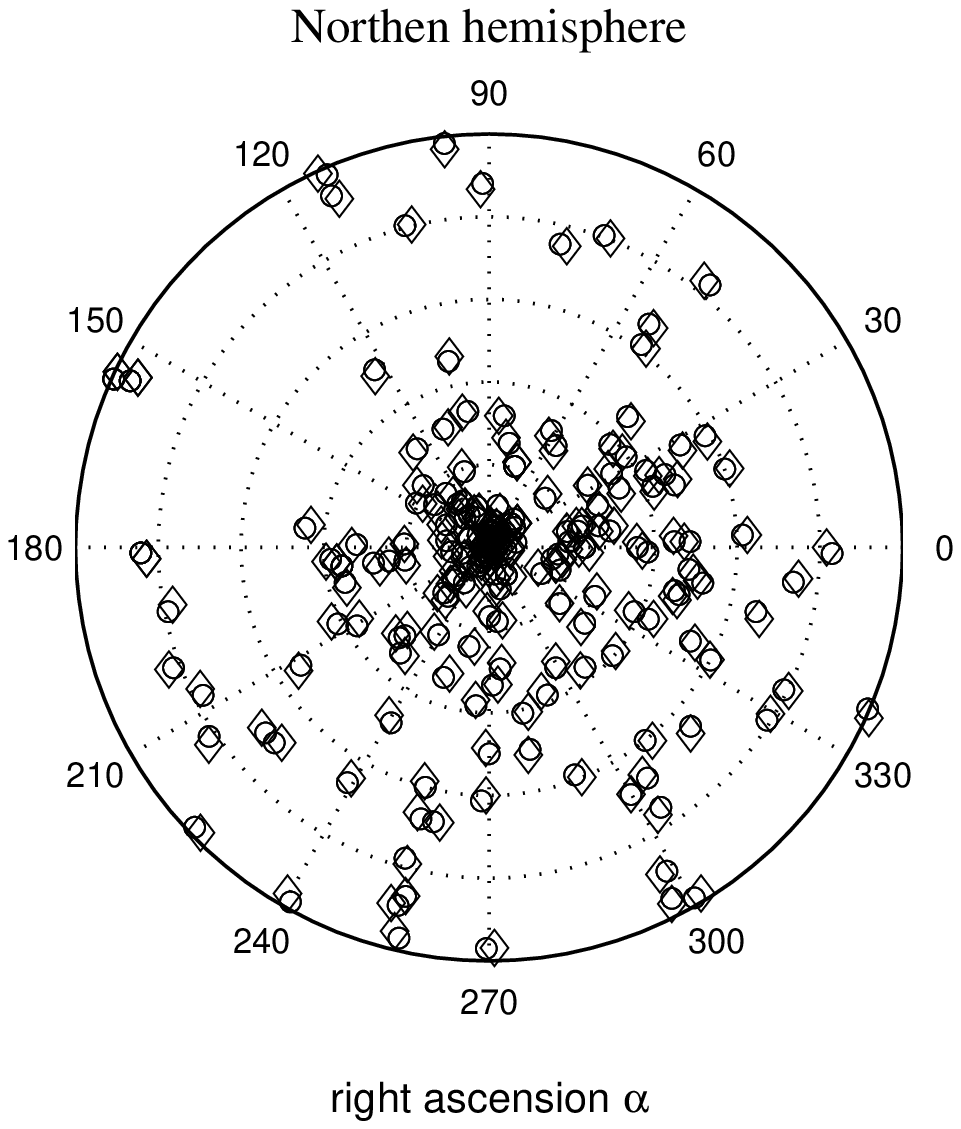} 
\centering \includegraphics[width=8.3cm]{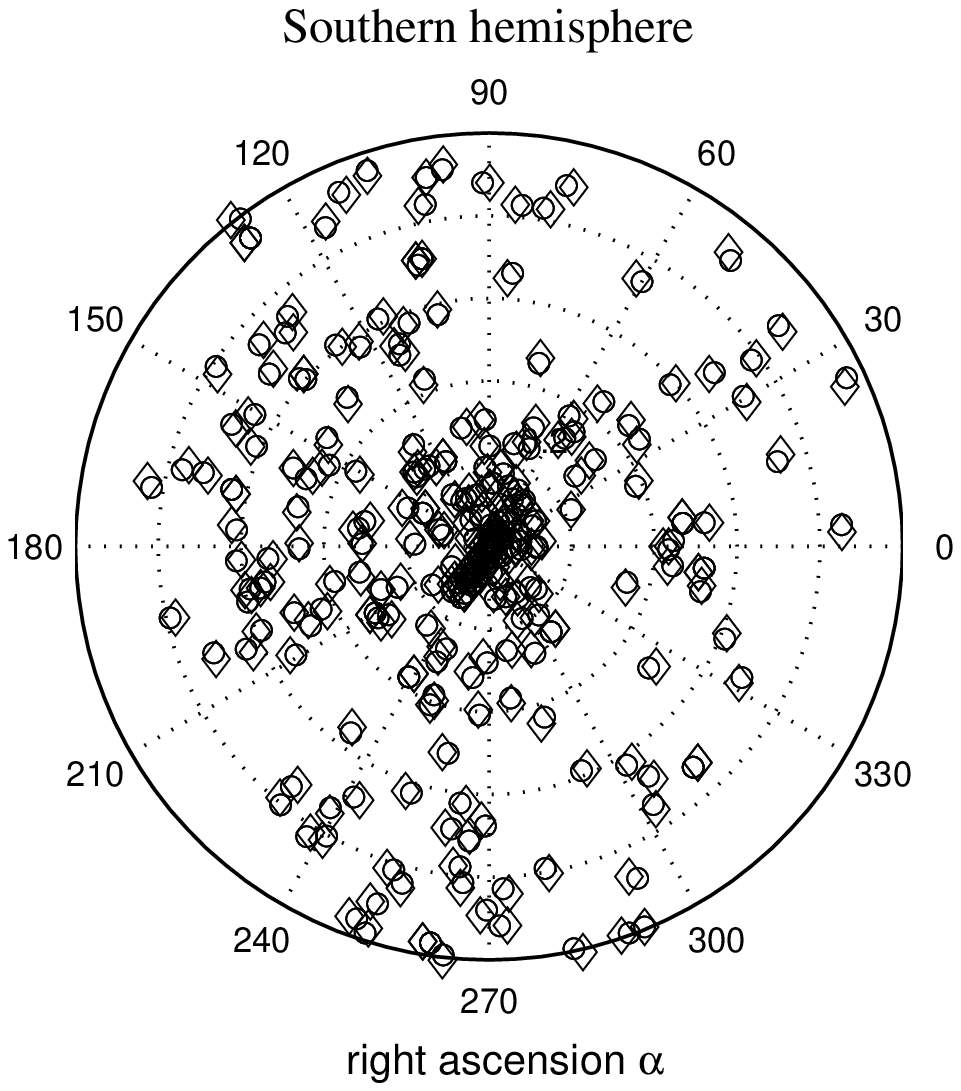} 
\caption{
These figures show the location in the sky of the loudest coincident
template in each 1.2~Hz sub-band. The circles show the templates found with
the L1 search and the diamonds show the coincident templates found with the H1 search.}
\label{f:LSky}
\end{figure}

Figures~\ref{f:Floudest} and \ref{f:Lcoinc} show the values of the detection statistic for the loudest events before and after coincidence, respectively. Figure~\ref{f:LcoincDistr} shows the distributions of $2\F$ for the loudest
coincident templates.
Figure~\ref{f:LSky} shows the distribution of loudest coincident 
templates over the entire sky for all the 1.2~Hz sub-bands. A higher concentration of templates is apparent at the poles. This is to be expected since the poles are the regions from where a monochromatic signal would be received by our detectors at a nearly constant frequency. In other words, spectral artifacts at fixed frequency are consistent with sources close to the poles, during our observation time. 

\subsubsection{Sco X-1}
\label{sss:coincidence_scox1}

\begin{figure}
\centering \includegraphics[width=8.3cm]{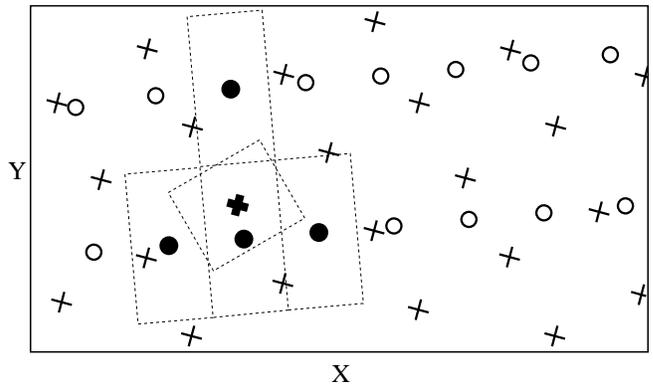} 
\caption{
Here we represent a small region of orbital parameter space.  
The crosses represent orbital
templates used in the L1 search and the circles
represent those used in the H1 search.  Note that the
template orientation and spacing are not the same in each template
bank.  The bold $\times$ represents the location of a candidate
event in the L1 detector, and the dashed rectangle surrounding it
represents the area within which a signal must lie if this template is
the closest to the signal.  The filled circles represent those
templates in the H1 detector that are possible coincident
candidates.  They are identified due to the overlap between their
respective dashed rectangular regions and the L1 candidate event dashed region.
In the S2 search there are on average $12$ orbital templates 
in H1 that are ``consistent'' with each template in L1.}
\label{f:coincidence_scox1}
\end{figure}

In the Sco X-1 analysis we identify coincident templates in L1 and H1
by using the metric on the relevant
parameter space; see Section~\ref{sss:bank_scox1}.  A real signal, if
present and of sufficient amplitude, will trigger templates in both
detectors.  These templates will be in close proximity in parameter
space but not necessarily identical.  The procedure starts by taking
each template above threshold from the L1 detector and first testing
for orbital parameter consistency with templates from the H1 detector.
We do this using a property inherent to the orbital template bank,
which is schematically illustrated in
Fig.~\ref{f:coincidence_scox1}. We know that a signal will return at
least $90\%$ of its optimal detection statistic when processed using
the ``closest'' template (in the absence of noise), where the distance
between templates is defined by the metric.  We can therefore
identify a rectangular region around the L1 filter in the
2-dimensional plane of orbital parameters. This region is easily
calculated from the metric used to place the templates, and we would
expect the true signal parameters to lie within it. We can now repeat
this process for each H1 template and construct, based on the metric
associated with H1, the region covered by the H1 filters (i.e. the
filters that are associated with a value of $2 \F$ above threshold).
The test now becomes a simple matter of checking for any overlap
between the region covered by the L1 filter under scrutiny and the H1
filters.  Overlap implies a possible consistent signal location (in
orbital parameters) capable of triggering a filter in L1 and H1 from
the two candidate events from a single common signal.  In this 
analysis there are on average 12 orbital
templates in H1 that are ``consistent'' with each template in L1. The
process that we have just described is then repeated for all the L1
filters.

So far we have considered only the orbital parameters. The second stage 
to identify filters in ``coincidence'' is to test for frequency consistency
amongst those filters that have survived the previous test.  The use of the
projected metric described in Sec.~\ref{sss:bank_scox1} exploits the 
correlations between the orbital templates and the
gravitational wave frequency to reduce the overall number of filters.  
Doing so allows greater differences between
the true and detected source orbital parameters and 
greater differences between the true and detected gravitational 
wave frequency.  Using Monte Carlo simulations it has been possible 
to define the maximum
separation between a signal's true and detected frequency.
This separation is very strongly governed by the spanned
observation time.  For the data sets chosen the maximum separation for
L1 was found to be $2.158\times10^{-4}$~Hz and for H1 was
$1.773\times10^{-4}$~Hz.  This corresponds to a maximum separation of
$3.931\times10^{-4}$~Hz between a candidate in L1 and H1 in order to
be consistent with a common signal. This is the frequency coincidence window that we have chosen.  Note that it is equivalent to
$\pm 40$ frequency bins in the H1 search.

If a pair of candidate events is found to be consistent in both
orbital parameter space and frequency space then they are classed as a
coincident event.  Note that a single candidate event in the L1
detector can have many possible coincident pairs in the H1 detector (and vice versa).

\begin{figure}
\centering
\includegraphics[width=8.0cm]{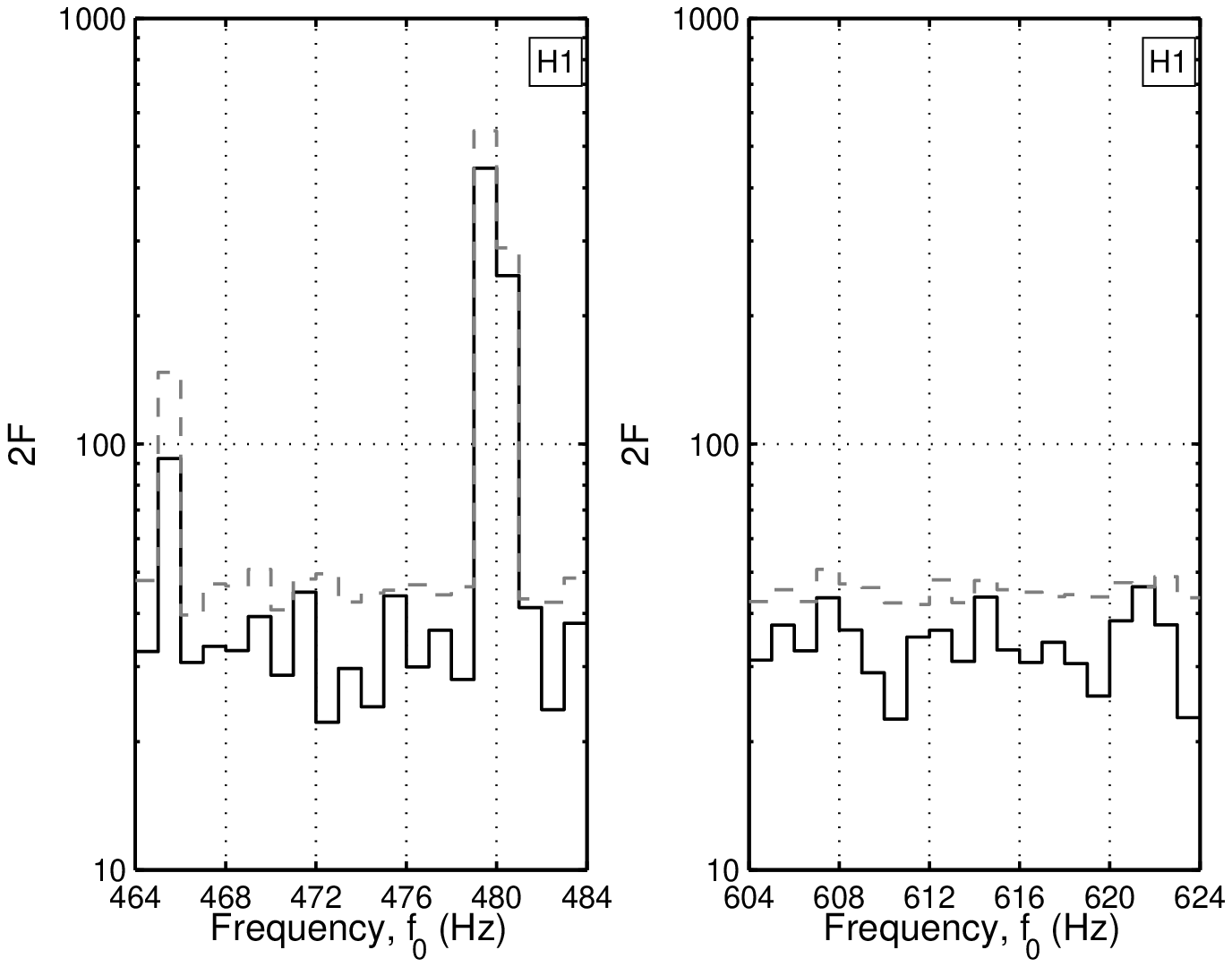}
\includegraphics[width=8.0cm]{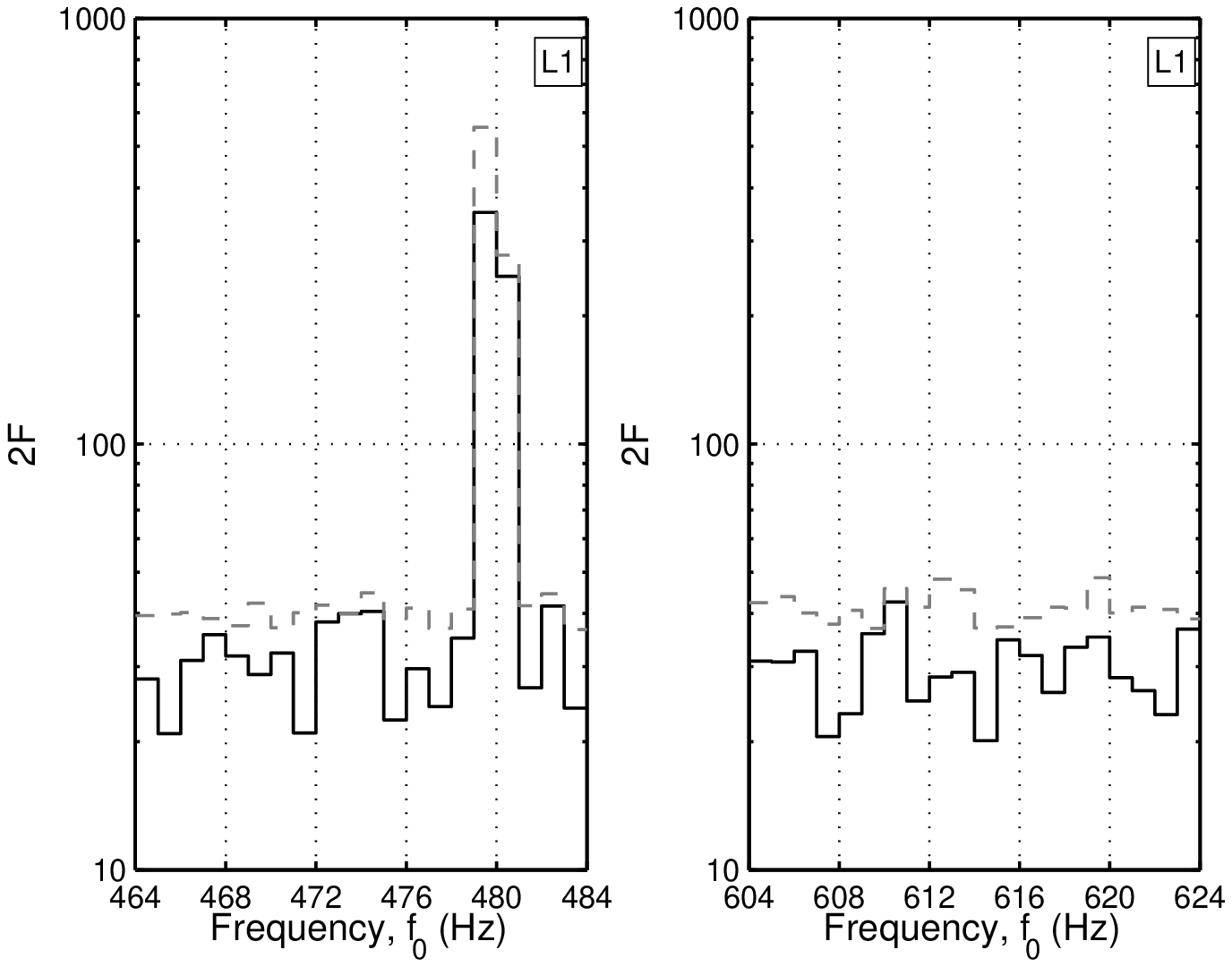}
\caption{Here we show the effect that the coincidence analysis has on the
loudest measured detection statistic in each $1$~Hz sub-band within the
Sco X-1 parameter space.  The solid black curves represent the loudest 
coincident $2\F$ values. The dashed gray curves represent
the $2\F$ values before the coincidence analysis.  Note that in clean sub-bands 
there is a reduction of $\sim 1.4$ in the loudest $2\F$ value.}
\label{f:scox1_coincidence_effect}
\end{figure}

\begin{figure}
\centering \includegraphics[width=8.3cm]{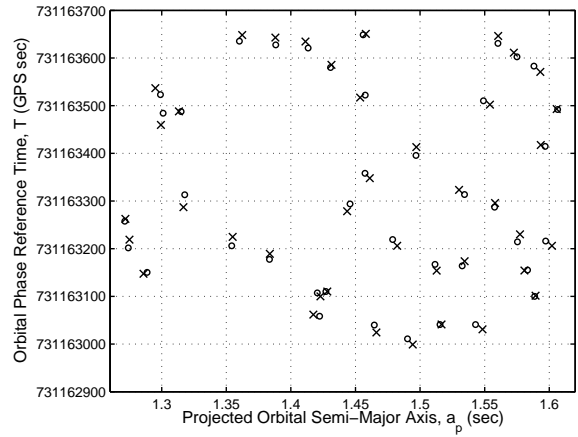}
\caption{The locations in orbital parameter space of the 
loudest events found in each $1$~Hz sub-band.  The crosses represent events
found in L1 and the circles represent events found in H1.}
\label{f:scox1_coincidence_orbital}
\end{figure}

The power of the coincidence analysis is shown in
Fig.~\ref{f:scox1_coincidence_effect} where the effect of the
coincidence constraint is seen to reduce the values of our loudest
events.  Before coincidence the average value of $2\mathcal{F}$ for the loudest events 
(excluding three 1 Hz sub-bands that contain major spectral disturbances: $465$--$466$~Hz and
$479$--$481$~Hz) was
$40.8$ for L1 and $45.4$ for H1.  After coincidence this becomes $28.6$ for
L1 and $33.5$ for H1 which corresponds to an improvement in $h_{0}$
sensitivity of $\sim 16\%$. This is broadly consistent with the results obtained
for the isolated neutron star analysis. In Fig.~\ref{f:scox1_coincidence_orbital} we
show the location of the coincident templates in the orbital parameter plane that produce the loudest
event in each of the 40 1-Hz sub-bands.

\subsection{Upper limits}
\label{ss:ULproc}

In every $\sim 1 $~Hz sub-band an upper limit on the amplitude of the
gravitational wave signal from either a population of isolated neutron
stars or from Sco X-1 is placed, based on the loudest coincident event
found in that band during the search.  The procedure employed is
conceptually identical to the one used in \cite{S1PulPaper} to set an
upper limit on the emission from J1939+2134, given the measured values
of the $\F$ statistic for that targeted search. In this section we
describe the Monte Carlo procedure in detail.

Let $s^*(f_0)$ indicate the measured value of the joint significance
of the loudest coincident event in the sub-band beginning at frequency $f_0$.
For every sub-band a set of $N$
injections of fake signals in the real data is performed at fixed amplitude
$h_0$. Each injection is searched for in the data and, if detected as
a coincident event, its significance is computed. A confidence
$C(h_0)$ is assigned to this set of injections
\beq
\label{eq:confidence}
C(h_0)=n(h_0)/N
\eeq
with $n(h_0)$ being the number of trials out of $N$ in which the
measured joint significance of the injected signal is greater than or equal
to
$s^*$. Eq.~(\ref{eq:confidence}) defines the $h_0$ upper limit  
value as a function of the confidence $C$.

For every injection in a set at fixed $h_0$ the remaining parameters
are chosen randomly from within the boundaries of our parameter space.
These include the orbital parameters (orbital semi-major axis and orbital 
phase reference time) for the Sco X-1 search, the sky position for the
all-sky search, and the frequency $f_{0}$ and the nuisance parameters
$\psi$, $\cos{\iota}$, and $\phi_{0}$ for both searches.  Uniform
distributions are used for $f_{0}$ in the sub-band, $\psi$ between $0$
and $2\pi$, $\cos{\iota}$ between $-1$ and $1$, and $\phi_{0}$ between
$0$ and $2\pi$. For the all-sky search the population of injected
signals is uniformly distributed on the celestial sphere, that is to
say that $\alpha$ is uniformly distributed between $0$ and $2\pi$ and
$\cos\delta$ uniformly distributed between $-1$ and $1$, with $\delta$
between $-\pi/2$ and $\pi/2$. For the Sco X-1 search the signal population
is uniformly distributed across the $2$ dimensional orbital parameter 
space (the parameters of which are given in Table~\ref{t:Sco-param}).  
The semi-major axis is selected from within the range $1.26$ to $1.62$ seconds 
and the orbital phase reference time is selected from within the GPS time range 
of $731163028$ and $731163626$.  The sky position and orbital period are 
held fixed at values corresponding to the center of their respective ranges.  
For each set of injections the orbital eccentricity is held fixed at one 
of the following discrete values: $e=0.0,10^{-4},5\times10^{-4},10^{-3},5\times10^{-3}$.

A search over the entire parameter space is {\it not} performed to
search for every injection---it is computationally
prohibitive. Rather, the detection statistic is computed {\it at the
nearest template} grid point with respect to the injected signal
parameters. The nearest template is chosen consistently with the
criteria used for laying out the template bank. For the Sco X-1 search
the closest grid point is defined by the metric governing the orbital
parameter space. For the isolated pulsar search a Euclidean measure is
used. In the actual search noise might conspire to produce a higher
value of the detection statistic at a template grid point which is
{\it not} the nearest to the actual signal's parameters. This means
that our Monte Carlo may slightly underestimate the detection
efficiency of the actual search, leading to an over-conservative (thus
still correct) upper limit. However, since our template bank has been
chosen so that at most a few percent of the detection statistic may
not be recovered at the nearest grid point due to signal-template
mismatch, we do not expect this effect to be severe. Furthermore, a
detection/coincidence scheme based on the global properties of the
detection statistic \cite{IP05} (far from the signal's true
parameters) remains to be understood.

A set of injections at fixed $h_0$ comprises at least $6000$ trials
for the isolated pulsar search and $5000$ trials for the Sco X-1
search.

\begin{figure}
\label{f:isolatedInj}
\centering \includegraphics[width=8.3cm]{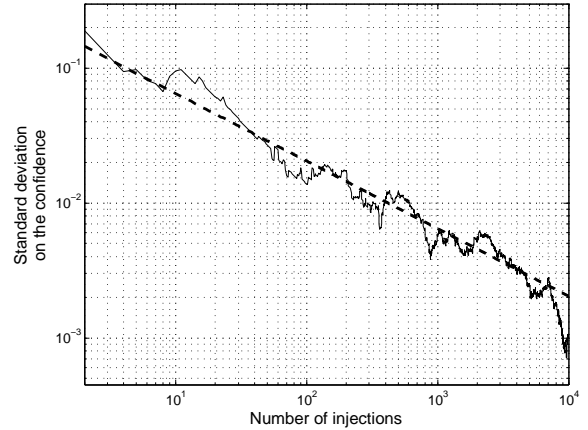} 
\caption{
A total of seven sets of ten thousand injections was performed in a
small band around 409~Hz at a strain $ h_0 = 1.2\times 10^{-22}$. For each set the confidence was estimated for different numbers of injections. The plot shows the standard deviation based on the seven estimates as a function of the number of injections. The dashed line shows the expected value of the standard deviation based on the binomial distribution with a single-trial probability of $95.6$, which corresponds to the measured mean confidence at $10000$ injections.}
\label{f:NoInj}
\end{figure}

To determine the number of injections, several sets of 10000
isolated pulsar injections have been analyzed. The injections were
performed at a fixed strain ($h_0 = 1.2\times 10^{-22}$) in a small band
around $409$~Hz, with sky locations and nuisance parameters
distributed as described above.  Figure~\ref{f:NoInj} shows
the results of this analysis. We plot the standard deviation on the confidence as a
function of the number of injections for seven sets of 
injections and compare it with the expected values. The plots show that above $5000$ injections the standard deviation on the confidence is less than $0.2\%$ and in agreement with the expectations even for small total number of injections.

\begin{figure}
\label{f:isolatedCvsh0Sets}
\centering \includegraphics[width=9.0cm]{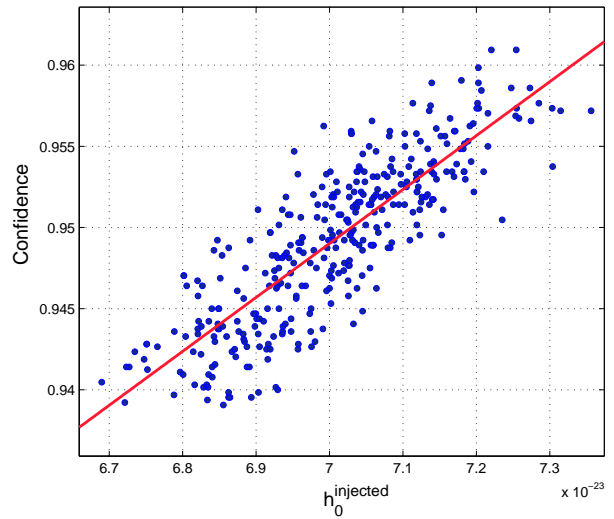} 
\caption{
Distribution of $h_0^{\rm{injected}}$ versus confidence values for sets of 6400 trials. The maximum spread in the confidence around the target $95\%$ value is about 2\% and the maximum spread in $h_0$ is about $10\%$. 360 points are plotted in this figure, corresponding to 360 sets, each comprising several thousands injections in the 247.6-248.8 Hz sub-band. The linear fit from these points is $C=3.3\times 10^{21} h_0^{\rm{injected}}+0.72$. For C=0.95 the estimated  $h_0^{\rm{injected}}$ is $6.9697\times 10^{-23}$. This is consistent with what is found in the actual analysis with a much smaller (order 10) set of injections: $(6.999\pm 0.04)\times 10^{-23}$.
}
\label{f:Cvsh0}
\end{figure}

For the isolated pulsar search the following approach has been adopted
to estimate the uncertainties related to a frequentist upper limit
based on signal injections.  The finite sample size of the population of signals 
that we construct by a Monte Carlo method 
results in fluctuations in the value of the confidence $C$
which we measure at fixed gravitational wave amplitude $h_0^{\rm{injected}}$. 
Figure~\ref{f:Cvsh0} shows a distribution
of ($h_0^{\rm{injected}}$,$C$) values for various sets of injections
around the target $95\%$ confidence value in a reasonably clean
sub-band of the data. Close to the target confidence the relationship
$C(h_0^{\rm{injected}})$ is well described by a linear
relationship. In order to estimate this relation we perform between 6
and 15 sets of injections. Each set is composed of at least 3200
injections and yields a value for ($h_0^{\rm{injected}}$, $C$). The
linear relation is then estimated from these ($h_0^{\rm{injected}},
C$) points with a standard best fit technique. We define $h_0^{95\%}$
as the value of $h_0^{\rm{injected}}$ yielding $C=95\%$ according to
the fitted linear relation.  From the fit we estimate the $\pm 1
\sigma$ ($h_0, C$) curves and from the intercepts of these with
$C=95\%$ the uncertainties on $h_0^{95\%}$, which we expect to be a 
few percent.

In the Sco X-1 search each set of $5000$ injections is divided into
$10$ subsets, each containing $500$ injections.  The confidence
$C_{i}(h_0^{\rm{injected}})$ is calculated for each subset of
injections where the index $i=1,\dots , 10$ labels each subset.  Values
of $h_{0,i}(C=95\%)$ are obtained by interpolation between the two
values of $C_{i}(h_0^{\rm{injected}})$ closest to $95\%$ within a
given subset.  The final value of $h_{0}^{95\%}$ is
calculated as the mean of $h_{0,i}(C=95\%)$ and the uncertainty in
this quantity is taken as the standard error in the mean,
$\sigma(h_{0}^{95\%})=\sqrt{\sigma^{2}/(10-1)}$, with $\sigma^2$ the variance of the $h_{0,i}(C=95\%)$ sample.  This approach has typically
yielded uncertainties in the values of the upper limit of $\sim
1-3\%$.

An additional and larger uncertainty arises from the instrument calibration, which varies with time and depends on the detector and the frequency band. Over the entire parameter space, we estimate that for the data sets used in the analyses presented here, the uncertainties amount to 11\% and 9\% for the isolated neutron star and Sco X-1 analysis, respectively. These estimates are conservative.

The tables in \cite{ULtables} detail all the upper limit results.
The uncertainties associated with the upper limit Monte Carlo procedure 
are reported separately from the calibration uncertainties and typically they are smaller than the latter.

\begin{table}
\begin{tabular}{lcc}\hline
       & P1 & P2 \\
\hline \hline
$f_0$ (Hz) & 1279.123457 & 1288.901235 \\
$\dot{f}$ (Hz/s) & 0 & $-10^{-8}$ \\
$\alpha$ (rad) & 5.147162 & 2.345679 \\
$\delta$ (rad) & 0.376696 & 1.234568 \\
$\psi$ (rad) & 0 & 0    \\
$\cos\iota$ & 0 & 0 \\
$\Phi_0$ & 0 & 0 \\
$T_0$ (sec) & 733967667.12611231 & 733967751.52249038\\
$h_0$ & $2 \times 10^{-21}$ & $2\times 10^{-21}$ \\
SNR H1 (exp./meas.) & $17/18$ & $34/35$ \\
SNR L1 (exp./meas.) & $20/22$ & $21/22$ \\
\hline
\end{tabular}
\caption{ Parameters of the two hardware injected pulsars.  See
  Eqs.~(\ref{eq:ht}) and (\ref{e:time}) for
  the definition of the parameters.  $T_0$ is the reference time for
  the initial phase, in GPS seconds in the SSB frame.  }\label{tab:S2hdw}
\end{table}

\subsection{Validation: Hardware injections}
\label{ss:hardinj}

Signals can be injected into the instrument via the actuator, by
physically moving the mirrors of a Fabry-Perot cavity to mimic a
gravitational wave signal. Hardware injections are designed to give an
end-to-end validation of the data analysis pipeline, including some,
but not all, components of the calibration. Toward the end of the S2
run, two simulated isolated pulsar signals were injected into the
data. We denote the two pulsars P1 and P2 and give their parameters in Table \ref{tab:S2hdw}.

The data sets were prepared using the final S2 calibration version
\cite{S2caldoc}, and consist of 17 30-minute SFTs for H1 and 14 30-minute SFTs
for L1.  We performed a targeted search to look for the pulsar
signals.

The expected SNRs are computed using a noise estimation technique that
accounts for the amplitude modulation of the signal throughout the
observation time. The results are shown in Table \ref{tab:S2hdw}. The
agreement is very good. The measured SNRs, however, are systematically
somewhat larger than expected. This is probably due to a small
systematic error in the calibration. The differences between the
expected SNR values shown here, and those quoted in
\cite{LSC:05-S2TD}, arise primarily from differences in the
lengths of observation times used to make the estimate.  In \cite{LSC:05-S2TD}, a nominal
observation time of 12 hours was used. This is the length of time during which the hardware injections were performed. Here we have 
used the actual science observation time which is shorter, reflecting science quality data and calibration quality flags based on which we discard data as not reliable enough to be included in an astrophysical search.

\section{Results}
\label{s:results}

In this section we present the results of the analysis performed using the 
pipeline shown in Fig.~\ref{f:pipeline} and described in Section~\ref{s:analysis}. We first discuss the results regarding the all-sky search for isolated
neutron stars and then turn to upper limits on radiation from Sco X-1.

\begin{figure*}
\centering \includegraphics[width=15.0cm]{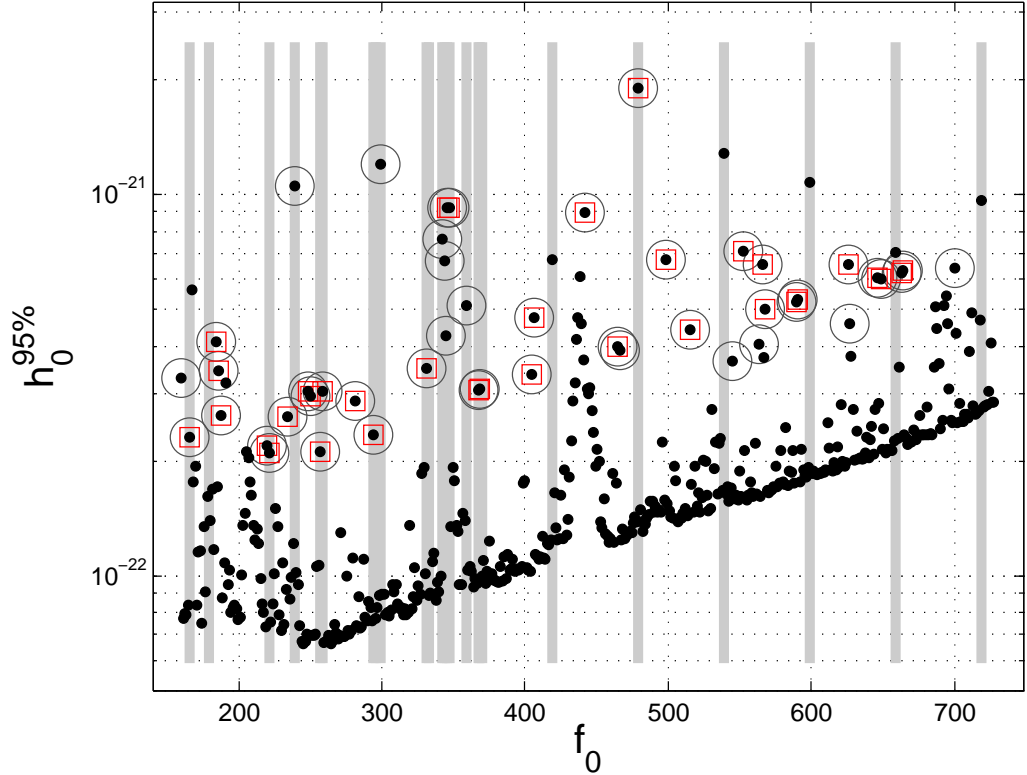} 
\caption{
Upper limits based on the loudest template over the whole sky in 1.2~Hz sub-bands. The vertical stripes mark the sub-bands containing known spectral disturbances. The circles mark the 90th percentile most significant results. The squares indicate that the values of the detection statistic in the two detectors are not consistent with what one would expect from an astrophysical signal.}
\label{f:isolatedULs}
\end{figure*}

\subsection{Isolated neutron stars}
\label{ss:isolatedResults}

Figure~\ref{f:isolatedULs} shows the $95\%$ upper limits on
$h_0$ for every $1.2$~Hz wide sub-band over the whole sky. The values of
the frequency refer to the lower extremum of each sub-band.
(The upper limit values, along with their estimated uncertainties,
may also be found in tabular form in ~\cite{ULtables}.)
The circles around
the upper limit dots mark points in the 90th percentile in joint significance. About 2/3 of these points are also in
the 90th percentile for $h_0^{95\%}$.

\begin{figure*}
\includegraphics[width=7.0cm]{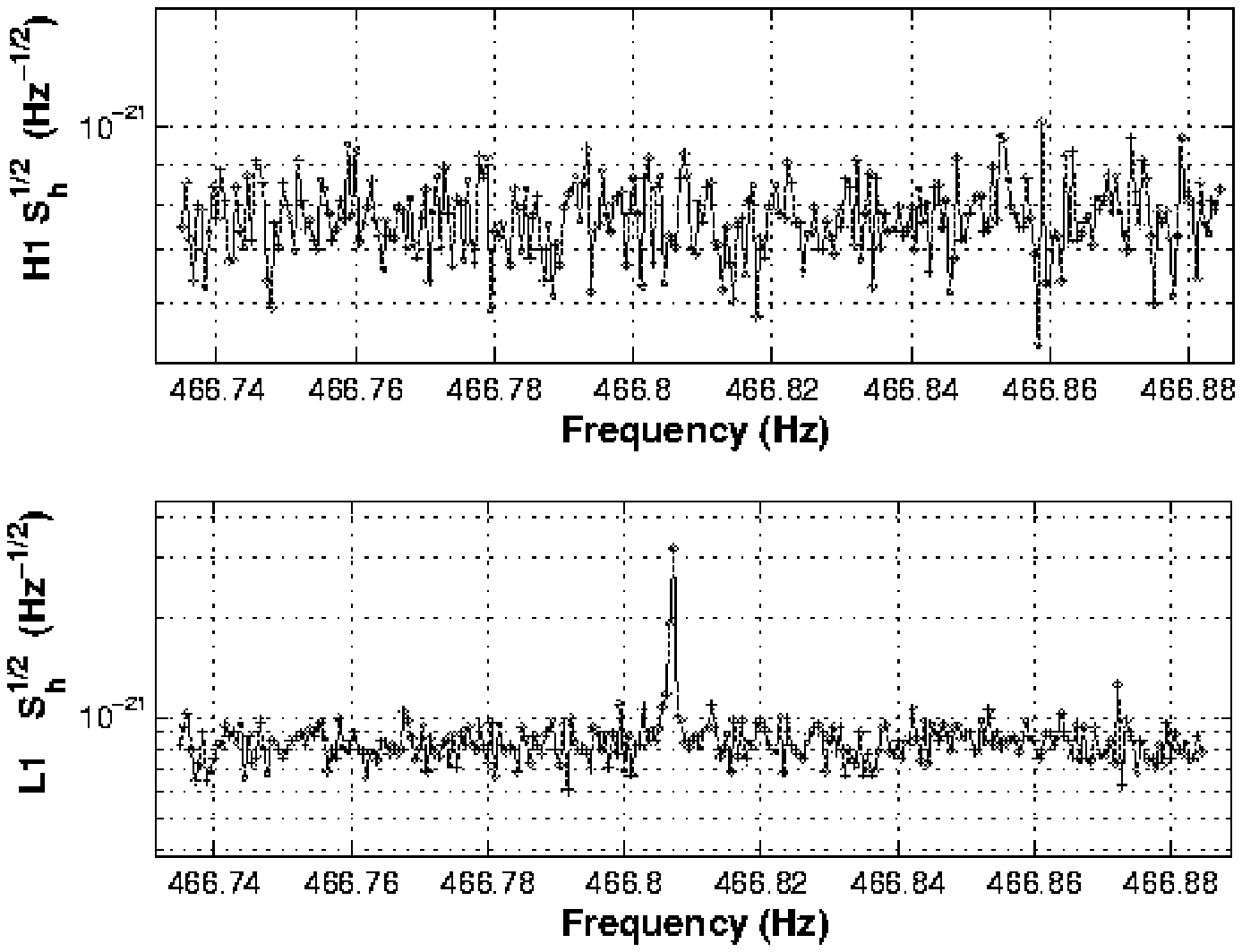}
\includegraphics[width=7.0cm]{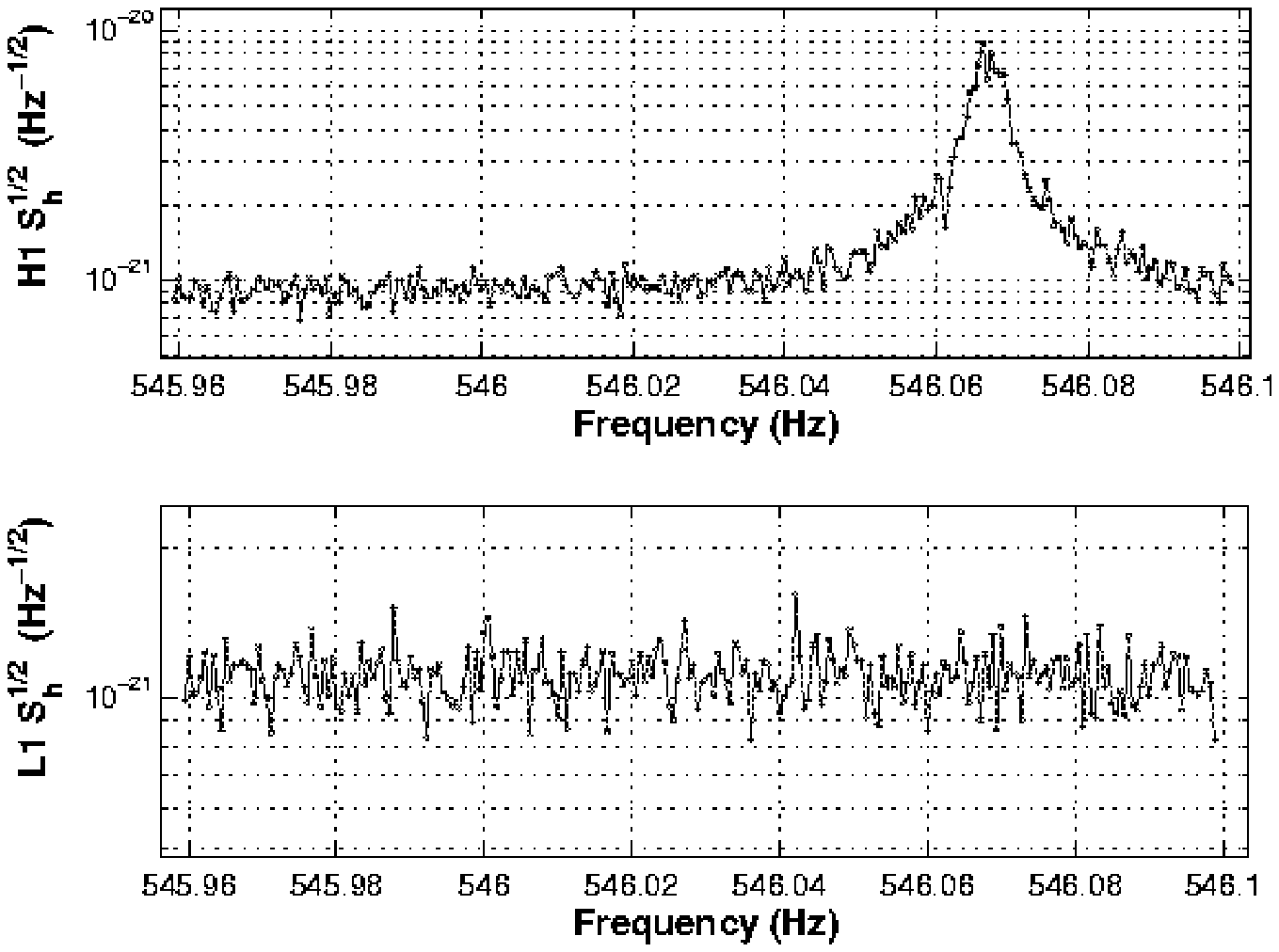} 
\includegraphics[width=7.0cm]{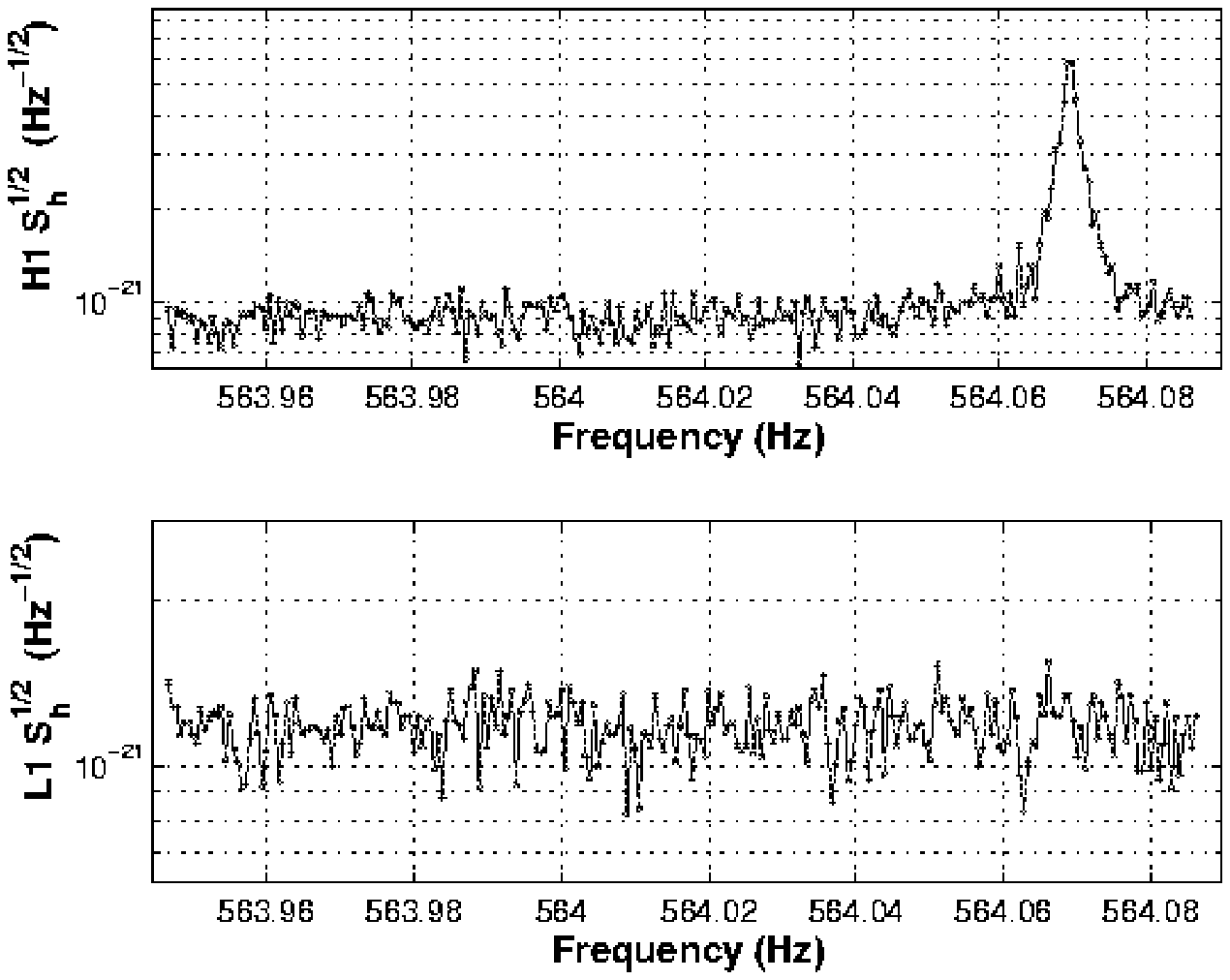} 
\includegraphics[width=7.0cm]{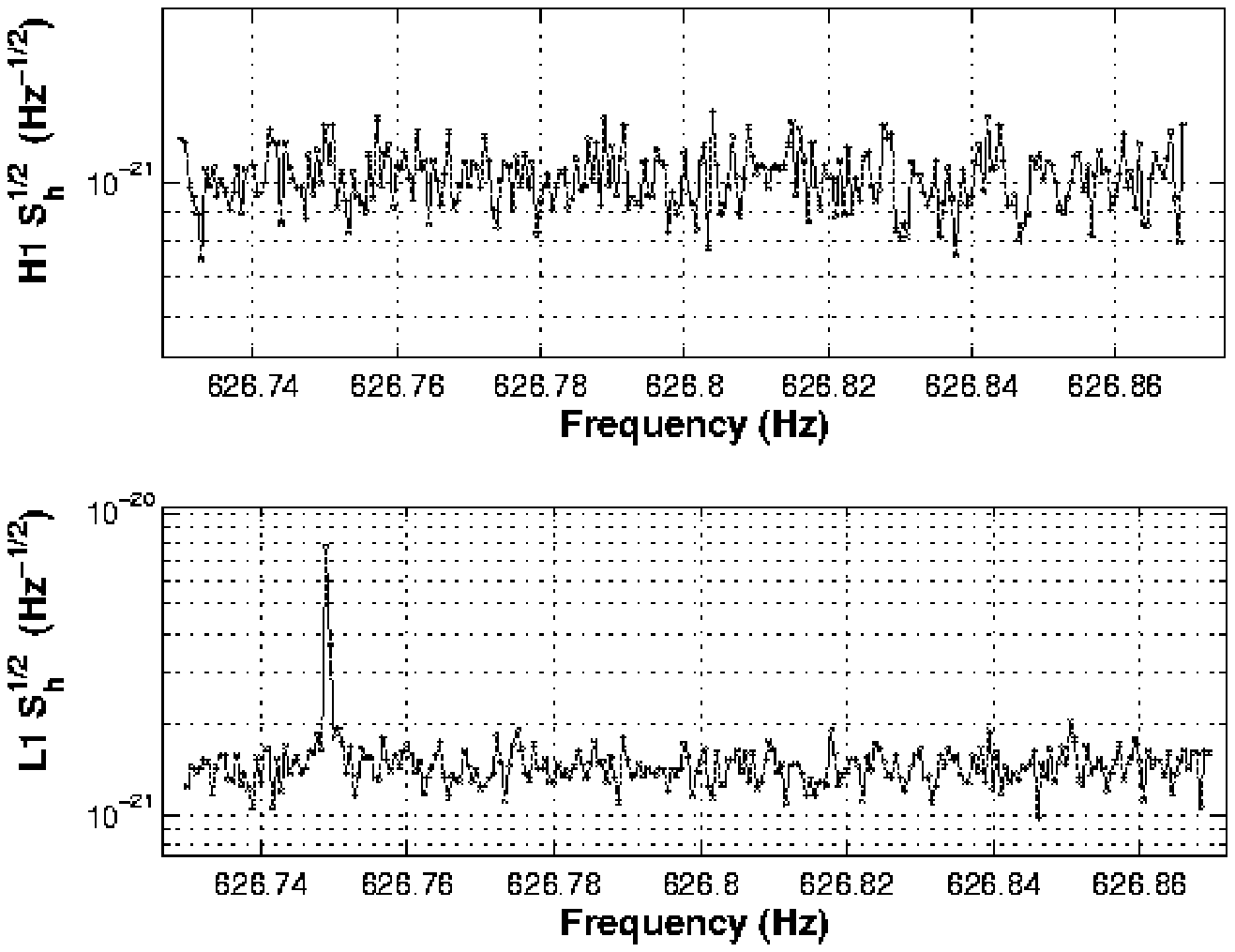} 
\includegraphics[width=7.0cm]{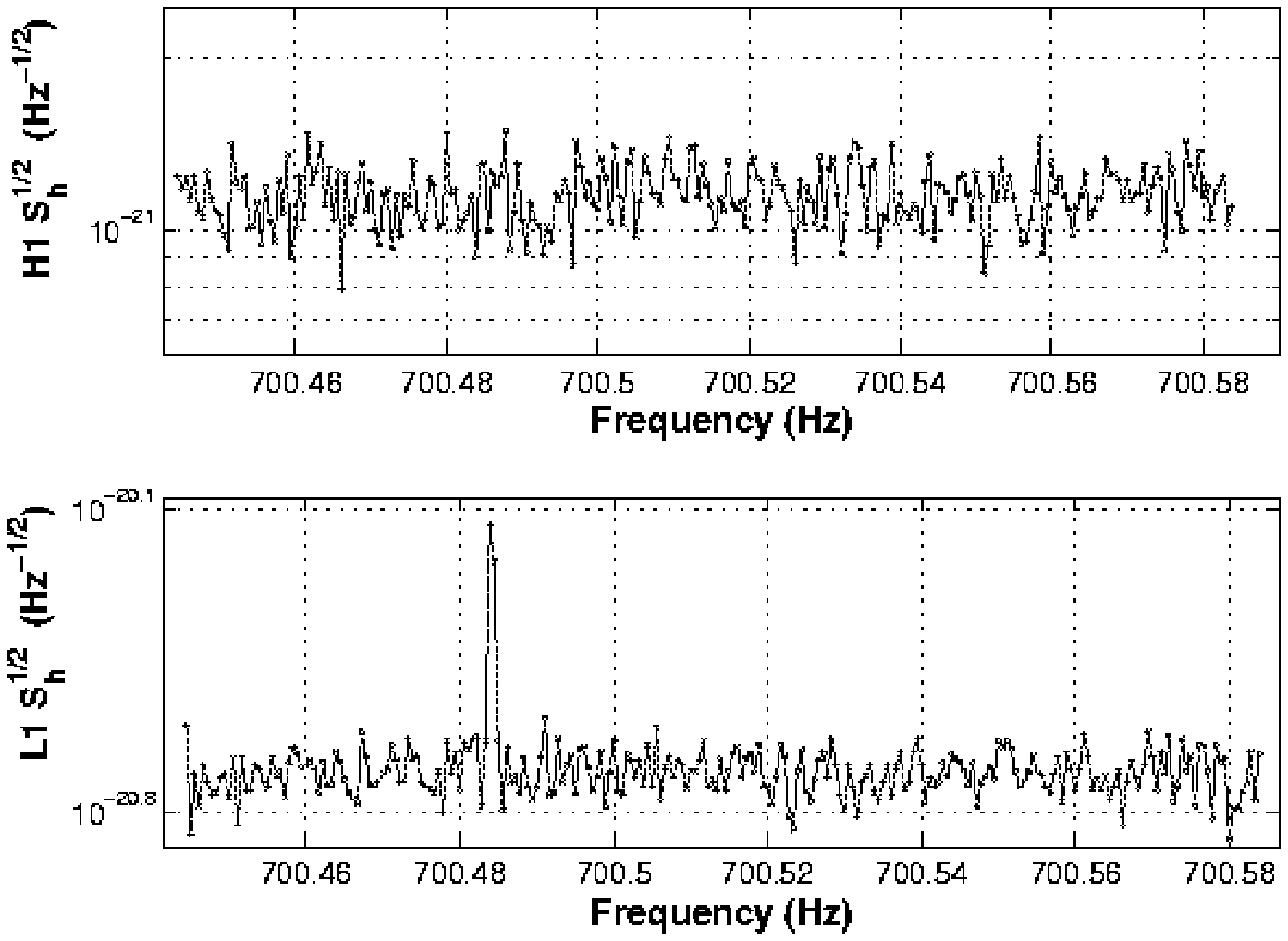}
\caption{The average power spectral density in the two
detectors in the frequency regions around the five not immediately
explained outliers from the search for isolated pulsars. The width of
these disturbances is sufficiently small that they could not be
discarded as of non-astrophysical origin based on this. A peak in
power spectral density is clearly visible only in one of the
detectors, but the measured values of the detection statistics are
not inconsistent with an astrophysical signal, albeit a
rare one. As explained in Section~\ref{ss:isolatedResults} we very
strongly suspect these excesses of power are not due to a continuous
wave source because of their inconsistent amplitudes in science runs
subsequent to S2.
}
\label{f:psdz4}
\end{figure*}

\begin{figure}
\centering \includegraphics[width=8.85cm]{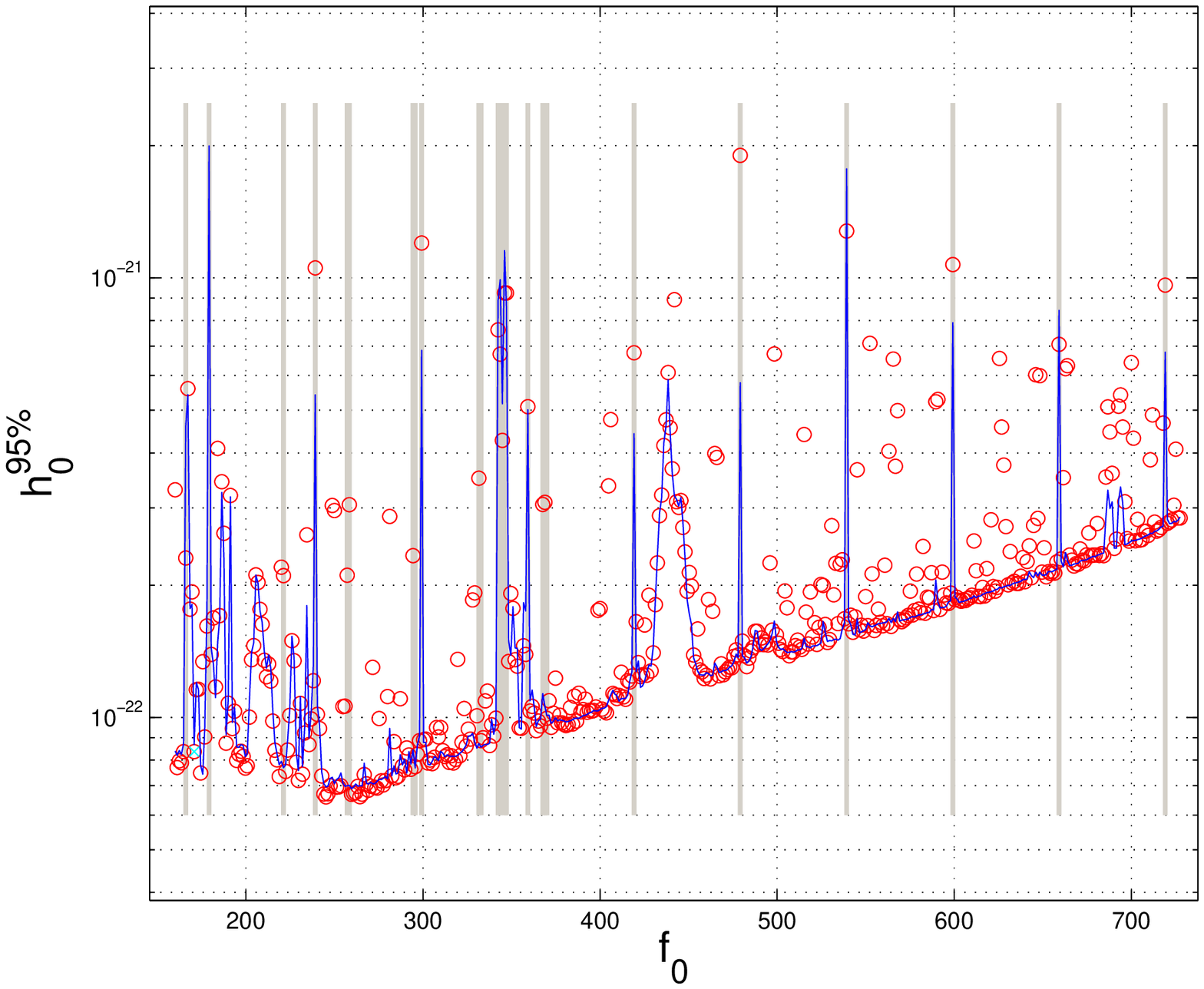} 
\centering \includegraphics[width=8.85cm]{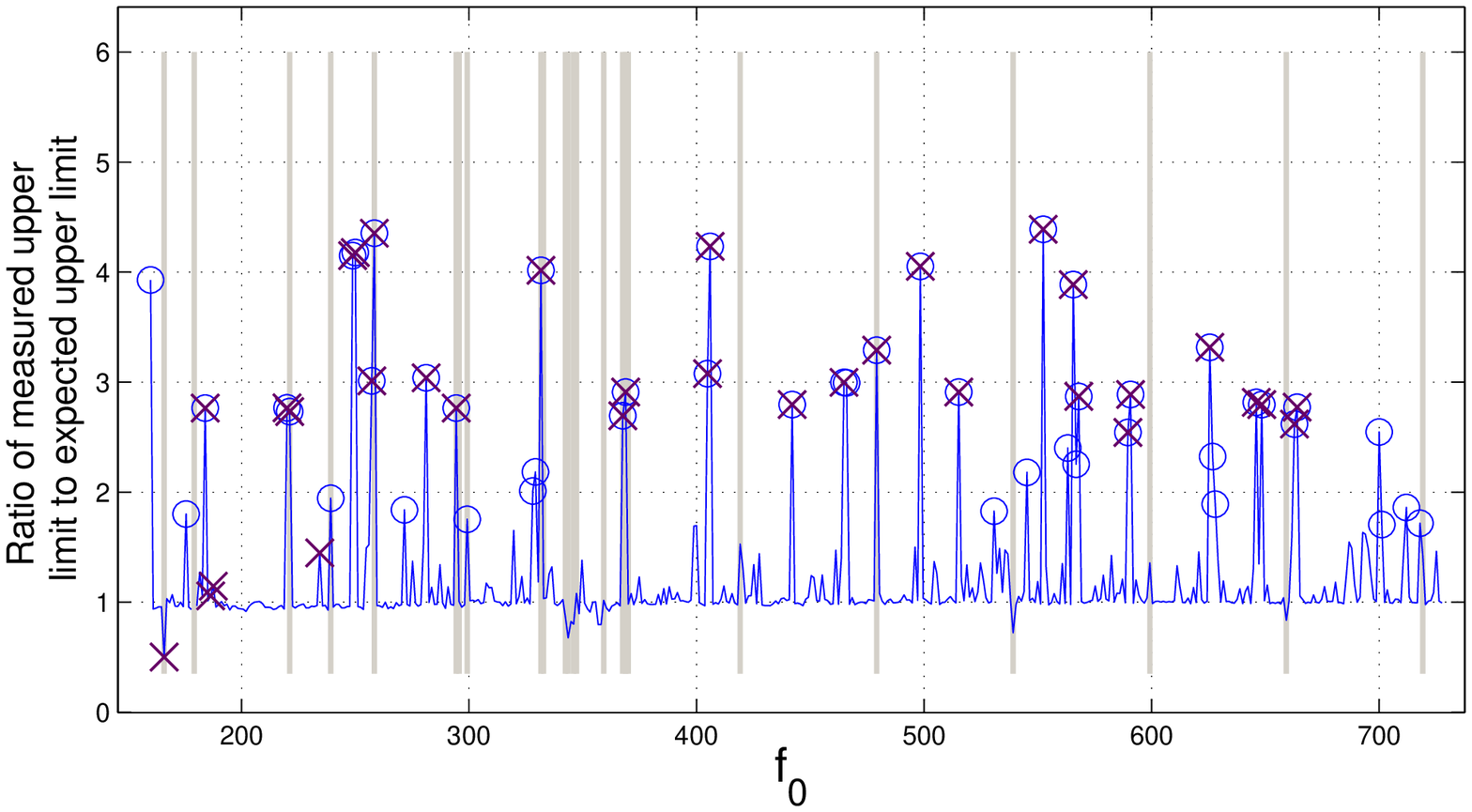} 
\caption{
Top plot: the circles are the upper limits based on the loudest template over the whole sky in 1.2~Hz sub-bands. The solid line is the expected upper limit under the assumption that the noise in the detectors is Gaussian and stationary. Bottom plot: ratio of the measured upper limit values to the expected upper limit values. The circles indicate the 90th percentile values of the measured to expected upper limit ratio. The crosses mark the upper limit values for which the ratio of the detection statistic values in the two detectors is not consistent with what we would expect from a signal. In both plots the shaded regions indicate frequency bands affected by known spectral disturbances.
}
\label{f:GSNexpectedULs}
\end{figure}

About one quarter of the 90th percentile significance points lie in sub-bands influenced by
spectral disturbances (points with circles on shaded bands in Fig.~\ref{f:isolatedULs}). 
Most of the remaining points can be immediately attributed 
to non-astrophysical sources because the ratio of the $\F$-statistic values
in the two detectors is
either too large or too small to be consistent with being due to the same
signal. These points are indicated in Fig.~\ref{f:isolatedULs} by a square. There remain 6 points which are in the 90th percentile in significance and cannot be excluded based on the ratios of the $\F$-statistic values. They appear at the frequencies 160.0~Hz, 466.79~Hz, 546.03~Hz,
564.02~Hz, 626.80~Hz, and 700.51~Hz. The 160.0~Hz frequency coincides with the 10-th harmonic of 16.0~Hz, a key operating frequency of the data acquisition system. We are thus confident that the origin of this outlier is instrumental. The points at 466.79~Hz, 626.80~Hz, and 700.51~Hz are due to lines only 
in the L1 instrument which have disappeared in science runs subsequent
to S2. This check suggests that the outliers are of instrumental origin. The 546.03~Hz and 564.02~Hz points are due to lines which clearly appear only in
H1. However the lines are present in the S2 run and in later science runs. The amplitude of both lines {\it decreases} with increasing sensitivity of the instrument, dropping by a factor of 10 (in noise power) as the sensitivity increases by a factor greater than 5. This indicates a behavior which is not consistent with the model of the signal that we are searching for here and suggests that these lines are of instrumental origin. Figure~\ref{f:psdz4} shows the average power spectral density in both detectors in the frequency regions where these outliers are located.

Unlike what is described in \cite{S2Hough} no frequency band is excluded from the upper limit analysis due to it being contaminated by known noise artifacts. This
results in extremely loud events in some sub-bands: those
containing the 60~Hz power line harmonics, the L1 calibration
line (at 167~Hz), the violin modes of the suspension wires of the test mass (in the 340--350~Hz region) and the various oscillator harmonics at multiples of 36~Hz together with the beating of the 0.74~Hz pendulum mode of a
test mass against the oscillator line (in the 220-335~Hz region). In the case of the 179.4~Hz sub-band 
containing the 180~Hz power line harmonic, the 
spectral disturbance is so strong
that the upper limit Monte Carlo does not converge to an upper limit $h_0^{95\%}$ value.  The sub-bands marked by shaded vertical stripes indicate frequencies where known spectral artifacts are present. 

The upper limit values presented here are in broad agreement with what is expected and we consider this a further validation of the analysis pipeline. We have run the pipeline presented here on Gaussian stationary noise and empirically derived a formula for the expected $h_0^{95\%}$ as a function of the noise level in the detectors:
\beq
h_0^{95\%} \sim 29.5 \sqrt{\langle S_h \rangle \over 10~{\rm hrs}}\,,
\label{eq:GSNexpectedULs}
\eeq
with $\langle S_h \rangle$ being the average noise level over 1.2 Hz and over the observation time in every detector and then averaged over the detectors: this quantity is shown in Fig.~\ref{f:PSDs}. We would like to stress that Eq.~(\ref{eq:GSNexpectedULs}) refers to this particular analysis and pipeline. The expected upper limits for Gaussian stationary noise are plotted against the measured ones in Fig.~\ref{f:GSNexpectedULs}. It is clear that in regions where the data is not Gaussian and stationary Eq.~(\ref{eq:GSNexpectedULs}) does not predict correctly the values that we measure and the discrepancy between the prediction and the measured value depends on the details of the spectral disturbance and of the method used for estimating the noise. This is particularly evident close to spectral disturbances, where clearly the noise is not white Gaussian and often not stationary and the  predictions can even be larger than the actually measured upper limit value (see the points below 1 in the lower plot of Fig.~\ref{f:GSNexpectedULs}). However the ratio of the measured upper limits to the expected one never exceeds $4.4$ and the 90th percentile level in this ratio is 1.7. 

\begin{figure*}
  \centering \includegraphics[width=12.5cm]{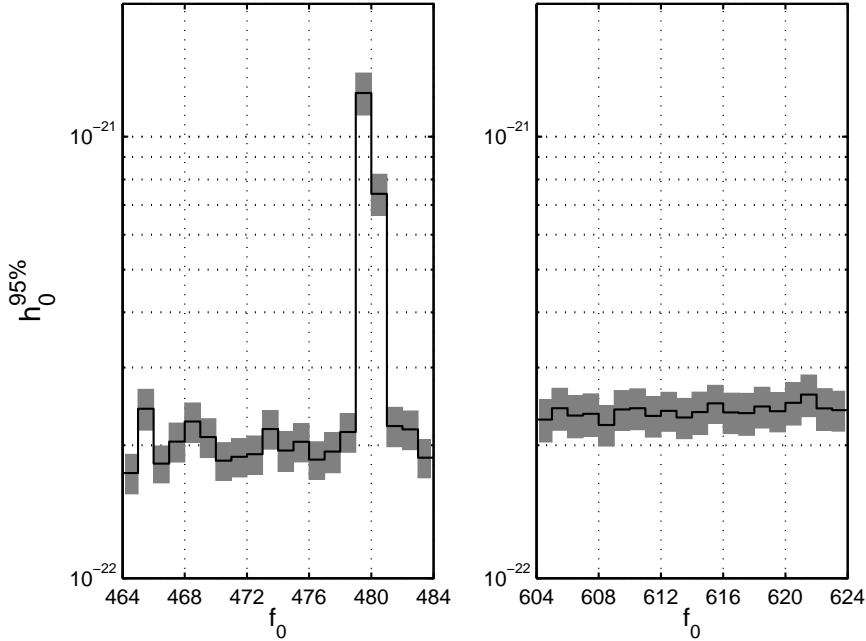}
\caption{The upper limits on the amplitude of gravitational wave radiation from Sco X-1. The plot shows 
the limits on $h_{0}$ at 95\% confidence (solid line) as a function of frequency. We report one limit
for every 1~Hz sub-band and the Sco X-1 orbit is assumed to be exactly circular. The shaded region shows
the combined errors on $h_{0}^{95\%}$ due to the injection process (typically in the range 
$\pm 2-5\times 10^{-24}$) and instrument calibration. The latter, estimated at the level $\approx 9\%$,
dominates the uncertainties.}
\label{f:SCO_UL}
\end{figure*}

\subsection{Sco X-1}

The upper limits on gravitational waves from Sco X-1 are summarized in Fig.~\ref{f:SCO_UL} 
(more details are provided in ~\cite{ULtables}): we show $h_0^{95\%}$
over 1~Hz sub-bands in the range $464$--$484\,
\mathrm{Hz}$ and $604$--$624\,\mathrm{Hz}$, assuming 
that the source is in an exactly circular orbit. We would like to stress that
these limits apply to a source whose orbital parameters lie in the region
reported in Table~\ref{t:Sco-param}, corresponding to 1-$\sigma$ errors.
The typical value of $h_0^{95\%}$ is $\approx 2\times 10^{-22}$ over the 
whole analyzed 40 Hz band, with the exception
of a band $\approx$ 2~Hz around 480~Hz which corresponds to one of the strong harmonics 
of the 60~Hz power-line, cf.\ Fig.~\ref{f:sconoise}. In this region the upper limit is $h_0^{95\%} 
\approx 10^{-21}$. Such values are consistent with the sensitivity estimates
shown in Fig.~\ref{f:ExpSens_sco}, 
which were derived under the assumption of Gaussian and stationary noise and include a number
of approximations to quantify the effects of each stage of the pipeline considered in
this search.  Through the statistical modelling of the pipeline we are able to express 
the expected $h_{0}^{95\%}$ upper limit as 
\be
h_{0}^{95\%}(f)\sim 28\sqrt{\frac{\langle S_{h}(f)\rangle}{6\,\rm hrs}},
\ee
where $\langle S_{h}(f)\rangle$ is the noise level over the 1 Hz 
sub-bands averaged over the 
observation time, the frequency band and the detectors.
  
\begin{figure}
\centering \includegraphics[width=8.3cm]{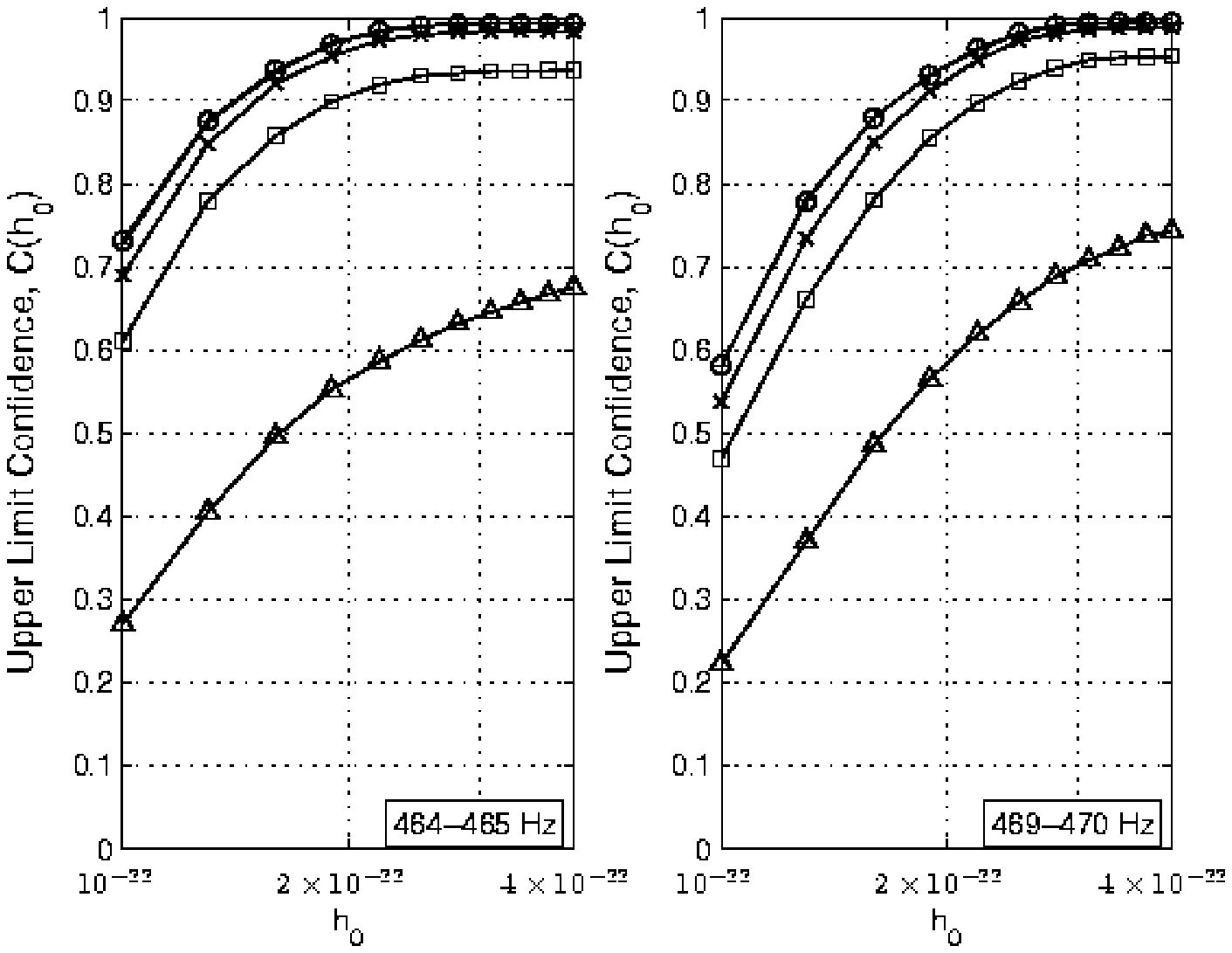}
 \centering \includegraphics[width=8.3cm]{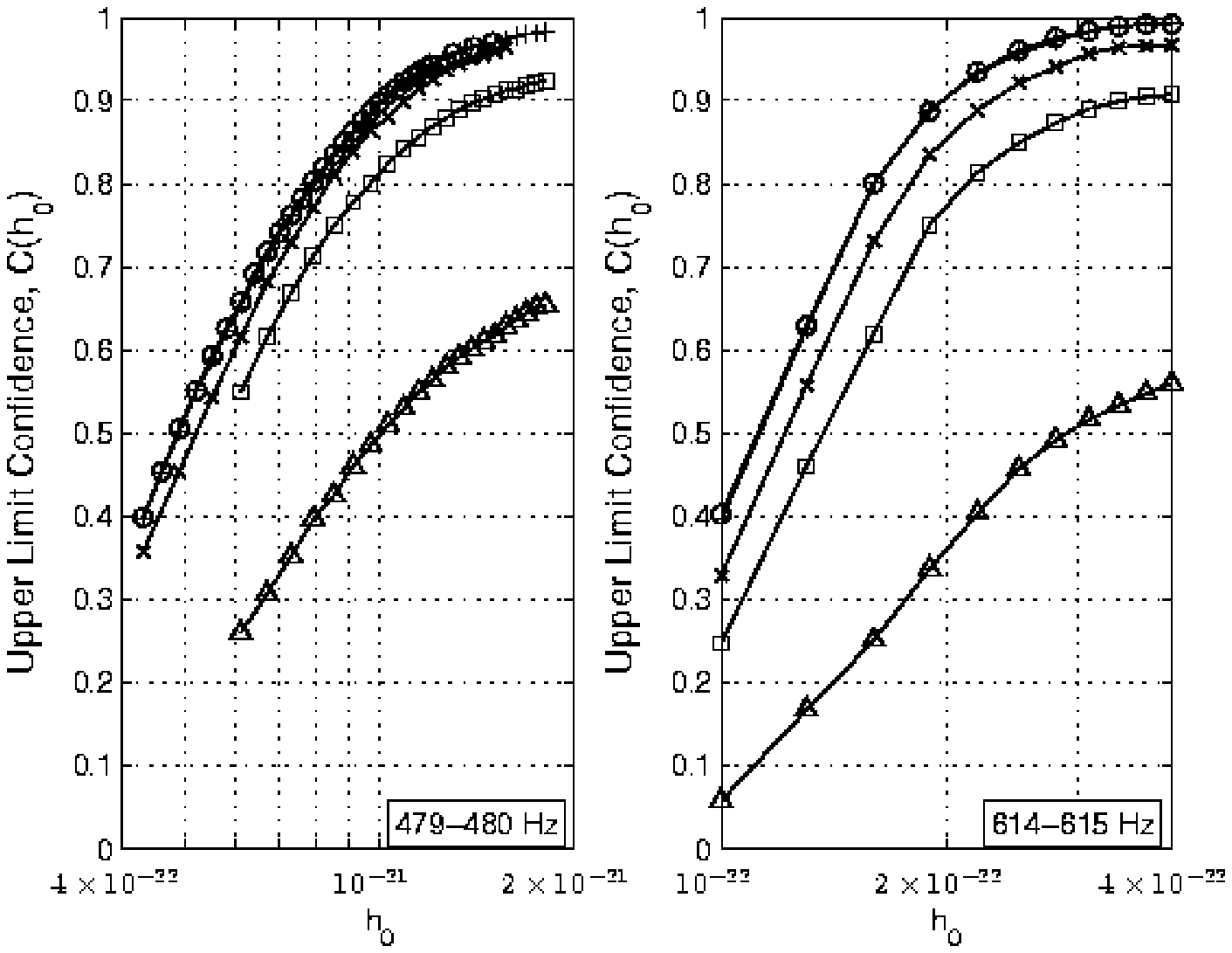}
\caption{Examples of how the confidence in the upper limit on $h_0$ from Sco X-1  scales
with the amplitude of injected signals.  Here we show $4$ plots 
each corresponding to a different $1$~Hz sub-band.  The sub-bands 
$464$--$465$~Hz, $469$--$470$~Hz, and $614$--$615$~Hz show the confidence
across the range of typical injected $h_{0}$ values 
($10^{-22}$ -- $4\times10^{-22}$).  The sub-band $479$--$480$~Hz contains a
large spectral disturbance and the injections are appropriately 
increased in amplitude in order to achieve the required confidence.
The 5 curves in each plot represent the confidence associated with different values 
of orbital eccentricity.  The circles represent $e=0$, the plusses represent $e=10^{-4}$,
the crosses represent $e=5\times10^{-4}$, the squares represent $e=10^{-3}$ and the triangles represent $e=5\times 10^{-3}$.
Note that for the $6$ hour observation, the pipeline is as sensitive 
to signals with values of $e<10^{-4}$ as it is to circular orbits.
For signals with $e=10^{-3}$ the pipeline has a maximum detection 
efficiency of $\sim 90$--$95\%$.  For $e=5\times10^{-3}$ the maximum 
detection efficiency reaches only $\sim 50$--$70\%$.}  
\label{f:confidence}
\end{figure}

 In Fig.~\ref{f:confidence} we show the value of $h_0^C$ as a function of
$C$ for selected frequency sub-bands. We have considered both
``quiet'' and ``noisy'' spectral intervals, but have restricted this
analysis to only four 1~Hz frequency sub-bands due to computational
burdens.  Figure~\ref{f:confidence} shows that for $C = 0.99\, (0.5)$
the upper limit on $h_0$ would be a factor $\approx 2$ larger
($\approx 2$ smaller) than $h_0^{95\%}$.

\begin{figure*}
  \centering \includegraphics[width=8.3cm]{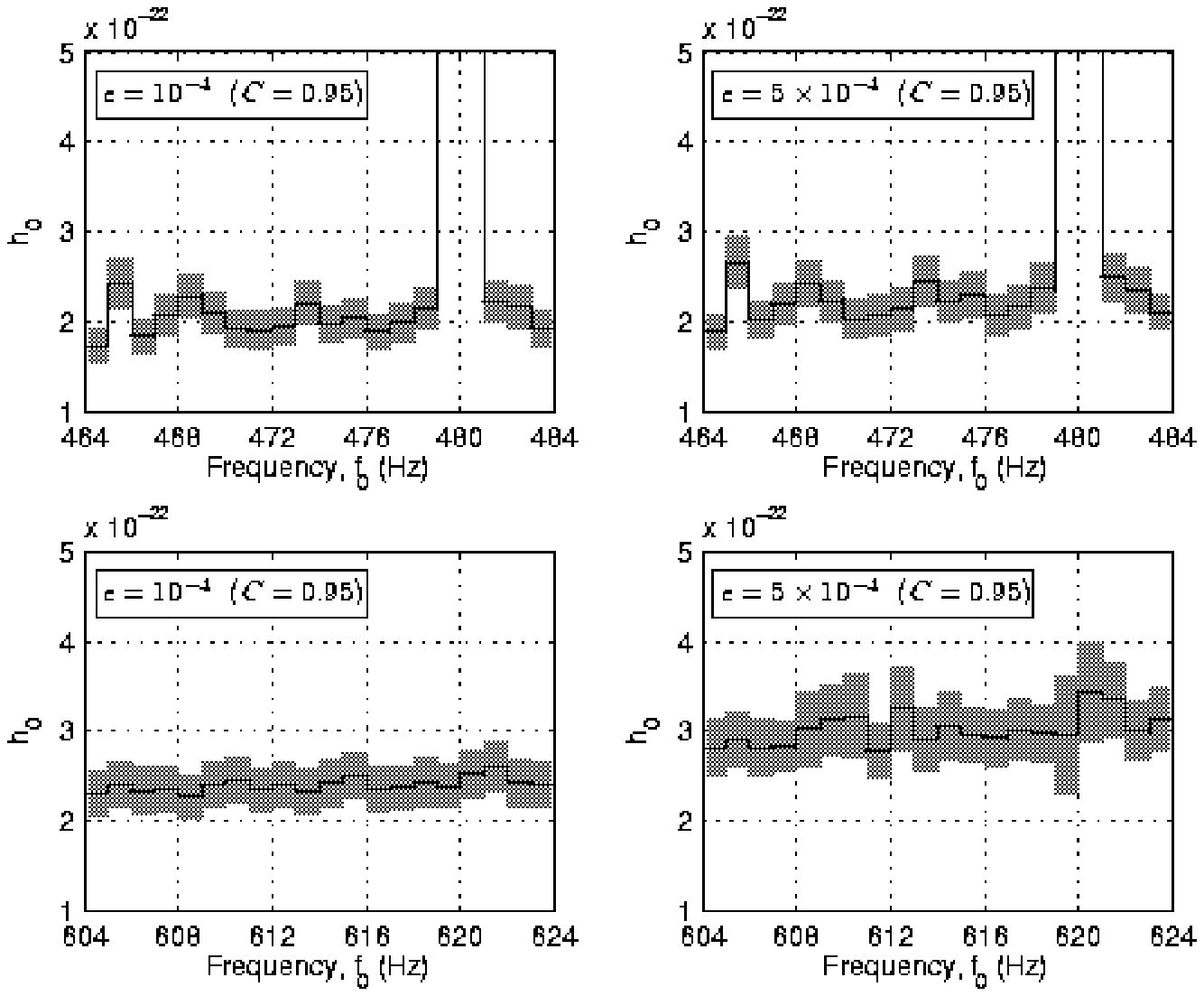}
  \centering \includegraphics[width=8.3cm]{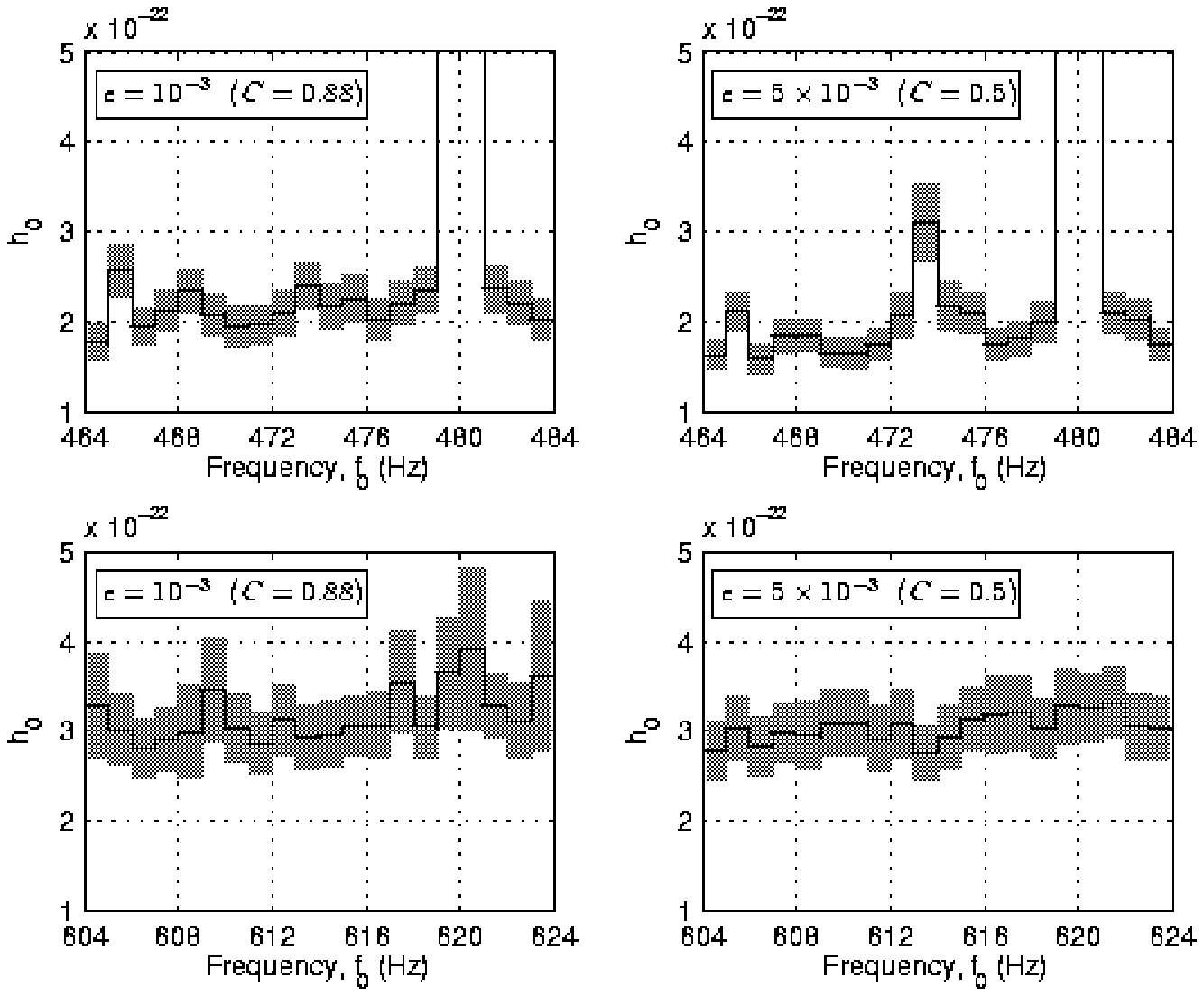}
\caption{The upper limits on $h_{0}$ as a function of frequency for different orbital eccentricities 
for the Sco X-1 search. Notice that for sake of clarity, we have adopted a different scale on
the vertical axis than that used in Fig.~\ref{f:SCO_UL}; as a consequence the upper limits 
in the frequency region 479 -- 481~Hz are not shown because they are off scale. 
Due to the systematic loss
of signal-to-noise ratio due to signal-template mismatch for populations of signals from sources in eccentric orbits, the confidence $C$ at which the upper limit $h_0^C$ is computed is different depending on the
value of eccentricity $e$; see text for a detailed justification. We report $h_0^{95\%}$ for $e = 10^{-4}$,
$h_0^{95\%}$ for $e = 5\times 10^{-4}$, $h_0^{88\%}$ for $e = 10^{-3}$, and $h_0^{50\%}$ for 
$e = 5\times 10^{-3}$. The shaded areas represent the combined errors due to the injection process
and the instrument calibration.}
\label{f:SCO_UL2}
\end{figure*}

So far we have restricted the discussion of the upper limits to the
case of an exactly circular orbit. This is the model that we have
assumed in building the templates used in the analysis. As we have
discussed in Sec.~\ref{s:s:sco}, the orbital fits of the optical data
are consistent with $e = 0$, which is in agreement with the
theoretical expectations inferred from evolutionary models. However
present observations are not in a position to constrain the
eccentricity to $e \simlt 10^{-4}$, which would introduce {\em
systematic} losses of signal-to-noise ratio smaller than $\approx
0.1$, the value of the mismatch adopted for this search. It is
therefore important to explore the consequences of a (unlikely but
possible) non-zero eccentricity of the Sco X-1 orbit on the results
reported so far. The pipeline that we have developed allows us to
quantify this effect in a fairly straightforward way: the Monte Carlo
software injections used to set upper limits are performed again by drawing
signals from a population of binaries where now the eccentricity is
set to a (constant) value $e \ne 0$. The orbital parameters and the
frequency are chosen randomly exactly as in the case for a circular
orbit. We detect the signals from eccentric orbits with the search
pipeline constructed with a bank of filters for a perfectly circular
orbital model. In this way we can quote consistently an upper limit on
$h_0$ for $e \ne 0$. We repeated this procedure for selected values of
the eccentricity, $e = 10^{-4}, 5\times 10^{-4}, 10^{-3}$ and $5\times
10^{-3}$. The dependence of the value of the upper limit on $h_0$ as a
function of the confidence for four representative frequency sub-bands
is shown in Fig.~\ref{f:confidence} and the upper limits over the
whole 40~Hz region (at fixed confidence) are summarized in
Fig.~\ref{f:SCO_UL2} and in ~\cite{ULtables}. 
Notice that in this case we choose
different values of the confidence depending on the eccentricity of
the orbit of the putative source population used for the
injection. This stems directly from the fact that the detection
efficiency of the search pipeline is progressively reduced as the
model of the injected signals differs more and more from that of the
detection templates. In other words we suffer from systematic losses
of signal-to-noise ratio due to the fact that the templates are not
properly matched to the signal: for $e \simgt 10^{-3}$ the fitting
factor of a filter generated by modelling Sco X-1 as a circular orbit
binary is $< 0.9$. Indeed, regardless of the strength of the injected
signals, the pipeline is unable to detect at least 95\% of them, see
Fig.~\ref{f:confidence}.  We find that for $e = 10^{-3}$ and $5\times
10^{-3}$ the pipeline has a maximum detection efficiency in the ranges $\sim
90$--$95\%$ and $\sim 50$--$70\%$, respectively. As a consequence, for $e =
10^{-3}$ we report $h_0^{88\%}$, and for $e =
5\times 10^{-3}$ we consider $h_0^{50\%}$, because across each of the 1 Hz
sub-bands we have achieved at worst an 88\% and 50\% confidence, respectively 
(see Fig.~\ref{f:SCO_UL2}). On the other hand, for $e \le 5\times 10^{-4}$
the systematic loss of signal-to-noise ratio is small or even
negligible and we can quote upper limits at 95\% confidence. We find
that, as expected, the values of $h_0^{95\%}$ for $e = 0$ and $e =
10^{-4}$ are essentially identical. For $e = 5\times 10^{-4}$,
$h_0^{95\%}$ is about 50\% bigger than in the case $e = 0$.

\begin{figure}
\centering \includegraphics[width=8.3cm]{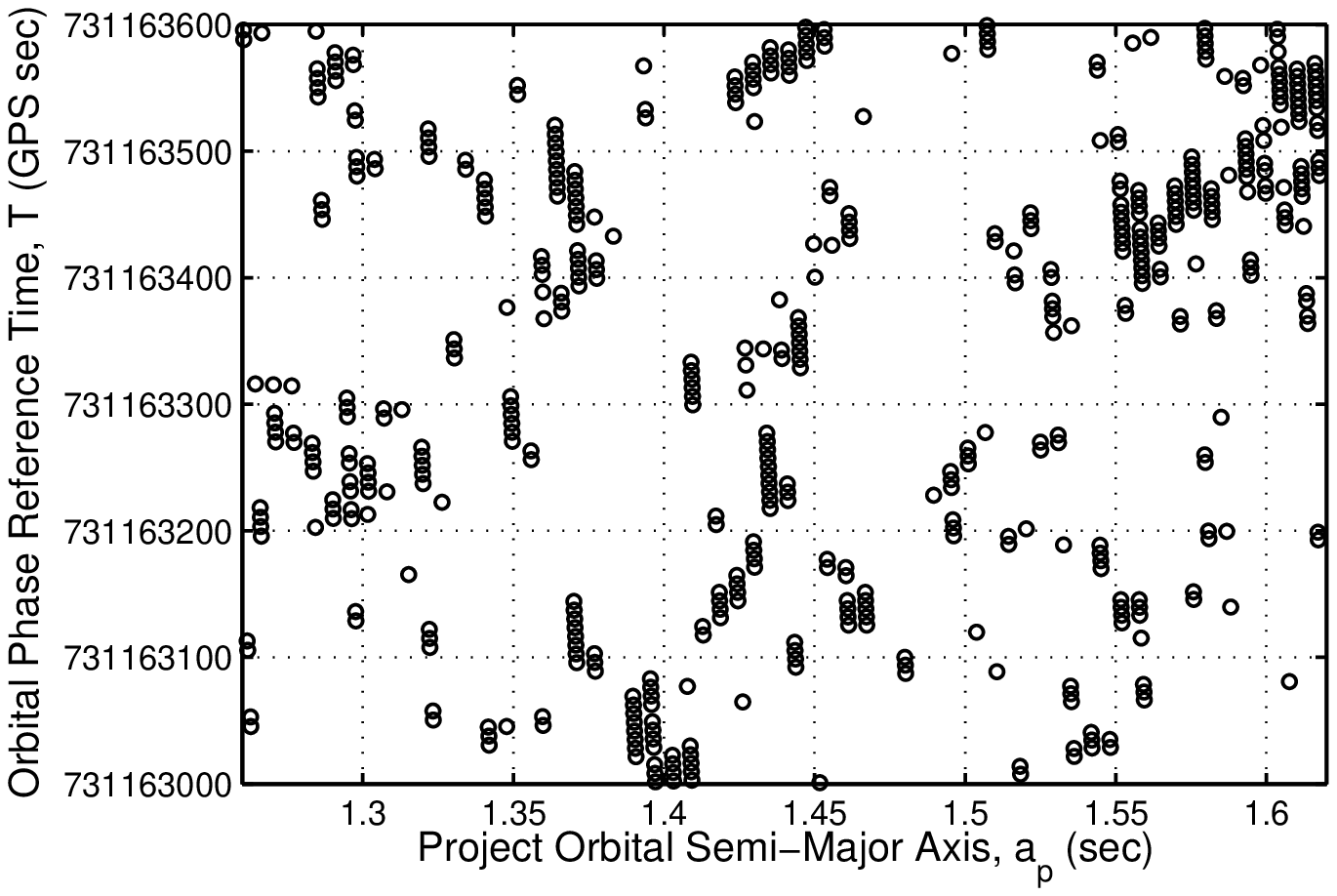}
\centering \includegraphics[width=8.3cm]{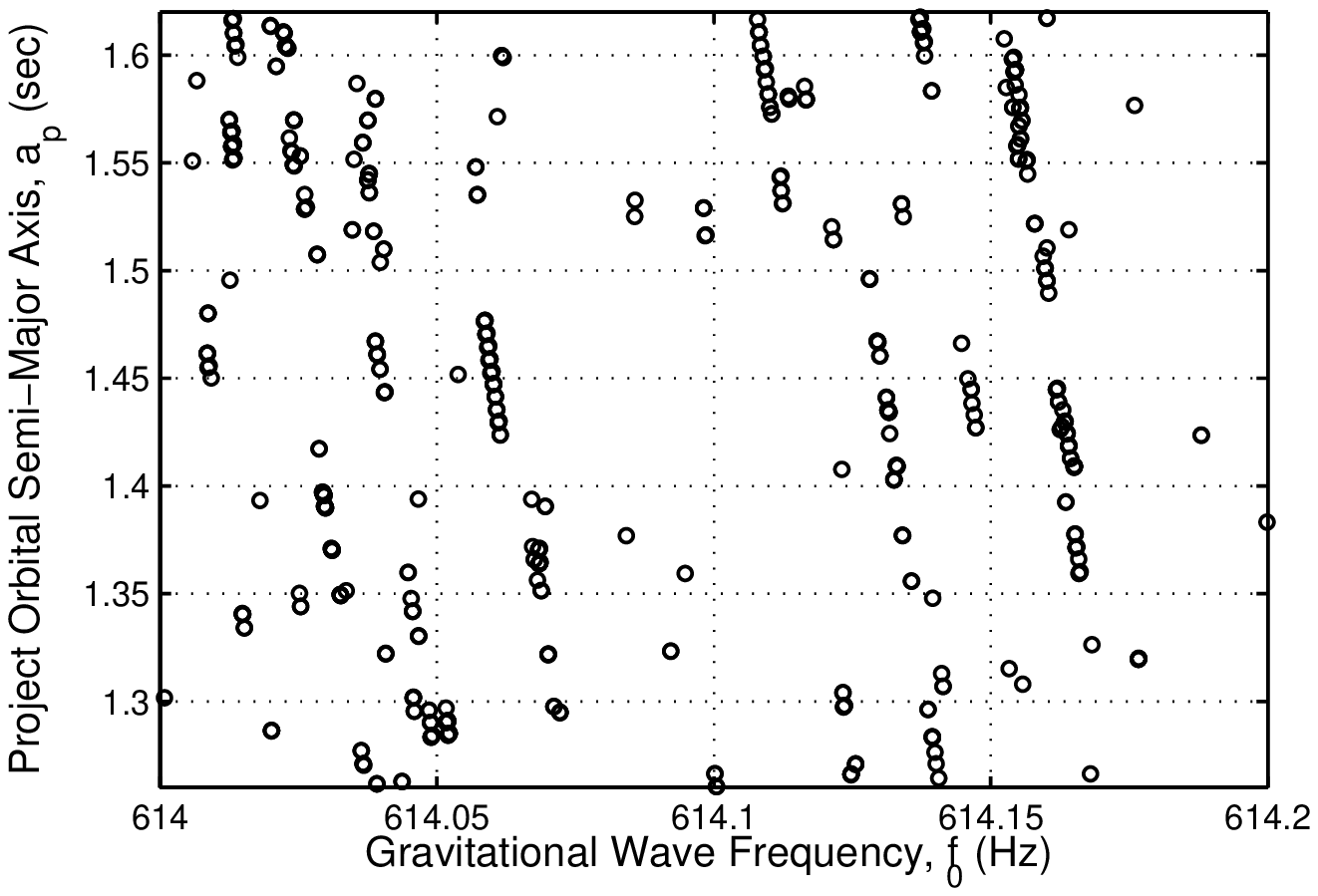}
\centering \includegraphics[width=8.3cm]{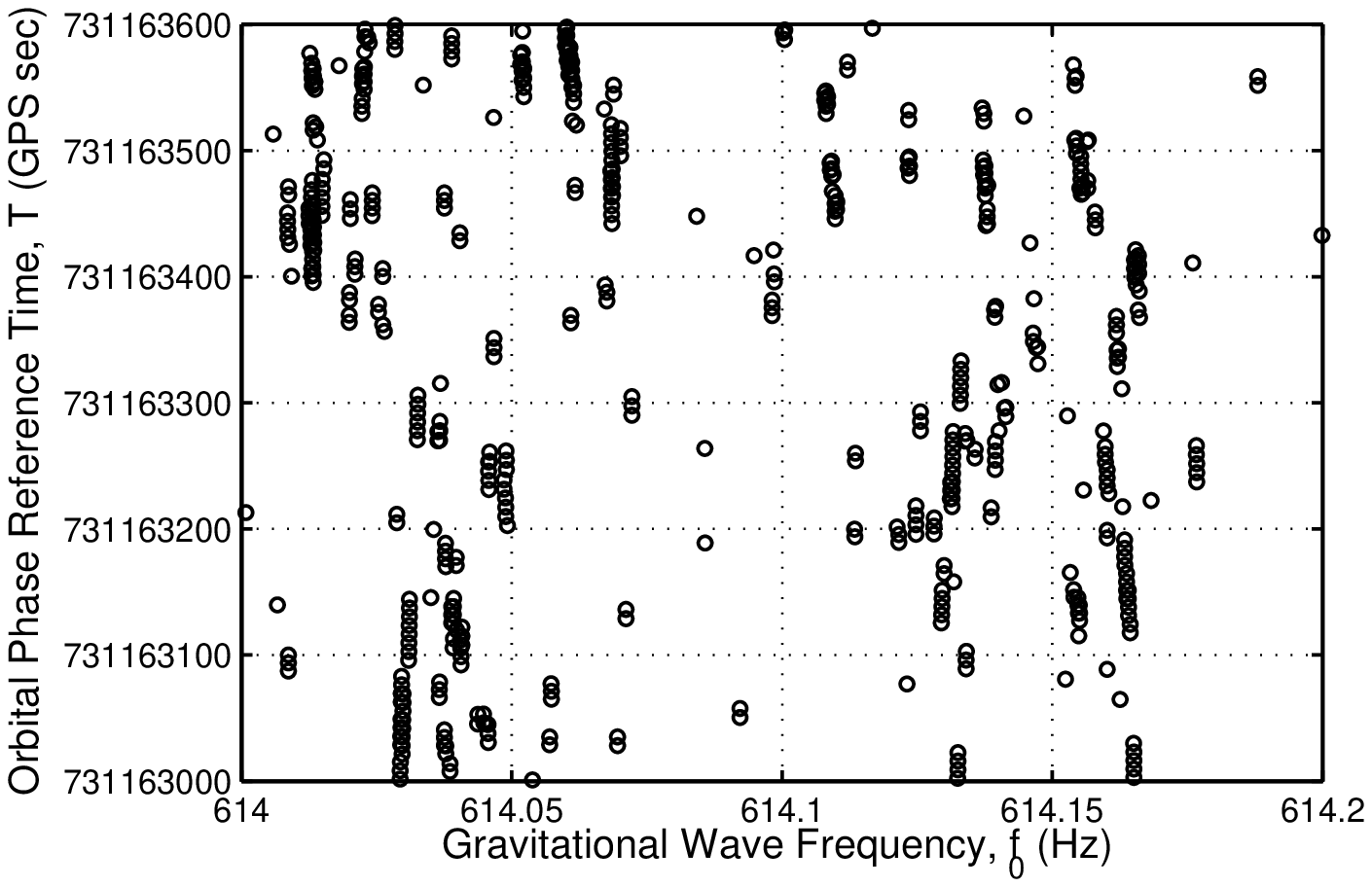}
\caption{The locations of coincident templates within a representative
subsection of the Sco X-1 parameter space. These plots contain events
found in the frequency band $614.0$--$614.2$~Hz and are representative
of a ``clean'' search band.  We show only events from the H1 detector
for clarity (L1 coincident events lie in approximately the same
locations).  We show three $2$-dimensional projections through the
$3$-dimensional search space.  The first plot shows the projected
orbital semi-major axis, $a_{\rm{p}}$, versus the orbital phase
reference time, $\bar{T}$, and shows no obvious structure.  The second
and third plots show $a_{\rm p}$ and $\bar{T}$, respectively, versus $f_{0}$.  
Here we do see structure caused by small narrow
lines present in the data that the demodulation process has failed to
smooth out over the relatively short $6$ hour observation time.}
\label{f:scoeventloc}
\end{figure}

Considering the limited sensitivity of the present analysis (see
Fig.s~\ref{f:ExpSens_sco}, \ref{f:SCO_UL} and \ref{f:SCO_UL2}) with
respect to the astrophysical predictions, Eq.~(\ref{scolimit}), we
have not followed up ({\em e.g.}\ using a longer integration time
$T_{\rm span}$) regions of the parameter space that yielded
particularly large values of $\F$.  Such a follow up would be
computationally very intensive and the fact that we are targeting a
continuous gravitational wave emitter allows to go back to the same
parameter space in the future, exploiting higher sensitivity, better
quality data, and a more sensitive search algorithm. This work is
already in progress.  It is however important to establish that the
results that we have obtained do not show any obvious unexplained
feature and are qualitatively consistent with the expectation that no
signal is present in the data set at the sensitivity level of the
search. In Fig.~\ref{f:scoeventloc} we show the distribution of the
parameters that characterize the filters in coincidence in L1 and H1
for a representative frequency band, 614 -- 614.2~Hz.  Due to the high
correlation of the templates used in the analysis one would expect a
cluster in parameter space of filters in coincidence, were a real
signal present.  Considering the coincident filters in the
3-dimensional search space ($f_0\,,a_{\rm{p}}\,,\bar{T}$) and
projecting them onto the plane $(a_{\rm{p}},\bar{T})$, no particular
structure is evident, with coincident templates evenly distributed
across the plane. This is also broadly consistent with the
distributions of coincident templates that we have obtained by
performing Monte Carlo simulations of the entire search pipeline on
stationary and Gaussian noise.  There is however some structure in the
plane $(a_{\rm{p}},\bar{T})$ that is determined by narrow spectral
disturbances and accounts for the large outliers in the values of the
detection statistic at the end of the analysis pipeline.  In order to
explore this, it is useful to project the same 3-dimensional parameter
space onto either the $(f_{0},a_{\rm{p}})$ or $(f_{0},\bar{T})$
plane. Structures are now clearly visible consisting of ``stripes'' of
events at approximately (but not exactly) constant frequency. These
are caused by small narrow spectral features present in the data that
produce relatively large values of the $\F$-statistic for a number of
orbital templates.  Due to the short coherent integration time, such
disturbances are not averaged out by the demodulation process and are
registered in the single detector search. We find that they are very
common in the output of the single detector search; the coincidence
stage of the analysis allows those ``stripes'' of events that exist in
both detectors at approximately the same frequency to survive the
entire pipeline.

We have so far reported the results of the analysis as upper limits on the signal amplitude $h_0$. We can now 
re-cast them as upper limits on the ellipticity $\epsilon$ of the neutron star, taking the distance of
Sco X-1 as $d = 2.8$ kpc; see Table~\ref{t:Sco-param}. Using 
\begin{equation}
  \epsilon \simeq 0.237\,\left(\frac{h_{0}}{10^{-24}}\right)\,
\left(\frac{d}{1~\,\mathrm{kpc}}\right)\,
\left(\frac{1~\,\mathrm{Hz}}{f}\right)^2\,
\left(\frac{10^{45}~\mathrm{g}\, ~\mathrm{cm}^{2}}{I_{\rm zz}}\right),
\end{equation}
and the canonical value for the principal moment of inertia $I_{\rm zz}=10^{45}\,\mathrm{g\,cm}^{2}$ we obtain:
\be
\begin{array}{lll}
\epsilon^{95\%} = & 4.0\times 10^{-4} - 3.6 \times 10^{-3} & {\rm {for}} \quad e \le 10^{-4}\,,
\end{array}
\label{e:epsilon_sco}
\ee
over the frequency band. The previous result can be generalised to the case of a more pronounced non zero eccentricity; {\em e.g.} we obtain
\be
\begin{array}{lll}
\epsilon^{88\%} = & 5.1 \times 10^{-4} - 3.7 \times 10^{-3} & {\rm {for}} \quad e = 10^{-3}\,.
\end{array}
\label{e:epsilon_sco_1}
\ee
Despite being far from 
astrophysically interesting, Eq.~(\ref{e:epsilon_sco}) and~(\ref{e:epsilon_sco_1}) represent the first direct measurements of the ellipticity of the neutron star in Sco X-1 in the relevant frequency band.

\section{Conclusions}
\label{s:concl}

We have presented here results from two coherent wide parameter space searches for continuous gravitational wave signals. A subset of data from the second science run of the LIGO instruments was analyzed, the data chosen to maximize the sensitivity of the search. Two different astrophysical searches were performed: an all-sky search aimed at signals from isolated neutron stars and an orbital parameter search aimed at signals from the neutron star in the binary system Sco X-1. Both searches also cover a wide range of possible emission frequencies: a 568.8-Hz band for the isolated pulsars search and two 20-Hz bands for the Sco X-1 search.

The sensitivity of these analyses makes the detection of a signal extremely unlikely. As a consequence the main goal of the paper is to 
demonstrate an analysis method using real data, with a pipeline considerably more complex than any other coherent searches previously performed. 
More importantly, this coherent search will be deployed in a hierarchical analysis scheme.  The first step of a hierarchical analysis sets the ultimate sensitivity of the search: candidates that do not survive the first threshold are lost. It is thus crucial to employ the most sensitive possible technique in this first step. The coherent method described in this paper provides an implementation of such a first step. 

Overall, hierarchical approaches are expected to achieve optimal sensitivity at constrained computational resources. We will employ such approaches for deep searches on long duration and high sensitivity data such as those that are now being recorded by the instruments. 
With one year of data at the design sensitivity of the detectors, the improvements that we can expect by means of hierarchical schemes that utilize this type of coherent analysis as one of the steps is of order $10$ with respect to what was presented here.

The most constraining 95\% confidence $h_0$ upper limit from the
all-sky search is $6.6\times 10^{-23}$ in the band $245.2$--$246.4$ Hz,
reflecting the highest sensitivity of the instruments at these
frequencies. This is still a factor of $\sim 16$ higher than the
strongest signal that we expect based on the optimistic (but not
unreasonable) assumptions of the statistical argument presented in
Sec.~\ref{ss:maximum_expected_amplitude}.

The 95\% confidence upper limits from the Sco X-1 search, assuming a non-eccentric orbit, are $h_0\approx 2 \times 10^{-22}$; the most stringent 95\% confidence upper limits from the Sco X-1 search, assuming a non-eccentric orbit, are $h_0  = 1.7\times 10^{-22}$ in the 464 -- 484~Hz band and $h_0 = 2.2\times 10^{-22}$ in the 604 -- 624~Hz frequency band. The Sco X-1 results presented here are the first direct gravitational wave upper limits placed on the system.

Coherent all-sky searches for continuous signals from isolated stars have been performed in the past, but over much smaller parameter space. In \cite{astone:05} an all-sky 0.76 Hz band search was performed around 921.38 Hz, including spin-down parameters in the range $-2.36 \times 10^{-8}$ Hz s$^{-1}$ to $+ 2.36 \times 10^{-8}$ Hz s$^{-1}$. Three data sets, each $48$ hours long, were coherently analyzed and a $90\%$ confidence upper limit was placed at the level of $1.0\times 10^{-22}$ based on the the cleanest of the data sets.

In the context of a hierarchical search aimed at detecting a signal, data cleaning procedures to remove noise artifacts are likely to be employed. This is in contrast to the approach used in the analysis reported in this paper where no cleaning at all was considered. Here we wanted to investigate the effect of noisy data segments with a variety of artifacts on the output of a search pipeline; we have purposely kept the ``bad'' data in the presentation of the results for illustration purposes. This is relevant to future searches because the upfront ``cleaning'' of known noise artifacts does not guarantee that longer observation times will not uncover unknown periodicities. These must then be either identified as of instrumental origin, as done here, or followed up.

Longer observation times mean a higher resolution in parameter space and higher computational costs. It is thus important to lay the template banks in a way that takes advantage of the correlations in parameter space. The Sco X-1 search presented here already does this. In future work the metric approach will be used also for template placement for the searches for signals from isolated sources. However, longer observation times and targeting different LMXB systems will require more sophisticated template placement strategies than the one presented here for Sco X-1. For example the parameter space may grow to include 
the orbital period, the eccentricity and the spin period derivatives.  For any specific 
source the number of search parameters will be defined by the precision
to which these source parameters have been measured via electromagnetic observations.

Redundant template grids produce redundant events. This increases the false alarm rate and in practice reduces the sensitivity of the search; in fact the threshold on signal-to-noise ratio that defines the candidates to follow up depends by how many follow-ups one can afford with given computational resources. It is therefore important for future work to develop techniques to recognize non-independent candidates in parameter space, rank them and keep the information on only the most significant. In this paper we have made the first moves in this direction with the algorithm that identifies as a single candidate values of the $\F$ statistic which are ``near'' to each other in search frequency. The concept must be generalized to the multi-dimensional space of the search parameters, and ultimately connected to the global properties of the detection statistic over the parameter space.

\section{Acknowledgments}

The authors gratefully acknowledge the support of the United States National 
Science Foundation for the construction and operation of the LIGO Laboratory 
and the Particle Physics and Astronomy Research Council of the United Kingdom, 
the Max-Planck-Society and the State of Niedersachsen/Germany for support of 
the construction and operation of the GEO600 detector. The authors also 
gratefully acknowledge the support of the research by these 
agencies and by the Australian Research Council, the Natural Sciences and Engineering Research 
Council of Canada, the Council of Scientific and Industrial Research of India, 
the Department of Science and Technology of India, the Spanish Ministerio de 
Educacion y Ciencia, The National Aeronautics and Space Administration, 
the John Simon Guggenheim Foundation, the Alexander von Humboldt Foundation, 
the Leverhulme Trust, the David and Lucile Packard Foundation, 
the Research Corporation, and the Alfred P. Sloan Foundation. This document has
been assigned LIGO Laboratory document number LIGO-P050008.

\end{document}

%% file: ourdefs.tex
\def\bea{\begin{eqnarray}}
\def\eea{\end{eqnarray}}
\def\beq{\begin{equation}}
\def\eeq{\end{equation}}
\def\be{\begin{equation}}
\def\ee{\end{equation}}

\newcommand{\ba}{\begin{eqnarray}}
\newcommand{\ea}{\end{eqnarray}}

\newcommand{\bml}{\begin{mathletters}}
\newcommand{\eml}{\end{mathletters}}

%% file: authorlist.tex
%
%
%
\newcommand*{\AG}{Albert-Einstein-Institut, Max-Planck-Institut f\"ur Gravitationsphysik, D-14476 Golm, Germany}
\affiliation{\AG}
\newcommand*{\AH}{Albert-Einstein-Institut, Max-Planck-Institut f\"ur Gravitationsphysik, D-30167 Hannover, Germany}
\affiliation{\AH}
\newcommand*{\AN}{Australian National University, Canberra, 0200, Australia}
\affiliation{\AN}
\newcommand*{\CH}{California Institute of Technology, Pasadena, CA  91125, USA}
\affiliation{\CH}
\newcommand*{\DO}{California State University Dominguez Hills, Carson, CA  90747, USA}
\affiliation{\DO}
\newcommand*{\CA}{Caltech-CaRT, Pasadena, CA  91125, USA}
\affiliation{\CA}
\newcommand*{\CU}{Cardiff University, Cardiff, CF2 3YB, United Kingdom}
\affiliation{\CU}
\newcommand*{\CL}{Carleton College, Northfield, MN  55057, USA}
\affiliation{\CL}
\newcommand*{\CO}{Columbia University, New York, NY  10027, USA}
\affiliation{\CO}
\newcommand*{\HC}{Hobart and William Smith Colleges, Geneva, NY  14456, USA}
\affiliation{\HC}
\newcommand*{\IU}{Inter-University Centre for Astronomy  and Astrophysics, Pune - 411007, India}
\affiliation{\IU}
\newcommand*{\CT}{LIGO - California Institute of Technology, Pasadena, CA  91125, USA}
\affiliation{\CT}
\newcommand*{\LM}{LIGO - Massachusetts Institute of Technology, Cambridge, MA 02139, USA}
\affiliation{\LM}
\newcommand*{\LO}{LIGO Hanford Observatory, Richland, WA  99352, USA}
\affiliation{\LO}
\newcommand*{\LV}{LIGO Livingston Observatory, Livingston, LA  70754, USA}
\affiliation{\LV}
\newcommand*{\LU}{Louisiana State University, Baton Rouge, LA  70803, USA}
\affiliation{\LU}
\newcommand*{\LE}{Louisiana Tech University, Ruston, LA  71272, USA}
\affiliation{\LE}
\newcommand*{\LL}{Loyola University, New Orleans, LA 70118, USA}
\affiliation{\LL}
\newcommand*{\MP}{Max Planck Institut f\"ur Quantenoptik, D-85748, Garching, Germany}
\affiliation{\MP}
\newcommand*{\MS}{Moscow State University, Moscow, 119992, Russia}
\affiliation{\MS}
\newcommand*{\ND}{NASA/Goddard Space Flight Center, Greenbelt, MD  20771, USA}
\affiliation{\ND}
\newcommand*{\NA}{National Astronomical Observatory of Japan, Tokyo  181-8588, Japan}
\affiliation{\NA}
\newcommand*{\NO}{Northwestern University, Evanston, IL  60208, USA}
\affiliation{\NO}
\newcommand*{\SC}{Salish Kootenai College, Pablo, MT  59855, USA}
\affiliation{\SC}
\newcommand*{\SE}{Southeastern Louisiana University, Hammond, LA  70402, USA}
\affiliation{\SE}
\newcommand*{\SA}{Stanford University, Stanford, CA  94305, USA}
\affiliation{\SA}
\newcommand*{\SR}{Syracuse University, Syracuse, NY  13244, USA}
\affiliation{\SR}
\newcommand*{\PU}{The Pennsylvania State University, University Park, PA  16802, USA}
\affiliation{\PU}
\newcommand*{\TC}{The University of Texas at Brownsville and Texas Southmost College, Brownsville, TX  78520, USA}
\affiliation{\TC}
\newcommand*{\TR}{Trinity University, San Antonio, TX  78212, USA}
\affiliation{\TR}
\newcommand*{\HU}{Universit{\"a}t Hannover, D-30167 Hannover, Germany}
\affiliation{\HU}
\newcommand*{\BB}{Universitat de les Illes Balears, E-07122 Palma de Mallorca, Spain}
\affiliation{\BB}
\newcommand*{\BR}{University of Birmingham, Birmingham, B15 2TT, United Kingdom}
\affiliation{\BR}
\newcommand*{\FA}{University of Florida, Gainesville, FL  32611, USA}
\affiliation{\FA}
\newcommand*{\GU}{University of Glasgow, Glasgow, G12 8QQ, United Kingdom}
\affiliation{\GU}
\newcommand*{\MA}{University of Maryland, College Park, MA 20742, USA }
\affiliation{\MA}
\newcommand*{\MU}{University of Michigan, Ann Arbor, MI  48109, USA}
\affiliation{\MU}
\newcommand*{\OU}{University of Oregon, Eugene, OR  97403, USA}
\affiliation{\OU}
\newcommand*{\RO}{University of Rochester, Rochester, NY  14627, USA}
\affiliation{\RO}
\newcommand*{\UW}{University of Wisconsin-Milwaukee, Milwaukee, WI  53201, USA}
\affiliation{\UW}
\newcommand*{\VC}{Vassar College, Poughkeepsie, NY 12604}
\affiliation{\VC}
\newcommand*{\WU}{Washington State University, Pullman, WA 99164, USA}
\affiliation{\WU}

\author{B.~Abbott}    \affiliation{\CT}
\author{R.~Abbott}    \affiliation{\LV}
\author{R.~Adhikari}    \affiliation{\CT}
\author{A.~Ageev}    \affiliation{\MS}  \affiliation{\SR}
\author{J.~Agresti}    \affiliation{\CT}
\author{B.~Allen}    \affiliation{\UW}
\author{J.~Allen}    \affiliation{\LM}
\author{R.~Amin}    \affiliation{\LU}
\author{S.~B.~Anderson}    \affiliation{\UW}
\author{W.~G.~Anderson}    \affiliation{\TC}
% ...

 \collaboration{The LIGO Scientific Collaboration, http://www.ligo.org}
 \noaffiliation
%
%